\documentclass[12pt]{article}
\usepackage{a4wide}
\usepackage{amsmath,amssymb,bbm}
\usepackage{multirow}

\begin{document}
\setcounter{page}{1}
%


%

\def\pct#1{(see Fig. #1.)}

\begin{titlepage}
\hbox{\hskip 12cm KCL-MTH-09-05  \hfil}
\vskip 1.4cm
\begin{center}  {\Large  \bf    The $E_{11}$ origin of all maximal
supergravities}\\
\vspace{.6cm} {\large  \bf    The hierarchy of field-strengths}
\vspace{1.8cm}

{\large \large Fabio Riccioni , \ Duncan Steele \ and \ Peter West}
\vspace{0.7cm}

{\sl Department of Mathematics\\
\vspace{0.3cm}
 King's College London  \\
\vspace{0.3cm}
Strand \ \  London \ \ WC2R 2LS \\
\vspace{0.3cm} UK}
\end{center}
\vskip 1.5cm

\abstract{Starting from $E_{11}$ and  the space-time translations we
construct an algebra that promotes the global $E_{11}$ symmetries to
local ones, and consider all its possible massive deformations. The
Jacobi identities imply that such deformations are uniquely
determined by a single tensor that belongs to the same
representation of the internal symmetry group as the $D-1$ forms
specified by $E_{11}$. The non-linear realisation of the deformed
algebra gives the field strengths of the theory which are those of
any possible gauged maximal supergravity theory in any dimension.
All the possible deformed algebras are in one to one correspondence
with all the possible massive maximal supergravity theories. The
hierarchy of fields inherent in the $E_{11}$ formulation plays an
important role in the derivation. The tensor that determines the
deformation can be identified with the embedding tensor used
previously to parameterise gauged supergravities. Thus we provide a
very efficient, simple and unified derivation of all the field
strengths and gauge transformations of all maximal gauged
supergravities from $E_{11}$. The dynamics arises as a set of first
order duality relations among these field strengths.}

\vfill
\end{titlepage}
\makeatletter \@addtoreset{equation}{section} \makeatother
\renewcommand{\theequation}{\thesection.\arabic{equation}}
\addtolength{\baselineskip}{0.3\baselineskip}

\section{Introduction}
The maximal supergravity theories have played a key role in our
understanding of string theory. The gauged supergravity theories
have been studied for 25 years beginning with the  first paper
\cite{dewitnicolai} which found an $SO(8)$ gauged theory within the
$D=4$, $N=8$ theory. These theories are sometimes called massive
theories in that they are a deformation of the massless theory by a
massive parameter. They have generally been  found by starting from
the massless supergravity theory in the dimension of interest and
adding a deformation to the action such as a cosmological constant
or a non-abelian interaction for the vectors and using supersymmetry
closure to complete the theory.  In relatively recent years  all
maximal gauged supergravity theories in each dimension $D$ has been
classified in terms of a single object called the embedding tensor
which can be thought of as belonging to a representation of the
internal symmetry group  of the supergravity theory in $D$
dimensions
\cite{nicolaisamtlebenD=3,dWSTgeneral,dWSTD=5,samtlebenweidnerD=7,dWSTmagnetic,dWSTD=4,embeddingtensorD=6}.
Thus for example all the gauged supergravity theories in five
dimensions are parametrised modulo further constraints by an
embedding tensor in the ${\overline{\bf 351}}$ of the symmetry group
$E_6$.
\par
Certain gauged supergravities have played an important role in more
recent developments. Two of the most important examples are the five
dimensional gauged supergravity theory which results from
dimensionally reducing the ten dimensional IIB supergravity theory
on $S^5$ which is central to the AdS/CFT conjecture and those
theories that occur in flux compactifications with a view to moduli
stabilisation. However, it is fair to say that  gauged
supergravities in general have not been fitted into any conventional
discussions of M theory.
\par
It was conjectured in 2001 that the theory underlying string theory
should possess an $E_{11}$ symmetry and indeed the non-linear
realisation of this symmetry contained the eleven dimensional
supergravity theory \cite{peterE11first}. By taking different
decompositions of $E_{11}$ one finds different supergravity
theories. In particular, to find the theory in $D$ dimensions one
performs the decomposition of $E_{11}$ into $GL(D,\mathbb{R})\otimes
G$ which corresponds to the algebra remaining after deleting the
$D$th node of the $E_{11}$ Dynkin diagram. In particular in  ten
dimensions one finds two theories which have at low levels precisely
the content of the IIA and IIB supergravity theories
\cite{peterE11first,igorpeterIIB}. Moreover, the Romans theory was
found to be a non-linear realisation \cite{igorpeterromans} which
includes all form fields up to and including a 9-form with a
corresponding set of generators. This 9-form is automatically
encoded in the non-linear realisation of $E_{11}$
\cite{axeligorpeter}, and its 10-form field-strength is dual to
Romans cosmological constant.

More recently the from fields, that is those field with only
completely anti-symmetrised Lorentz indices, were found in all
dimensions, $D$ \cite{fabiopeterE11origin,ericembeddingtensor}.
These include the $D-1$ forms in the $D$-dimensional theory whose
equation of motion generically leads to a cosmological constant. As
such the number of such forms should correspond to the number of
gauged supergravity theories and indeed the representation of the
$D-1$ forms is precisely the same as that of the embedding tensor
used to classify the gauged supergravity theories. It  was therefore
apparent that $E_{11}$ encoded all the possible maximal gauged
supergravity theories. Thus for the first time the gauged
supergravities were included in some underlying unifying formulation
rather that found as the possible massive deformations in each
dimension.
\par
A feature that  is always present in the $E_{11}$ theories in
different dimensions is that every form field has a corresponding
dual field, indeed if the $n$-form fields belong to the
representation ${\bf R_n}$ then we also find $D-n-2$-dual form
fields in the complex conjugate representation, {\it i.e.}  ${\bf
R_{D-n-2} = \overline{R}_n}$. This was already apparent in the case
of eleven dimensions and the IIA and IIB theories
\cite{peterE11first,igorpeterIIB}. As mentioned above the rank $D-1$
forms are dual to a cosmological constant while the rank $D$ forms
are not dual to anything but play an important role in brane
associated dynamics. This can be thought of  as a hierarchy of
fields of ascending rank.
\par
The results in the two paragraphs above are of a purely kinematical
nature, however, progress has been made in constructing the dynamics
of gauged supergravity theories using $E_{11}$. Initially this was
achieved using the so called $l_1$ representation
\cite{peterl1multiplet} to provide an $E_{11}$ covariant generalised
space-time \cite{fabiopeterextendedspacetime}. While wishing to
continue with this approach at a future date we also pursued an
alternative more bottom up approach introducing only the usual
$D$-dimensional space-time, with its corresponding space-time
translations operator and at the same time extending the $E_{11}$
algebra to include generators that  had the effect of making local
all the rigid Borel $E_{11}$ transformations
\cite{fabiopeterogievetsky}. These so called Ogievetsky generators
lead in the non-linear realisation to fields that can be eliminated
covariantly and do not appear in the final dynamics. Nonetheless
they play a crucial role in determining the field strengths of all
the fields. Therefore, the algebra formed by the non-negative level
$E_{11}$ generators, the $D$-dimensional space-time translation
generator and the above mentioned Ogievetsky generators, called
$E_{11,D}^{local}$ in \cite{fabiopeterogievetsky}, determines the
field strengths of the massless maximal supergravity theories in any
dimension. It also emerged in \cite{fabiopeterogievetsky} that in
the case of gauged supergravities the final dynamics is controlled
by a massive deformation $\tilde{E}_{11,D}^{local}$ of the algebra
$E_{11,D}^{local}$, in which the deformed $E_{11}$ generators have a
non-trivial commutation relation with the momentum operator. This
was shown in detail for the case of the Scherk-Schwarz reduction of
IIB to nine dimensions, the five-dimensional gauged maximal
supergravity and Romans massive IIA theory. In the first case it was
also shown that $\tilde{E}_{11,9}^{local}$ is a subalgebra of the
algebra $E_{11,10B}^{local}$ that describes the IIB theory in ten
dimensions, while the last case reproduced the results of
\cite{igorpeterromans}, where the field strengths of the Romans
theory were constructed adopting a non-trivial commutator between
the $E_{11}$ generators and momentum.

We note that gauged supergravities have also been discussed from the
$E_{10}$ \cite{E10original} viewpoint. In particular the case of
Romans IIA was discussed in \cite{E10romans} while the case of
maximal gauged supergravity in three dimensions was analysed in
\cite{E103dim}.
\par
In this paper we continue the analysis of
\cite{fabiopeterogievetsky} and construct all the massive
deformations in each dimension. We find that the underlying $E_{11}$
algebra and the Jacobi identities imply that the deformations in a
given dimension are uniquely determined by one object that belongs
to the same representation as the $D-1$ form generators and so can
be identified with the embedding tensor used previously to classify
gauged supergravity theories.  We use the algebra to construct in a
simple way all the fields strengths of all the gauged supergravities
in all dimensions. The dynamics then arises as first order equations
that are duality relations among these field strengths. In
particular, the scalar equation results from the curl of the duality
relation between the $D-2$-form fields and the scalars, using also
the fact that the $D-1$-form field is dual to the embedding tensor.
In general there is more than one gauge covariant quantity that one
can construct contracting the scalars with the embedding tensor, and
this procedure does not determine their relative coefficient, and
therefore does not determine the exact form of the scalar potential.
We analyse each dimension from three to nine, and these results,
together with the ten-dimensional deformation corresponding to the
Romans theory analysed in \cite{fabiopeterogievetsky}, give the
field strengths of all possible massive maximal supergravities in
any dimension.

The paper is organised as follows. In section 2 we derive the
general method of constructing the deformed $E_{11}$ algebra in any
dimension. In section $D$, with $D=3, ...., 9$, we explicitly derive
the deformed algebra in a given dimension $D$. In section 10 we
discuss the form of the duality relations that the various field
strengths must satisfy in any dimension, and section 11 contains the
conclusions. The paper also contains three appendices. In appendix A
we review some group theoretic techniques and we derive from
$E_{11}$ the relevant projection formulae used in the paper. In
appendix B we explicitly evaluate the field strengths in the general
notation of section 2. In appendix C we derive the field strengths
of the four-dimensional theory using a different method, that is
based on the non-linear realisation of $E_{11} \otimes_s l_1$ and
applies in four dimensions the analysis that was carried out in
\cite{fabiopeterextendedspacetime} in the five-dimensional case.

\section{The general method}
We wish to consider the formulation of the $E_{11}$ algebra
appropriate to $D$ dimensions which can be found by decomposing
$E_{11}$ with respect to the algebra that results from deleting the
$D$th node of the $E_{11}$ Dynkin diagram. This resulting algebra is
$SL(D, \mathbb{R})\otimes G$ where $SL(D,\mathbb{R})$ is associated
with $D$-dimensional gravity and $G$ is the internal symmetry
algebra. The resulting form generators, that is those with only
anti-symmetric Lorentz  indices, are explicitly given in the later
sections. In this section,  we are interested in a universal
treatment valid for every dimension and so we introduce a
corresponding notation. We denote  the form generators  as
$R^{a_1\ldots a_n, M_n}$ and the generators with no Lorentz indices
are written as $R^\alpha$. The latter generators are those of  the
internal symmetry algebra,  $G$ and the generators $R^{a_1\ldots
a_n, M_n}$ carry the representation ${\bf R_n}$ of $G$ which
transforms the $M_n$ indices.  We note that in this notation ${\bf
R_0}$ is the adjoint representation. For example, in the case of
five dimensions $G=E_6$ and the form generators are given in eq.
(\ref{listofgeninD=5}).
\par
The $E_{11}$ algebra involving the form generators is then given by
  \begin{equation}
  [R^{a_1\ldots a_m, M_m},R^{b_1\ldots b_n,
  N_n}]=f^{M_mN_n}{}_{P_{n+m}} R^{a_1\ldots a_{m}b_1 \ldots b_n,{P_{n+m}} }
  \label{2.1}
  \end{equation}
and
  \begin{equation}
  [ R^\alpha, R^{a_1\ldots a_m, M_m} ]= (D^\alpha)_{N_m}{}^{M_m}
  R^{a_1\ldots a_m, N_m},\ \
  [R^\alpha,R^\beta]=f^{\alpha\beta}{}_\gamma R^\gamma \quad ,\label{2.2}
  \end{equation}
where $f^{M_mN_n}{}_{P_{n+m}}$ are generalised structure constants
whose form will be shown in the following in several examples. By
studying the table of \cite{fabiopeterE11origin} of forms contained
in $E_{11}$, which is table \ref{alwaysthesametable} in this paper
(observe that the table contains the representations of the fields,
which are the complex conjugate of the representations of the
corresponding generators), one finds that in all dimensions the
representation of the 4-form generators is in the anti-symmetric
tensor representation formed from two representations ${\bf R_2}$
and so it has  indices $[ M_2 N_2 ]$. As such we may write the
4-form generators as $R^{a_1a_2b_1b_2 , M_2N_2}=R^{a_1a_2b_1b_2 , [
M_2N_2 ]}$; in terms of our general notation the indices $M_4$ are
for this generators represented by $[M_2N_2]$. As a result the
commutator of two 2-form generators can be written as
 \begin{equation}
  [ R^{a_1a_2 M_2}, R^{b_1b_2 N_2}]= R^{a_1a_2b_1b_2 , M_2N_2}
  \quad ,  \label{2.3}
  \end{equation}
where we have taken the constant of proportionality in the
commutator to be one as this commutation relation can be taken to
define the way the four form generator appears in the $E_{11}$
algebra. The particularly simple form of this commutator will prove
to be useful in this paper. Some other related observations that
will be useful are that ${\bf R_1\otimes R_{n}}$ contains the
representation ${\bf R_{n+1}}$ and that ${\bf
R_n=\overline{R}_{D-n-2}}$ for $n\not=D-1,D$. The first implies that
one can find all form generators by taking repeated commutators of
the one form generators and the second reflects that in the $E_{11}$
formulation one finds dual fields for all the form fields  usually
associated with the physical degrees of freedom of the theory.
Taking $n=0$ we find that ${\bf R_{D-2}=R_0}$ which is the adjoint
representation and so it is real.
\par
In fact the above algebra contains  the $E_{11}$ form generators
that have positive level with respect to the level associated with
node deletion discussed above. It is  a truncation of the $E_{11}$
algebra to contain just these generators. Clearly,  eqs. (\ref{2.1})
and (\ref{2.2}) obey certain Jacobi identities which imply, for
example, that the structure constants are invariant tensors of the
internal symmetry group $G$. However, the structure constants also
obey restrictions resulting from their $E_{11}$ origin. These result
from the Jacobi identities, but also from the construction of
$E_{11}$ from its Chevalley generators. In particular, the left hand
side of the commutator of eq. (\ref{2.1}) implies that the form
generators on the right hand side must belong to the ${\bf
R_n\otimes R_m}$ representation of $G$, however,  only the ${\bf
R_{n+m}}$ representation arises. As a result, the structure constant
$ f^{M_mN_n}{}_{P_{n+m}}$ must obey the conditions that project onto
only this latter representation. A particular example, that will be
important for what follows,  is the case for $m=1$ and $n=D-2$ whose
corresponding commutator has the form
  \begin{equation}
  [R^{a_1, N_1},R^{a_2\ldots a_{D-1}, \alpha}]= f^{N_1
  \alpha}{}_{M_{D-1}}R^{a_1 a_2\ldots a_{D-1}, M_{D-1}} \label{2.4}
  \end{equation}
where the generator on the right hand side corresponds to the next
to space-filling form fields that give rise in the non-linear
realisation to   the cosmological constant. Here we have used that
${\bf R_{D-2}}$ is the adjoint representation and so is labelled by
$\alpha, \beta,\ldots$. For the cases of $D=4,5,6$ {\it i.e.} $E_7$
and $E_6$ and $E_5\equiv D_5$ the ${\bf R_1\otimes R_{adj}}$
contains three irreducible representations, only one of which is the
representation to which the next to space-filling generators belong.
For the other dimensions one finds more representations in the
tensor product, but in all cases there are two or more
representations in the tensor product that must be projected out to
find the representation, or for $D=9,8,7,3$ the two presentations to
which the next to space-filling generators belong ( see table
\ref{tablewithembeddingtensorinvariousdim}). As such the structure
constants $f^{N_1 \alpha}{}_{N_{D-1}}$ must obey at least two
projections conditions that turn out to be of the form
  \begin{equation}
  (D_\alpha)_{N_1}{}^{M_1} f^{N_1 \alpha}{}_{N_{D-1}}=0,\ \ \ (D^\beta
  D_\alpha)_{N_1}{}^{M_1} f^{N_1 \alpha}{}_{N_{D-1}}=c f^{M_1
  \beta}{}_{N_{D-1}} \label{2.5}
  \end{equation}
for a suitable constant $c$. Such projector conditions are discussed
in more detail in appendix A.
\par
To the  $E_{11}$ algebra we add, as explained in reference
\cite{fabiopeterogievetsky}, the space-time  translation operator
$P_a$ and an infinite number of so called  Ogievetsky generators. In
fact for our purposes we need only add the lowest order such
generators, $K^{a,b_1\ldots b_n, M_n}$, which by definition obey the
commutator
  \begin{equation}
  [ K^{a,b_1\ldots b_n, M_n},P_c ]= \delta_c^{a}R^{b_1\ldots b_n, M_n}
  -\delta_c^{[a}R^{b_1\ldots b_n ], M_n}\quad . \label{2.6}
  \end{equation}
As this equation makes clear the generator $K^{a,b_1\ldots b_n,
M_n}$ is associated with the $E_{11}$ generators $R^{b_1\ldots b_n,
M_n}$ and carries the same internal symmetry representation, ${\bf
R_n}$. It also satisfies $K^{ [a,b_1\ldots b_n ], M_n }=0$.  The
Ogievetsky generators rotate into  themselves under the action of
the $E_{11}$ generators and the commutator of two Ogievetsky
generators gives another Ogievetsky generator. We take the
space-time translation operator $P_a$ to commute with the positive
level generators of $E_{11}$. Indeed, it is this requirement that
forces us to consider only the positive level generators of
$E_{11}$.
\par
We now consider a massive deformation of the above algebra which is
parameterised by the symbol $g$ and given by
  \begin{eqnarray}
  & & [R^{a_1\ldots a_m, M_m},R^{b_1\ldots b_n,
  M_n}]=f^{M_mN_n}{}_{P_{n+m}} R^{a_1\ldots a_{m}b_1 \ldots
  b_n,{P_{n+m}}
  }\nonumber \\
  & & \quad \qquad +g L^{M_m N_n }{}_{P_{n+m-1}}K^{[a_1,a_2\ldots a_m]
  b_1\ldots b_n, P_{n+m-1}} \label{2.7}  \\
  & &  [R^\alpha, R^{a_1\ldots a_m, M_m} ]= (D^\alpha)_{N_m}{}^{M_m}
  R^{a_1\ldots a_m, N_m} \label{2.8}  \\
  & &  [ R^{a_1\ldots a_m, M_m}, P_c ]= -g
  W^{M_m}{}_{M_{m-1}}\delta^{[a_1}R^{a_2\ldots a_m ], M_{m-1}}
  \label{2.9} \\
  & &  [ K^{a,b_1\ldots b_n, M_n},P_c ]= \delta_c^{a}R^{b_1\ldots b_n,
  M_n} -\delta_c^{[a}R^{b_1\ldots b_n ], M_n} +g
  U^{M_n}{}_{M_{n-1}}\delta_c^{[b_1} K^{ |a| b_2\ldots b_n], M_{n-1}}
  \label{2.10}  ,
  \end{eqnarray}
while the deformation of eq. (\ref{2.3}) for the commutator of two
two forms is given by
 \begin{equation}
[ R^{a_1a_2 M_2}, R^{b_1b_2 N_2}]= R^{a_1a_2b_1b_2 , M_2N_2}+ g
V^{M_2N_2}{}_{P_3}K^{[a_1 , a_2]b_1b_2,P_3} \quad .\label{2.11}
\end{equation}
For the case of the 4-form generator  eq. (\ref{2.9}) can be written
as
 \begin{equation}
[ R^{a_1 \ldots a_4, M_2N_2}, P_c ]= -g
W^{M_2N_2}{}_{P_3}\delta^{[a_1}R^{a_2\ldots a_4], P_{3}} \quad .
\label{2.12}
\end{equation}
The above commutators preserve  the grading $[R]=0$, $[P]=-1$,
$[K]=1$ provided we also assign $[g]=-1$ to the constant $g$. For
each set of objects $W$ we find a different deformation of the
$E_{11}$ algebra. The deformed algebra of eq. (\ref{2.7}) to
(\ref{2.12}) is the general version of that given  in
\cite{fabiopeterogievetsky} for special cases such as that for the
the gauged nine-dimensional supergravity that arises from
Scherk-Schwarz reduction of IIB, gauged five-dimensional maximal
supergravity and Romans massive IIA.
\par
We define $W^{N_1}_{Q_0}\equiv\Theta^{N_1}_{Q_0}=
\Theta^{N_1}_{\alpha}$ as the index $Q_0$ is  the index on $
R^{Q_0}$ which is just the index $\alpha$. In terms of this notation
the lowest order example of eq. (\ref{2.12}) is given by
  \begin{equation}
  [R^{aN_1},P_b]= - g \delta_b^a \Theta^{N_1}_{\alpha}R^\alpha \quad
  .
  \label{definitionofthetainalldimensions}
  \end{equation}
 We will see that $ \Theta^{N_1}_{\alpha}$ will turn out
to be the embedding tensor discussed so much in the literature on
gauged supergravities.
\par
The  $\Theta^{N_1}_{\alpha}$, like all the objects $W$,  are not
invariant tensors of the internal symmetry group $G$. One can think
of them as a kind of spurion; for each allowed value of
$\Theta^{N_1}_{\alpha}$ one finds a different gauged supergravity,
for example the local gauge group  is determined  by the value of
$\Theta^{N_1}_{\alpha}$.
\par
We will now work out the consequences of the Jacobi identities for
the deformed algebra of eqs. (\ref{2.7}) to (\ref{2.11}). We begin
with the Jacobi identities that arise from taking two $E_{11}$
generators and $P_c$. These will place linear conditions on
$W^{P_{n+1}}{}_{S_n}$ as we have only one $P_c$.  In particular we
first consider the identity
  \begin{equation}
  [[R^{a_1M_1}, R^{b_1\ldots b_n, N_n} ],
P_c] =[R^{a_1M_1},[ R^{b_1\ldots b_n, N_n} , P_c]] +[[R^{a_1M_1},
  P_c],R^{b_1\ldots b_n, N_n} ] \quad .\label{2.14}
  \end{equation} We evaluate this using
eqs. (\ref{2.7}) to (\ref{definitionofthetainalldimensions}). Not
all structures of Lorentz indices that arise are independent due to
the identity
 \begin{equation}
  n\delta_c^{[b_1|} R^{a|b_2\ldots b_n]}=\delta_c^{a} R^{b_1b_2\ldots
  b_n}-(n+1)\delta_c^{[a} R^{b_1b_2\ldots b_n]} \quad .\label{2.15}
  \end{equation}
As such it suffices to consider the coefficients of only the terms
involving $\delta_c^ {[a} R^{b_1b_2\ldots b_n]}$ and those of the
form $\delta_c^ {a} R^{b_1b_2\ldots b_n}$ and use the above equation
to express any other contributions in terms of these two forms. We
find that at order $g$ this leads, respectively, to the two
equations
 \begin{equation}
f^{M_1N_n}{}_{P_{n+1}}W^{P_{n+1}}{}_{S_n}=X_n{}^{M_1}{}_{S_n}{}^{N_n}
-f^{M_1Q_{n-1}}{}_{S_{n}}W^{N_{n}}{}_{Q_{n-1}} \label{2.16}
\end{equation}
and
  \begin{equation}
  L^{M_1N_n}{}_{P_{n}}=-{1\over
  n}f^{M_1Q_{n-1}}{}_{P_{n}} W^{N_{n}}{}_{P_n}
  -X_n{}^{M_1}{}_{P_n}{}^{N_n} \label{2.17}
  \end{equation}
where $X_n{}^{M_1}{}_{S_n}{}^{N_n}=\Theta^{M_1}_\alpha
(D^\alpha){}_{S_n}{}^{N_n}$.
\par
As noted above,  the representation ${\bf R_1\otimes R_n}$ always
contains the representation ${\bf R_{n+1}}$ and so the structure
constant $f^{M_1N_n}{}_{P_{n+1}}$ can be inverted to leave  only
$W^{P_{n+1}}{}_{S_n}$ on the right-hand side of eq. (\ref{2.14}).
Thus this equation solves for $W^{P_{n+1}}{}_{S_n}$ in terms of
$\Theta^{M_1}_\alpha$ and the lower level $W^{P_{n}}{}_{S_{n-1}}$
and these equations provide a set of recursion relations that allow
one to solve for all the $W^{P_{n+1}}{}_{S_n}$'s in terms of
$\Theta^{M_1}_\alpha$. Eq. (\ref{2.12}) then just gives
$L^{M_1N_n}{}_{P_{n}}$ in terms of $\Theta^{M_1}_\alpha$.
\par
At order $g^2$ we find that the Jacobi identity of eq. (\ref{2.16})
implies the relation
 \begin{equation}
L^{M_1N_n}{}_{P_{n}}U^{P_n}{}_{S_{n-1}}=
L^{M_1Q_{n-1}}{}_{S_{n-1}}W^{N_n}{}_{Q_{n-1}} \quad .\label{2.18}
\end{equation}
\par
At lowest order eq. (\ref{2.14}) implies that
 \begin{equation}
  f^{M_1N_1}{}_{P_{2}}W^{P_{2}}{}_{S_1}=2X_1{}^{(M_1}{}_{S_1}{}^{N_1)}
  \label{2.19} \quad .\end{equation}
In deriving this relation we have used that
 \begin{equation}
[R^{aM_1}, \Theta^{N_1}_\alpha R^\alpha ]=
f^{M_1Q_0}{}_{S_{1}}W^{N_{1}}{}_{Q_0}R^{aS_1}=
-X_1^{N_1}{}_{S_1}{}^{M_1}R^{aS_1} \label{2.20}
  \end{equation}
since in terms of our notation $f^{M_1Q_0}{}_{S_{1}}= -f^{Q_0
M_1}{}_{S_{1}}= -(D^\alpha)_{S_1}{}^{M_{1}}$ and using our earlier
definition $\Theta^{N_1}_\alpha = W^{N_{1}}{}_{Q_0}$.
\par
While we have solved for all the $W$'s in terms of $\Theta$ using
the above equations it is more practical to do this step for the $W$
involved with the 4-form generator using the Jacobi identity
 \begin{equation}
[[R^{a_1a_2 M_2}, R^{b_1 b_2, N_2} ], P_c] =[R^{a_1a_2 M_2},[ R^{b_1
b_2, N_2} , P_c]] +[[R^{a_1a_2 M_2}, P_c],R^{b_1 b_2, N_2} ]
\label{2.21}
  \end{equation}
and eqs. (\ref{2.11}) and (\ref{2.12}) rather than the Jacobi
identity of eq. (\ref{2.14}) for the case of   $m=1$ and $n=3$.
Using similar arguments to those deployed above,   we find at order
$g^1$ the two equations
 \begin{equation}
W^{M_2N_2} {}_{R_3}= -W^{N_2}{}_{R_1}{}f^{R_1M_2}{}_{R_3}
+W^{M_2}{}_{R_1}{}f^{R_1N_2}{}_{R_3} \label{2.22}\end{equation} and
 \begin{equation}
  V^{M_2N_2} {}_{R_3}= W^{M_2N_2} {}_{R_3}\quad . \label{2.23}
  \end{equation}
Clearly, these solve for $W^{M_2N_2} {}_{R_3}$ and $V^{M_2N_2}
{}_{R_3}$ in terms of $W^{N_2}{}_{R_1}$ and so in terms of
$\Theta^{N_1}_\alpha R^\alpha  $, while at order $g^2$ we find that
eq. (\ref{2.21}) implies that
  \begin{equation}
{1\over 3}V^{M_2N_2} {}_{P_3}U^{P_3}{}_{P_2}=-L^{R_1M_2}{}_{P_2}
W^{N_2}_{R_1} \quad .\label{2.24}\end{equation}
\par
It will be useful to also  consider the Jacobi identity
  \begin{equation}
[\Theta_\alpha^{N_1}R^\alpha, [R^{a_1\ldots a_n, P_n}, P_c]]=
[\Theta_\alpha^{N_1}R^\alpha, R^{a_1\ldots a_n, P_n}], P_c]
\label{2.25}
\end{equation} since $[R^\alpha ,P_c]=0$. It implies that
 \begin{equation}
X_{n-1}^{N_1}{}_{R_{n-1}}{}^{Q_{n-1}}W^{P_n}{}_{Q_{n-1}}=
X_{n}^{N_1}{}_{Q_{n}}{}^{P_{n}}W^{Q_n}{}_{R_{n-1}} \quad .
\label{2.26}\end{equation}
\par
We now consider the consequences of the Jacobi identities that
involve one $E_{11}$ generator $R^{a_1\ldots a_n, P_n}$ and the
generators $P_c$ and $P_d$. This implies a quadratic constraint on
$W^{P_n}{}_{Q_{n-1}}$'s that is given by
 \begin{equation}
  W^{P_n}{}_{Q_{n-1}}W^{Q_{n-1}}{}_{R_{n-2}}=0 \quad .\label{2.27}
  \end{equation}
At the lowest order, {\it i.e.} $n=1$, eq. (\ref{2.26}) implies that
  \begin{equation}
X_1^{N_1}{}_{Q_1}{}^{P_1}\Theta_\alpha^{Q_1}=\Theta_\epsilon^{N_1}
f^{\epsilon\gamma}{}_\alpha \Theta_\gamma^{P_1} \quad .
\label{2.28}\end{equation}
\par
Finally we consider the Jacobi identity with $K^{a,b_1\ldots b_n,
M_n}$ and $P_c$ and $P_d$, namely
 \begin{equation}
  [ [K^{a,b_1\ldots b_n, M_n},P_c ], P_d]-[ [K^{a,b_1\ldots b_n,
  M_n},P_d ], P_c] =0 \quad ,\label{2.29}
  \end{equation}
as $[ P_c , P_d]=0$. At order $g^1$ we find that
  \begin{equation}
  U^{P_n}{}_{P_{n-1}}=-{n\over n+1}W^{P_n}{}_{P_{n-1}} \quad ,\label{2.30}
  \end{equation}
while at order $g^2$ we find that
  \begin{equation}
  U^{P_n}{}_{P_{n-1}} U^{P_{n-1}}{}_{P_{n-2}}=0 \quad .\label{2.31}
  \end{equation}
The first equation solves for $U^{P_n}{}_{P_{n-1}}$ in terms of
$W^{P_n}{}_{P_{n-1}}$ and so in terms of $\Theta^{N_1}_\alpha $.
Using eq. (\ref{2.28}) and eq. (\ref{2.30}) we observe that eq.
(\ref{2.31}) is automatically solved. Furthermore substituting eq.
(\ref{2.30}) into eq.  (\ref{2.24}) we find it is automatically
satisfied using eq. (\ref{2.17}).
\par
We now summarise the content of this section so far. The deformation
of the algebra of eqs. (\ref{2.7}) to (\ref{2.12}) involves a number
of  the constants, namely $L^{M_m N_n }{}_{P_{n+m-1}}$,
$W^{M_m}{}_{M_{m-1}}$, $ U^{P_n}{}_{P_{n-1}}$, $ V^{M_2N_2}{}_{P_3}$
and $ W^{M_2N_2}{}_{P_3}$. However,   the Jacobi identities imply
that  all of these may be solved in terms of the $W$'s and these are
in turn determined in terms of the single object
$\Theta^{N_1}_\alpha$. {\bf Thus the entire deformation is
determined in terms of $\Theta^{N_1}_\alpha $,  or equivalently eq.
(\ref{definitionofthetainalldimensions})}.
\par
However, the above equations also impose constraints on
$\Theta^{N_1}_\alpha $. Clearly, there are the  quadratic
constraints of eq.  (\ref{2.28}) which are a set of constraints on
$\Theta^{N_1}_\alpha $  once we have substituted for the $W$'s in
terms of $\Theta^{N_1}_\alpha $. However, we also have a set of
linear constraints that originate from eq. (\ref{2.16}) whose right
hand side can be expressed entirely in terms of $\Theta^{N_1}_\alpha
$, a variable in which it is linear. As explained above the
structure constant that occurs in the commutator of eq. (\ref{2.1})
obeys projector conditions arising from the fact that the form
generators on the right hand side do not belong to the
representation ${\bf R_m\otimes R_n}$, but only to the
representation ${\bf R_{n+m}}$ that it contains. The number of
projection conditions correspond to the number of irreducible
representation in ${\bf R_m\otimes R_n}$ which are not contained in
the representation ${\bf R_{n+m}}$. However, certain of these
structure constants, {\it i.e.} $f^{M_1N_n}{}_{P_{n+1}}$ appear on
the left hand side of eq. (\ref{2.16}) and so the object
$\Theta^{N_1}_\alpha $ that appears on the right hand side of this
equation will satisfy corresponding constraints. In particular,
taking $ n=D-2$ in eq. (\ref{2.16}) we find the structure constant
$f^{M_1\alpha}{}_{P_{D-1}}$ on the left hand side which obeys the
constraints of eq. (\ref{2.5})  for the cases of dimensions four,
five and six. This is evident from table
\ref{tablewithembeddingtensorinvariousdim} where we find that in
these dimensions the representation ${\bf R_{1} \otimes R_0}$
contains three irreducible representations only one of which is
${\bf R_{D-1}}$. As explained in appendix A this is a consequence of
the fact that for these dimensions ${\bf R_{1}}$ is the fundamental
representation of the internal symmetry group. In other dimensions
one has to project out more than two irreducible representations
from ${\bf R_{1}\otimes R_0 }$ to leave the representation ${\bf
R_{D-1}}$ (see table \ref{tablewithembeddingtensorinvariousdim}) and
so one has more projection conditions on the structure constant
$f^{M_1\alpha}{}_{P_{D-1}}$ and so on $\Theta^{N_1}_\alpha $. Thus
in dimensions four, five and six we will find  two linear conditions
on $\Theta^{N_1}_\alpha$ which are evaluated in detail later in this
paper and are found to be
  \begin{equation}
  (D^\alpha)_{N_1}{}^{M_1}\Theta^{N_1}_\alpha=0 \label{2.32}
  \end{equation} and
  \begin{equation}
  (D_\beta
  D^\alpha)_{N_1}{}^{M_1}\Theta^{N_1}_\alpha=c\Theta^{M_1}_\beta
\label{2.33}\end{equation} where $c$ is a constant plus possible
further constraints. In  dimension other than  four, five and six we
find these constraints as well as further constraints.  However, in
dimensions other than three, four, five and six one finds that all
the    conditions on $\Theta^{N_1}_\alpha$ already arise at lower
levels than $n=D-2$  from similar conditions on the corresponding
lower level structure constants. Hence, a priori although
$\Theta^{N_1}_\alpha$  could belong to the representation ${\bf
R_1\otimes R_0}$ the constraints discussed  above, and derived in
detail in each dimension in later sections,  restrict it  to
actually belong to the same representation as the $D-1$ form
generators {\it i.e.} the ${\bf R_{D-1}}$ representation in all
dimensions.
\par
To summarise this section so far.  {\bf We have found that  the
deformation is uniquely determined in terms of $\Theta^{N_1}_\alpha$
and will find, taking account of results in later sections,  that
this object obeys constraints that imply that it belongs to the same
representation as the $D-1$ form generators}.
\par
We turn our attention to the construction of the field strengths
from the Cartan forms. We write the group element of the algebra of
eqs. (\ref{2.7}) to (\ref{2.12}) in the form
 \begin{equation}
  g=e^{x^aP_a}e^{\Phi\cdot K}e^{A\cdot R} \label{2.34}\end{equation} where
 \begin{eqnarray}
  & & e^{A\cdot R}= \ldots e^{A_{a_1\ldots a_m, M_m}R^{a_1\ldots a_m,
  M_m}} e^{A_{a_1\ldots a_{m-1}, M_{m-1}}R^{a_1\ldots a_{m-1},
  M_{m-1}}}\ldots \nonumber\\
& & \quad \qquad e^{A_{a_1 a_2, M_2}R^{a_1 a_2, M_2}}e^{A_{a_1,
M_1}R^{a_1 , M_1}}g_\varphi \label{2.35}\end{eqnarray} where
$g_\varphi=e^{\varphi_\alpha R^\alpha}$ and $e^{\Phi\cdot K}$ is a
similar expression involving the Ogievetsky fields and generators.
The field strengths are contained in the Cartan forms which we can
write as
  \begin{equation}
  g^{-1}dg= {\cal V}^{(0)}+{\cal V}^{(1)}+\ldots \label{2.36}
  \end{equation} where
${\cal V}^{(n)}$ is the contribution at $g^n$. The full calculation
involves many terms but we are only interested in the field
strengths and so we will only keep terms that contain $E_{11}$
generators. These contain terms of the form $dx^\mu G_{\mu a_1\ldots
a_n,M_n}R^{a_1\ldots a_n,M_n}$ . The coefficients $G_{\mu a_1\ldots
a_n,M_n}$ are not totally anti-symmetrised in all their $\mu
a_1\ldots a_n$ indices, but the terms that are not are set to zero
using the inverse Higgs mechanism which solves for the corresponding
Ogievetsky field. This mechanism is discussed in detail in reference
\cite{fabiopeterogievetsky}. The term that is totally
anti-symmetrised  is the field strength and as this is what is
needed for the dynamics we will compute only this term. To carry out
this task we only need the commutation relation of eqs. (\ref{2.7})
and (\ref{2.9}) and need not include the Ogievetsky fields in our
computations.
\par
Let us denote  the totally anti-symmetric part of ${\cal V}$ by
${\cal V}_A$ and write it as
  \begin{equation}
  {\cal V}_A= \sum _m {1 \over m+1} \tilde F_{\mu a_1\ldots a_m, M_m}R^{a_1\ldots
  a_m, M_m}= g_\varphi ^{-1}\sum _m {1 \over m+1} F_{\mu a_1\ldots a_m,
  M_m}R^{a_1\ldots a_m, M_m} g_\varphi \label{2.37}
  \end{equation}
where $F_{\mu a_1\ldots a_m, M_m}= F_{[\mu a_1\ldots a_m], M_m}$. We
denote the order $g^p$ contribution by $F^{(p)}_{\mu a_1\ldots a_m,
M_m}$, and the structure of the algebra is such that only the order
zero and the first order in $g$ occur. The factors of $g_\varphi$
lead to the matrix functions $(e^{-\varphi_\alpha
D^\alpha})_{M_n}{}^{N_n}$ where $(D^\alpha )_{M_n}{}^{N_n}$ is in
the corresponding representation ${\bf R_m}$. That is $\tilde F_{\mu
a_1\ldots a_m, M_m}= (e^{-\varphi_\alpha D^\alpha})_{M_m}{}^{N_m}
F_{\mu a_1\ldots a_m, N_m}$. As these extra factors involving the
scalars just complicate the formulae we will only explicitly compute
the $F_{\mu a_1\ldots a_m, M_m}$.  The scalar factor just converts $
F_{\mu a_1\ldots a_m, M_m}$ which is in the linear representation
${\bf R_m}$ into $\tilde F_{\mu a_1\ldots a_m, M_m}$ which is in a
non-linear representation of the internal symmetry, in fact
transforming by a non-linear local subgroup rotation.
\par
The  terms in ${\cal V}_A^{(0)}$ are just the field strengths found
from  the $E_{11}$ algebra without any deformation and these have
been computed in several cases before. We will begin by computing
these  terms for any  dimension making use of the notation developed
above. We use the well known relation
  \begin{equation}
e^{-A}d e^{A}= {1-e^{-A}\over A }\star d A \quad ,
\label{2.38}\end{equation} where $A$ is a generic operator and the
$\star$-product is defined by
  \begin{equation}
  A\star B=[A,B] \qquad , \quad \quad A^p\star B= [A, A^{p-1}\star B] \quad .
  \end{equation}
We note that
  \begin{equation}
  [A_{a_1\ldots a_m, M_m}R^{a_1\ldots a_m, M_m}, R^{b_1\ldots b_n,
  N_n}] =L_{a_1\ldots a_m}{}_{P_{m+n}}{}^ {N_n} R^{a_1\ldots a_m
  b_1\ldots b_n, P_{m+n}} \quad , \label{2.39}
  \end{equation}
where
  \begin{equation}
  L_{a_1\ldots a_m}{}_{P_{m+n}}{}^ {N_n}= A_{a_1\ldots a_m, M_m}
  f^{M_m N_n}{}_{P_{m+n}}\quad .
  \end{equation}
Using these conventions we may also evaluate
  \begin{eqnarray}
  & & A_{a_1\ldots a_m, M_m}R^{a_1\ldots a_m, M_m}\star A_{b_1\ldots
  b_n, M_n}R^{b_1\ldots b_n, M_n}\star R^{c_1\ldots c_p, P_p}
  \nonumber \\
  & &  = \{L_{a_1\ldots a_m}L_{b_1\ldots b_n}\}_{P_{m+n+p}}{}^ {P_p}
  R^{a_1\ldots a_m b_1\ldots b_n,c_1\ldots c_p P_{m+n+p}} \quad ,
  \label{2.40} \end{eqnarray}
where the two $L$ factors are multiplied using matrix multiplication
on their internal symmetry indices.
\par
Denoting with
  \begin{equation}
  g_A^m = e^{A_{a_1\ldots a_{m}, M_{m}}R^{a_1\ldots a_{m},
  M_{m}}}
  \end{equation}
and
  \begin{equation}
  g_A^{< m}= e^{A_{a_1\ldots a_{m-1}, M_{m-1}}R^{a_1\ldots a_{m-1},
  M_{m-1}}}\ldots e^{A_{a_1 a_2, M_2}R^{a_1 a_2, M_2}}e^{A_{a_1,
  M_1}R^{a_1 , M_1}}g_\varphi \quad ,\label{2.43}
  \end{equation}
we write
  \begin{equation}
  {\cal V}_A^{(0)} = \sum_m (g_A^{< m})^{-1} (g_A^m )^{-1} d g_A^m \ g_A^{<
  m} \quad .
  \end{equation}
This expression can be evaluated using eq. (\ref{2.38}). The result
is further simplified by using the language of forms. We find that
  \begin{eqnarray}
  & & F^{(0)}_{\mu a_1\ldots a_m,N_m} dx^\mu\wedge dx^{a_1}\wedge
  \ldots \wedge dx^{a_m} = (m+1) \sum_{n_1,\ldots ,n_r}
  {(-1)^{n_1}\over n_1 !} \ldots {(-1)^{n_{r-1}}\over n_{r-1}
  !}{(-1)^{n_{r}}\over (n_{r}+1) !}
  \nonumber \\
  & & d x^\mu \wedge
  \{(L^{(1)}\wedge \ldots\wedge L^{(1)})(L^{(2)}\wedge \ldots \wedge
  L^{(2)})\ldots  (L^{(r)}\wedge \ldots \wedge
  L^{(r)})\}_{N_m}{}^{N_r} \partial_\mu A_{N_r} \quad
  ,\label{2.41}\end{eqnarray}
where $\partial_\mu A_{N_r}=
\partial_\mu A_{a_1\ldots a_r, N_r} dx^{a_1} \wedge
\ldots \wedge dx^{a_r}$ and $L^{(m)}{}_{\bullet}{}^\bullet =
A_{a_1\ldots a_m, M_m} dx^{a_1}\wedge \ldots \wedge dx^{a_m}f^{M_m
\bullet}{}_{\bullet}$. The sum being over all integers $n_p$ such
that $m=n_1+2n_2+\ldots +(r-1)n_{r-1}+r(n_r+1)$.
\par
We now compute the analogous terms at order $g^1$, that is those
involving totally anti-symmetric indices and $E_{11}$ generators.
Using eq. (\ref{2.9}) we find that
 \begin{eqnarray}
& & {\cal V}^{(1)}_A= e^{-A\cdot R}dx^\mu P_\mu e^{A\cdot R} \nonumber \\
& &  = g \sum_m (g_A^{< m})^{-1}{1- (g_A^m )^{-1} \over A_{a_1\ldots
a_m,M_m} R^{a_1\ldots a_m,M_m}} W^{N_m}{}_{N_{m-1}} dx^\mu A_{\mu
a_1\ldots a_{m-1},N_m}
R^{a_1\ldots a_{m-1},N_{m-1}} \ g_A^{< m} \nonumber \\
& & + dx^\mu e_\mu{}^a P_a  \quad .\label{2.42}
\end{eqnarray}
Further evaluating this expression using the above notation we find
that
  \begin{eqnarray}
  & & F^{(1)}_{\mu a_1\ldots a_m,N_m} dx^\mu\wedge dx^{a_1} \ldots \wedge
  dx^{a_m} =(m+1)g \Big(\sum_{n_1,\ldots ,n_r} {(-1)^{n_1}\over n_1
  !}\ldots {(-1)^{n_{r-1}}\over n_{r-1} !}{(-1)^{n_{r}}\over (n_{r}+1)
  !} \nonumber \\
  & &  dx^\mu\wedge \{(L^{(1)}\wedge \ldots\wedge
  L^{(1)})(L^{(2)}\wedge \ldots\wedge L^{(2)})\ldots  (L^{(r)}\wedge
  \ldots\wedge
  L^{(r)})\}_{N_m}{}^{N_{r-1}}W^{N_r}{}_{N_{r-1}}A^{(r)}_{\mu N_r}
  \nonumber \\
  &  &-{(-1)^{m}\over m+1 !}dx^\mu\wedge \{L^{(1)}\wedge \ldots\wedge
  L^{(1)}\}_{N_m}{}^{R_1} X_1^{N_1}{}_{R_1}{}^{M_1}
  A^{(1)}_{M_1}A^{(1)}_{\mu N_1} \Big) \quad ,\label{2.44}
  \end{eqnarray}
where $A^{(r)}_{\mu ,N_r}=  A_{\mu a_1\ldots a_{r-1}, N_r}
dx^{a_1}\wedge \ldots \wedge dx^{a_{r-1}}$ and  $A^{(1)}_{M_1}=  A_{
a , M_1} dx^{a}$. The sum is such that $m=n_1+2n_2+\ldots
+(r-1)(n_{r-1}+ 1) +r n_r$, there being $n_p$ factors of $L^{(p)}$
in the first term, where $r$ must be greater than 1, and $m-1$
factors of $L^{(1)}$ in the second term. The contribution to the one
form field strength consists of the term  $g A_{\mu M_1} dx^\mu
\Theta_\alpha^{M_1}R^\alpha $.
\par
To summarise this section. {\bf We have found all deformations of
the form of eqs. (\ref{2.7}) to (\ref{2.12}) are determined by the
one variable $\Theta^{N_1}_\alpha $ and this belongs to the same
representation as the $D-1$ forms, i.e. ${\bf R_{D-1}}$, as well as
satisfying certain quadratic constraints. We have computed the
fields strengths that occur in the non-linear realisation of the
deformed algebra. Thus we have found all the field strengths of all
the maximal supergravities in all dimensions.} We have therefore
reduced the computation of the field strengths and gauge
transformations of gauged supergravities to a purely algebraic
construction based on $E_{11}$.
\par
To conclude this section, we discuss the gauge transformations of
the fields. These arise in the non-linear realisation as rigid
transformations of the group element, $g \rightarrow g_0 g$, as long
as one includes the Ogievetsky generators
\cite{fabiopeterogievetsky}. In particular, in the massless theory
the action of
  \begin{equation}
  g_0 ={\rm exp} ( a_{a_1 ...a_n , N_n} R^{a_1
  ...a_n , N_n} )\label{gzeroactingongeneric}
  \end{equation}
generates a global transformation of the fields of parameter $a_{a_1
... a_n , N_n}$, and the net effect of including the Ogievetsky
generators is to promote this global symmetry to a local one via the
identification
  \begin{equation}
  a_{a_1 ...a_n , N_n} \rightarrow \partial_{[a_1} \Lambda_{a_2
  ...a_n ], N_n} \quad . \label{masslessidentification}
  \end{equation}
In the massive theory, this is modified due to the fact that the
$E_{11}$ generators have non-trivial commutation relations with the
momentum operator. If one acts with $g_0$ as in eq.
(\ref{gzeroactingongeneric}) on the group element of eq.
(\ref{2.34}) and uses eq. (\ref{2.9}), passing through $e^{x^a P_a}$
generates the term
  \begin{equation}
  {\rm exp}( - g W^{N_n}{}_{N_{n-1}} x^{a_1} a_{a_1 ...a_n, N_n}
R^{a_2 ...a_n , N_{n-1}}) \quad .
\label{gzeropassingthroughmomentum}
\end{equation}
Therefore, together with the constant transformation generated by
the action of the term in eq. (\ref{gzeroactingongeneric}), the
massive theory develops a transformation that is linear in $x$. The
inclusion of the Og generators then has the net effect of promoting
$a_{a_1 ... a_n , N_n}$ to a local parameter, and the gauge
transformation of the fields is obtained by taking the global
$a_{a_1 ... a_n , N_n}$ of the massless theory and making the
identification
  \begin{equation}
  a_{a_1 ...a_n , N_n} \rightarrow \partial_{[a_1} \Lambda_{a_2
  ...a_n ], N_n} - g W^{N_{n+1}}{}_{N_n} \Lambda_{a_1 .. .a_n,
  N_{n+1}} \label{massiveidentification}
  \end{equation}
instead of that of eq. (\ref{masslessidentification}). Indeed taking
$\Lambda_{a_1 ...a_{n-1}, N_n}$ to be at most linear in $x$ this
identification reproduces the transformations generated by eqs.
(\ref{gzeroactingongeneric}) and
(\ref{gzeropassingthroughmomentum}). Therefore the gauge
transformations of all the fields in the massive theory are given by
the ones of the massless theory, provided that one makes the change
  \begin{equation}
  \partial_{[a_1} \Lambda_{a_2
  ...a_n ], N_n}\rightarrow \partial_{[a_1} \Lambda_{a_2
  ...a_n ], N_n} - g W^{N_{n+1}}{}_{N_n} \Lambda_{a_1 .. .a_n,
  N_{n+1}}\quad .
  \end{equation}
A special case is the case $n=0$ in eq.
(\ref{massiveidentification}), for which although the first term on
the right-hand side is not present, the second term gives a gauge
transformation of parameter
  \begin{equation}
  - g \Theta^{M_1}_\alpha \Lambda_{M_1} \quad .
  \end{equation}
This determines the way in which all the fields transform under the
gauge parameter $\Lambda_{M_1}$ at order $g$, the field $A_{a_1
...a_n, N_n}$ transforming as
  \begin{equation}
  \delta A_{a_1 ... a_n , N_n } = - g \Theta^{M_1}_\alpha
  \Lambda_{M_1} D^\alpha_{N_n}{}^{M_n} A_{a_1 ... a_n , M_n } \quad
  . \label{thisisthegaugetransf}
  \end{equation}

In sections from 3 to 9 we will apply the results of this section to
all dimensions from 3 to 9, showing that in all cases
$\Theta^{M_1}_\alpha$ and the $D-1$ forms belong to the same
representation and determining the field strengths and gauge
transformations of all the form fields of any maximal supergravity
theory in any dimension.

\section{D=3}
\begin{figure}[h]
\begin{center}
\begin{picture}(380,70)
\multiput(10,10)(40,0){6}{\circle{10}}
\multiput(250,10)(40,0){3}{\circle{10}} \put(370,10){\circle{10}}
\multiput(15,10)(40,0){9}{\line(1,0){30}} \put(290,50){\circle{10}}
\put(290,15){\line(0,1){30}} \put(8,-8){$1$} \put(48,-8){$2$}
\put(88,-8){$3$} \put(128,-8){$4$} \put(168,-8){$5$}
\put(208,-8){$6$} \put(248,-8){$7$} \put(288,-8){$8$}
\put(328,-8){$9$} \put(365,-8){$10$} \put(300,47){$11$}
\put(85,5){\line(1,1){10}} \put(85,15){\line(1,-1){10}}
\end{picture}
\caption{\sl The $E_{11}$ Dynkin diagram corresponding to
3-dimensional supergravity. The internal symmetry group is
$E_{8(+8)}$. \label{Dynkinthree}}
\end{center}
\end{figure}
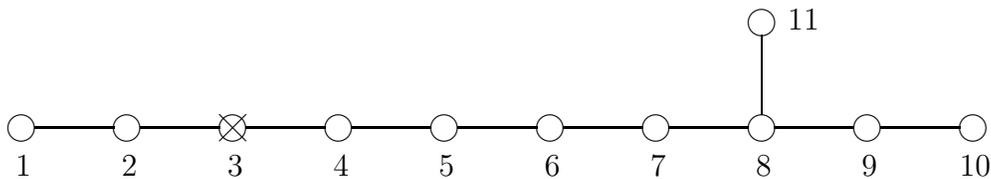

The bosonic sector of the massless maximal supergravity theory in
three dimensions \cite{D=3masslesssugra} describes 128 scalars
parametrising the manifold $E_{8(+8)}/SO(16)$ and the metric. This
theory arises from the $E_{11}$ decomposition appropriate to three
dimensions, corresponding to the deletion of node 3 as shown in the
Dynkin diagram of fig. \ref{Dynkinthree}.
The 1-form generators of $E_{11}$ that arise in this decomposition
belong to the ${\bf 248}$ of $E_8$. The corresponding fields are
dual to the scalars. There are also 2-form generators in the ${\bf 1
\oplus 3875}$, the corresponding fields having vanishing
field-strength in the massless theory. We will not consider in our
analysis the 3-forms and all the generators with mixed symmetry. To
summarise, we consider the form generators
   \begin{equation}
   R^{\alpha} \quad ({\bf {248}}) \qquad R^{a , \alpha}  \quad ({\bf 248} )
  \qquad R^{a_1 a_2, M}  \quad ( {\bf 3875} ) \qquad R^{a_1 a_2}
   \quad   ({\bf 1} )
  \quad , \label{generatorsinthreedimensions}
  \end{equation}
where $\alpha = 1, ..., 248$ denotes the adjoint and $M = 1,...,
3875$ denotes the ${\bf 3875}$ of $E_8$.

The $E_{11}$ commutation relations involving the generators in eq.
(\ref{generatorsinthreedimensions}) are
  \begin{eqnarray}
  & & [ R^\alpha , R^\beta ] = f^{\alpha \beta}{}_\gamma R^\gamma
  \nonumber \\
  & & [ R^\alpha , R^{a , \beta} ] = f^{\alpha \beta}{}_\gamma R^{ a,
  \gamma}
  \nonumber \\
  & & [ R^\alpha , R^{a_1 a_2 } ] = 0 \nonumber \\
  & & [ R^\alpha , R^{a_1 a_2 , M} ] = D^\alpha_N{}^M R^{a_1 a_2 ,
  N} \nonumber \\
  & & [ R^{a_1 , \alpha} , R^{a_2 ,\beta} ] = g^{\alpha \beta}
  R^{a_1 a_2} + S^{\alpha \beta}_M R^{a_1 a_2 , M} \quad ,
  \label{3dimalgebramassless}
  \end{eqnarray}
where $g^{\alpha \beta}$ is proportional to the Cartan-Killing
metric and it is the metric we use to raise $E_8$ indices in the
adjoint, $D^\alpha_N{}^M$ are the $E_8$ generators in the ${\bf
3875}$ and $S^{\alpha \beta}_M$ is an $E_8$ invariant tensor. This
invariant tensor is such that $S^{\alpha \beta}_M R^{a_1 a_2 , M}$
belongs to the ${\bf 3875}$, and using the $E_8$ conventions and
projection formulae of \cite{E8conventions} one deduces that
$S^{\alpha \beta}_M$ must satisfy the further conditions
  \begin{eqnarray}
  & & g_{\alpha \beta} S^{\alpha \beta}_M = 0 \nonumber \\
  & & S^{\alpha \beta}_M = - {1 \over 12}
  f^\epsilon{}_\gamma{}^\alpha f_{\epsilon \delta}{}^\beta
  S^{\gamma \delta}_M \quad . \label{conditionsonSinthree}
  \end{eqnarray}
Indeed $S^{\alpha \beta}_M$ is symmetric in $\alpha \beta$, and the
symmetric product of two ${\bf 248}$ representations is
  \begin{equation}
  {\bf [ 248 \otimes 248 ]_{\rm S} = 1 \oplus 3875 \oplus 27000 }
  \quad . \label{248symmproduct}
  \end{equation}
The conditions of eq. (\ref{conditionsonSinthree}) project out ${\bf
1 \oplus 27000}$ to ensure that $S^{\alpha \beta}_M R^{a_1 a_2 , M}$
belongs to the ${\bf 3875}$. The $E_8$ metric is related to the
structure constant by \cite{E8conventions}
  \begin{equation}
  g^{\alpha \beta} = -{1 \over 60} f^{\alpha}{}_{\gamma \delta}
  f^{\beta \gamma \delta} \quad ,
  \end{equation}
while another useful $E_8$ identity is \cite{E8conventions}
  \begin{equation}
  f_{\alpha \epsilon \tau} f_{\beta}{}^\epsilon{}_\sigma
  f^{\gamma}{}_\rho{}^\tau f^{\delta \rho \sigma}= 24
  \delta^\gamma_{(\alpha} \delta^\delta_{\beta )}  + 12 g_{\alpha
  \beta} g^{\gamma \delta} - 20 f^\epsilon{}_\alpha{}^\gamma
  f_{\epsilon\beta}{}^\delta + 10 f^\epsilon{}_\alpha{}^\delta
  f_{\epsilon\beta}{}^\gamma \quad .
  \end{equation}

From the group element
  \begin{equation}
  g =  e^{x \cdot P} e^{A_{a_1 a_2} R^{a_1 a_2}} e^{A_{a_1 a_2 ,M} R^{a_1 a_2
  ,M}}
  e^{A_{a
  , \alpha} R^{a , \alpha}} e^{\phi_{\alpha} R^{\alpha}}
  \label{groupelementD=3}
  \end{equation}
one derives the field-strengths of the 1-forms and 2-forms. These
indeed result from antisymmetrising the various terms in the
Maurer-Cartan form, which is computed imposing that the generators
in eq. (\ref{generatorsinthreedimensions}) commute with momentum. We
now consider the deformations of the algebra of eq.
(\ref{3dimalgebramassless}) resulting from imposing that the
generators have a non-trivial commutation relation with momentum
compatibly with the Jacobi identities.

We consider the general analysis of the previous section, applied to
the three-dimensional case via the identifications
  \begin{eqnarray}
  & & R^{a_1 , M_1} \rightarrow R^{a_1 , \alpha} \nonumber \\
  & & R^{a_1 a_2 , M_2} \rightarrow R^{a_1 a_2} \ , \ \ R^{a_1 a_2 , M}
  \nonumber \\
  & & \Theta^{M_1}{}_\alpha \rightarrow \Theta^\beta{}_\alpha \nonumber
  \\
  & & W^{M_2}{}_{M_1} \rightarrow W_\alpha \ , \ \  W^M{}_\alpha
  \quad .
  \end{eqnarray}
Eq. (\ref{2.19}), resulting from the Jacobi identity of two 1-forms
and momentum, reads
  \begin{equation}
  W^M{}_\delta S^{\alpha \beta}_M + W_\delta g^{\alpha \beta} =
  \Theta^\alpha{}_\gamma f^{\gamma \beta}{}_\delta +
  \Theta^\beta{}_\gamma f^{\gamma \alpha}{}_\delta
  \quad . \label{WThetarelationinD=3}
  \end{equation}
The embedding tensor $\Theta^\alpha{}_\beta$ has no a priori
symmetry, and thus is in the representations generated by the
symmetric product of two ${\bf 248}$ given in eq.
(\ref{248symmproduct}) together with those generated in the
antisymmetric product
  \begin{equation}
  {\bf [ 248 \otimes 248 ]_{\rm A} = 248 \oplus 30380 }
  \quad . \label{248antisymmproduct}
  \end{equation}

We now show that eq. (\ref{WThetarelationinD=3}) rules out the
possibility that the embedding tensor is antisymmetric. Using eq.
(\ref{conditionsonSinthree}) one derives from eq.
(\ref{WThetarelationinD=3}) the condition
  \begin{equation}
  \Theta^\alpha{}_\gamma f^{\gamma \beta}{}_\delta +
  \Theta^\beta{}_\gamma f^{\gamma \alpha}{}_\delta - {1 \over 31}
  g^{\alpha \beta} \Theta_{\gamma \rho} f^{\gamma \rho}{}_\delta
  + {1 \over 12} \Theta^\gamma{}_\sigma f^{\sigma \rho}{}_\delta
  [ f^\epsilon{}_\gamma{}^\alpha f_{\epsilon \rho}{}^\beta +
  f^\epsilon{}_\rho{}^\alpha f_{\epsilon \gamma}{}^\beta ] = 0 \
  . \label{WThetaprojectedinthreedim}
  \end{equation}
Taking $\Theta$ antisymmetric and contracting $\beta$ and $\delta$
this equation gives
  \begin{equation}
  f^{\alpha \beta \gamma} \Theta_{\beta \gamma} = 0 \quad ,
  \end{equation}
which rules out the ${\bf 248}$. Using this and contracting eq.
(\ref{WThetaprojectedinthreedim}) with $f_{\tau \beta}{}^\delta$ one
then shows that the antisymmetric part of $\Theta$ vanishes
completely, thus ruling out the ${\bf 30380}$ too. The fact that the
${\bf 248}$ is ruled out also implies
  \begin{equation}
  W_\alpha = 0 \quad , \label{walphaiserointhree}
  \end{equation}
as can be seen contracting $\alpha$ and $\beta$ in eq.
(\ref{WThetarelationinD=3}).

We thus take $\Theta$ to be symmetric, which corresponds to the
representations in eq. (\ref{248symmproduct}). The tensor
$W^M{}_\alpha$ has indices in ${\bf 3875 \otimes 248}$, and this
leads to the irreducible representations
  \begin{equation}
  {\bf 3875 \otimes 248 = 779247 \oplus 147250 \oplus 30380 \oplus
  3875 \oplus 248} \quad .
  \end{equation}
Therefore $W^M{}_\alpha$ is not along the ${\bf 27000}$. From eq.
(\ref{WThetarelationinD=3}) it then follows that taking $\Theta$ to
be in the ${\bf 27000}$ one gets
  \begin{equation}
   (\Theta^\alpha{}_\gamma )_{\bf 27000} f^{\gamma \beta}{}_\delta +
  ( \Theta^\beta{}_\gamma )_{\bf 27000} f^{\gamma \alpha}{}_\delta =0
  \quad ,
  \end{equation}
which is inconsistent because it is the condition of invariance of
$\Theta$. Therefore the ${\bf 27000}$ is also ruled out. The
invariant tensor $S^{\alpha \beta}_M$ satisfies
  \begin{equation}
  S^{\alpha \beta}_M S_{\alpha \beta N} = \delta_{MN} \quad ,
  \end{equation}
where $\delta_{MN}$ is the invariant tensor in the product ${\bf [
3875 \otimes 3875 ]_{\rm S}}$, and using this and eq.
(\ref{walphaiserointhree}) one can invert eq.
(\ref{WThetarelationinD=3}) to get
  \begin{equation}
  W^M{}_\delta  = 2
  \Theta^\alpha{}_\gamma f^{\gamma \beta}{}_\delta S_{\alpha \beta}^M
  \quad ,
  \end{equation}
that implies that $W^M{}_\alpha$ is in the ${\bf 3875}$.

We have thus shown that the algebra can only be consistently
deformed if the embedding tensor belongs to ${\bf 1 \oplus 3875}$.
In the case of the singlet deformation, $W^M_\alpha$ vanishes and
indeed eq. (\ref{WThetarelationinD=3}) becomes the invariance of
$\Theta$, which is the Cartan-Killing metric in this case. Therefore
our results reproduce the constraints on the embedding tensor found
using supersymmetry in \cite{nicolaisamtlebenD=3}. We now show that
also the quadratic constraints of \cite{nicolaisamtlebenD=3} follow
from the consistency of the deformed $E_{11}$ algebra. These come
again from the general analysis of the previous section. In
particular, given that $\Theta$ is symmetric, both eq. (\ref{2.27})
for $n=2$ and eq. (\ref{2.28})  give the same constraint, which is
  \begin{equation}
  \Theta^\alpha{}_\beta [ f^{\beta \epsilon}{}_\delta \Theta^{\gamma
  \delta} + f^{\beta \gamma}{}_\delta \Theta^{
  \delta \epsilon} ] =0 \quad .
  \end{equation}
This is the condition that the embedding tensor is invariant when
projected by the embedding tensor itself, and corresponds to the
condition that the embedding tensor is invariant under the subgroup
of $E_8$ which is gauged.

Here we have considered the Jacobi identities involving the 1-form
and 2-form generators, but one can show that also the Jacobi
identities involving the 3-forms close if one considers the
deformations arising from the embedding tensor in the ${\bf 1 \oplus
3875}$, and more generally the whole $E_{11}$ algebra can be
deformed consistently introducing this embedding tensor.

In section 2 we have given a general procedure to compute the field
strengths in any dimension. This is expanded in appendix B. In the
three-dimensional case, from the group element in eq.
(\ref{groupelementD=3}) and the commutators derived in this section
one then obtains the field strength for the 1-form,
  \begin{equation}
  F_{ab, \alpha} = 2 [\partial_{[a} A_{b ] , \alpha} + {1 \over 2}
  g \Theta^\beta{}_\delta f^{\delta \gamma}{}_\alpha A_{[a , \beta}
  A_{b] , \gamma} + g W^M{}_\alpha  A_{ab, M} ] \quad ,
  \end{equation}
transforming covariantly under the gauge transformations
  \begin{eqnarray}
  & & \delta A_{a ,\alpha } = \partial_a \Lambda_\alpha - g
  \Theta^\beta{}_\delta f^{\delta \gamma}{}_\alpha \Lambda_\beta
  A_{a , \gamma} - g W^M{}_\alpha \Lambda_{a , M} \nonumber \\
  & & \delta A_{ab , M} = \partial_{[a} \Lambda_{b ] , M} + {1 \over
  2} S^{\alpha \beta}_M \partial_{[a} \Lambda_\alpha A_{b ] , \beta}
  - g \Theta^\beta{}_\alpha D^\alpha_M{}^N \Lambda_\beta A_{ab , N} \nonumber \\
  & & \qquad \quad -
  {g \over 2} S^{\alpha \beta}_M W^N{}_\alpha \Lambda_{[a , N} A_{b
  ] , \beta} \quad ,
  \end{eqnarray}
where $D^\alpha_M{}^N$ are the generators in the ${\bf 3875}$. Given
the results in this section, we can also compute the field strength
of the 2-forms up to the term involving the 3-form. The result is
  \begin{eqnarray}
  & & F_{a_1 a_2 a_3 , M} = 3 [\partial_{[ a_1} A_{a_2 a_3 ] , M} + {1
  \over 2} S^{\alpha \beta}_M \partial_{[a_1} A_{a_2 , \alpha }
  A_{a_3 ] , \beta} + g A_{[a_1 a_2 , N} A_{a_3 ],\alpha }
  W^N{}_\beta S^{\alpha \beta}_M \nonumber \\
  & & \quad \qquad + {g \over 6} A_{[a_1 , \alpha}
  A_{a_2 , \beta} A_{a_3 ] , \gamma } \Theta^\alpha{}_\delta
  f^{\delta \beta}{}_\sigma S^{\gamma \sigma}_M  ] \nonumber \\
  & & F_{a_1 a_2 a_3 } = 3 [\partial_{[ a_1} A_{a_2 a_3 ] } + {1
  \over 2} g^{\alpha \beta} \partial_{[a_1} A_{a_2 , \alpha }
  A_{a_3 ] , \beta} + g A_{[a_1 a_2 , N} A_{a_3 ],\alpha }
  W^N{}_\beta g^{\alpha \beta}\nonumber \\
  & & \qquad \quad  + {g \over 6} A_{[a_1 , \alpha}
  A_{a_2 , \beta} A_{a_3 ] , \gamma } \Theta^\alpha{}_\delta
  f^{\delta \beta \gamma} ] \quad .
  \end{eqnarray}
To prove the gauge covariance of these field strengths of the
2-forms one must include the 3-forms and determine their gauge
transformations.

To summarise, we have obtained the field strengths and gauge
transformations of any gauged maximal supergravity theory in three
dimensions. These field strengths satisfy duality conditions. In
particular, the field strengths of the 1-forms are related to the
derivative of the scalars, while the field strengths of the 2-forms
are related to the embedding tensor.

\section{D=4}
\begin{figure}[h]
\begin{center}
\begin{picture}(380,70)
\multiput(10,10)(40,0){6}{\circle{10}}
\multiput(250,10)(40,0){3}{\circle{10}} \put(370,10){\circle{10}}
\multiput(15,10)(40,0){9}{\line(1,0){30}} \put(290,50){\circle{10}}
\put(290,15){\line(0,1){30}} \put(8,-8){$1$} \put(48,-8){$2$}
\put(88,-8){$3$} \put(128,-8){$4$} \put(168,-8){$5$}
\put(208,-8){$6$} \put(248,-8){$7$} \put(288,-8){$8$}
\put(328,-8){$9$} \put(365,-8){$10$} \put(300,47){$11$}
\put(125,5){\line(1,1){10}} \put(125,15){\line(1,-1){10}}
\end{picture}
\caption{\sl  The $E_{11}$ Dynkin diagram corresponding to
4-dimensional supergravity. The internal symmetry group is
$E_{7(+7)}$. \label{Dynkinfour}}
\end{center}
\end{figure}
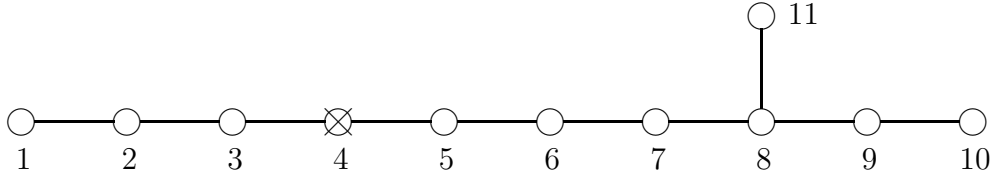

In this section we consider the $E_{11}$ decomposition relevant for
the four-dimensional theory. The corresponding Dynkin diagram is
shown in fig. \ref{Dynkinfour}. The global symmetry of
four-dimensional massless maximal supergravity
\cite{cremmerjuliaD=4} is $E_{7(+7)}$. This symmetry rotates
electric and magnetic vectors, and as such it is not a symmetry of
the lagrangian, but only of the equations of motion. This is in
agreement with $E_{11}$, in which fields and their magnetic duals
are treated on the same footing.

The bosonic field content of the supergravity theory contains 70
scalars parametrising the manifold $E_{7(+7)}/SU(8)$, the metric and
28 vectors, that together with their magnetic duals make the ${\bf
56}$ of $E_7$. $E_{11}$ contains the corresponding generators,
together with 2-form generators in the ${\bf 133}$ of $E_7$, whose
corresponding fields are dual to the scalars, 3-form generators in
the ${\bf 912}$, together with 4-form generators in the ${\bf 8645
\oplus 133}$ and an infinite number of generators with mixed
symmetry in the spacetime indices. Summarising, the form generators
are
  \begin{equation}
   R^{\alpha} \ ({\bf {133}}) \quad R^{a , M}  \ ({\bf 56} )
  \quad R^{a_1 a_2, \alpha}  \ ( {\bf 133} ) \quad R^{a_1 a_2 a_3 , A}
   \   ({\bf 912} ) \quad R^{a_1 ...a_4 , \alpha \beta} \ ({\bf 8645
   \oplus 133})
  \quad , \label{genlistD=4}
  \end{equation}
where $\alpha = 1 , ...,133$, $M=1 ,..., 56$ and $A=1, ... , 912$.
The $\alpha \beta$ indices of the 4-form are antisymmetric, which
indeed corresponds to the reducible representation ${\bf 8645 \oplus
133}$.

The $E_7$ algebra is
  \begin{equation}
  [ R^{\alpha} , R^{\beta} ] = f^{\alpha \beta}{}_\gamma R^{\gamma}
  \quad ,
  \end{equation}
where $f^{\alpha \beta}{}_\gamma $ are the $E_7$ structure
constants. We also introduce the generators $D^\alpha_M{}^N$ in the
${\bf 56}$, that satisfy the commutation relation
  \begin{equation}
  D^\alpha_M{}^P  D^\beta_P{}^N -  D^\beta_M{}^P  D^\alpha_P{}^N =
  f^{\alpha \beta}{}_\gamma  D^\gamma_M{}^N \quad .
  \end{equation}
The $M$ indices are raised and lowered by the antisymmetric
invariant metric $\Omega^{MN}$, that is for a generic object $V^M$
in the ${\bf 56}$ we have
  \begin{equation}
  V^M = \Omega^{MN} V_N \qquad V_M = V^N \Omega_{NM} \quad ,
  \label{raiseandlowerinfourdim}
  \end{equation}
which implies
  \begin{equation}
  \Omega^{MN} \Omega_{NP} = - \delta^M_P \quad .
  \end{equation}
Raising one index of the generator $D^\alpha_M{}^N$ one gets
  \begin{equation}
  D^{\alpha , MN} = \Omega^{MP} D^\alpha_P{}^N \quad ,
  \end{equation}
which is symmetric in $MN$.

We now write down the rest of the algebra. The commutators between
the scalars and the other generators are dictated by the $E_7$
representation that the generators carry. In particular for the
1-form one has
  \begin{equation}
  [ R^\alpha , R^{a, M} ] = D^\alpha_N{}^M R^{a , N}
  \end{equation}
and for the 2-form
  \begin{equation}
  [ R^\alpha , R^{a_1 a_2 , \beta} ] = f^{\alpha \beta}{}_\gamma
  R^{a_1 a_2, \gamma} \quad .
  \end{equation}
The other commutators are
  \begin{eqnarray}
  & & [ R^{a_1 , M} , R^{A_2 , N} ] = D_\alpha^{MN} R^{a_1 a_2 ,
  \alpha} \nonumber \\
  & & [ R^{a_1 , M} , R^{a_2 a_3 , \alpha} ] = S^{M \alpha}_A R^{a_1
  a_2 a_3 , A} \nonumber \\
  & & [ R^{a_1 a_2 , \alpha} , R^{a_3 a_4 , \beta} ] = R^{a_1 ...
  a_4, \alpha \beta} \nonumber \\
  & & [ R^{a_1 , M} , R^{a_2 a_3 a_4 , A} ] = C^{MA}_{\alpha \beta}
  R^{a_1 ... a_4 , \alpha \beta} \label{E11algebrainfourdimensions}
  \quad ,
  \end{eqnarray}
where we have introduced the two $E_7$ invariant tensors
$S^{M\alpha}_A$ and $C^{MA}_{\alpha \beta}$, the last one being
antisymmetric in $\alpha \beta$. Following \cite{dWSTgeneral}, we
are using the metric
  \begin{equation}
  g^{\alpha \beta} = D^\alpha_M{}^N D^\beta_N{}^M
  \label{E7metricproptokilling}
  \end{equation}
to raise and lower indices in the adjoint. This metric is
proportional to the Cartan-Killing metric, as can be seen from
  \begin{equation}
  f_{\alpha \beta\gamma} f^{\alpha \beta \delta} = - 3
  \delta^\delta_\gamma
  \quad .
  \end{equation}
A summary of the conventions for $E_7$ and $E_6$ is given in
appendix A. The Jacobi identity involving three 1-forms produces the
condition
  \begin{equation}
  D_\alpha^{(MN} S^{P) \alpha}_A = 0 \quad ,
  \label{otherconditiononS}
  \end{equation}
and $S^{M\alpha}_A$ also satisfies
  \begin{equation}
  D_{\alpha , M}{}^N S^{M \alpha}_A = 0 \quad ,
  \label{gravitinoconditiononS}
  \end{equation}
which can be deduced from the fact that there is no singlet in the
tensor product ${\bf 56 \otimes 912}$. Contracting eq.
(\ref{otherconditiononS}) with $D^\beta_{NP}$ gives
  \begin{equation}
  S^{M\alpha}_A + 2 (D^\alpha D_\beta )_N{}^M S^{N \beta}_A = 0
  \quad . \label{S+DDS=0infourdimensions}
  \end{equation}
As will be described in detail in appendix A, the conditions of eqs.
(\ref{gravitinoconditiononS}) and (\ref{S+DDS=0infourdimensions})
project the $M\alpha$ indices of $S^{M \alpha}_A$ along the ${\bf
912}$. Indeed, the only way of building an invariant from tensoring
a ${\bf 912}$ index with the product ${\bf 56 \otimes 133}$ is that
this product is projected on the ${\bf 912}$. The Jacobi identity
between $R^{a ,M}$, $R^{b , N}$ and $R^{cd , \alpha}$ gives the
condition
  \begin{equation}
  S^{M \alpha}_A C^{NA}_{\beta \gamma} + S^{N \alpha}_A C^{MA}_{\beta
  \gamma} + \delta^\alpha_{[\beta} D_{\gamma ]}^{MN} = 0 \quad .
  \end{equation}
One can check that also all the other Jacobi identities are
satisfied. We also define the invariant tensor $\Omega_{AB}$ in the
antisymmetric product of two ${\bf 912}$ representations, using the
relation
  \begin{equation}
  S^{M\alpha}_A S_{M \alpha B} = \Omega_{AB} \quad ,
  \end{equation}
and we use $\Omega_{AB}$ to raise and lower indices in the ${\bf
912}$, adopting conventions analogous to those of eq.
(\ref{raiseandlowerinfourdim}).

Writing down the group element
  \begin{equation}
  g =  e^{x \cdot P} e^{A_{a_1
  ...a_4, \alpha\beta} R^{a_1 ... a_4, \alpha\beta}} ...
  e^{A_{a
  , M} R^{a , M}} e^{\phi_{\alpha} R^{\alpha}}
  \label{fourdimensionalgroupelement}
  \end{equation}
one determines the field strengths of the massless theory by
antisymmetrising the spacetime indices of the various terms in the
Maurer-Cartan form, and the field equations of the supergravity
theory arise as duality relations. In particular, the field-strength
of the vector satisfies self-duality conditions, while the
field-strength of the 2-form in dual to the scalar derivative. The
field-strengths of the 3-forms vanish in the massless theory. In
deriving the field strengths of the massless theory one takes the
positive level $E_{11}$ generators to commute with momentum. In the
following we will consider the deformation of the $E_{11}$ algebra
which results from modifying the commutation relations of the
$E_{11}$ generators with momentum compatibly with the Jacobi
identities, following the general results of section 2.

Applying the general analysis of section 2 to the four-dimensional
case, one makes the identifications
  \begin{eqnarray}
  & & R^{a_1 , M_1}\rightarrow R^{a_1 , M} \qquad \qquad \ \ \Theta^{M_1}{}_\alpha
  \rightarrow \Theta^M_\alpha
  \nonumber \\
   & & R^{a_1 a_2 , M_2}  \rightarrow R^{a_1 a_2 , \alpha} \qquad
   \quad \
  W^{M_2}{}_{M_1} \rightarrow W_{(2)}^\alpha{}_M
  \nonumber
  \\
  & & R^{a_1 a_2 a_3 , M_3}  \rightarrow R^{a_1 a_2 a_3 , A} \quad \quad  W^{M_3}{}_{M_2}
  \rightarrow W_{(3)}^A{}_\alpha
  \nonumber \\
  & & R^{a_1 ...a_4 , M_4}   \rightarrow R^{a_1 ...a_4 , \alpha
  \beta} \quad  \quad W^{M_4}{}_{M_3} \rightarrow W_{(4)}^{\alpha \beta}{}_A
  \quad .
  \end{eqnarray}
Eq. (\ref{2.19}), arising from the Jacobi identity between two
1-forms and momentum, reads in this case
  \begin{equation}
  W_{(2)}^\alpha{}_P D_\alpha^{MN} = 2 X^{(MN)}{}_P \quad ,
  \label{WThetarelationinD=4}
  \end{equation}
where
  \begin{equation}
  X^{MN}{}_P = \Theta^M_\alpha D^\alpha_P{}^N \quad .
  \end{equation}
Using eq. (\ref{E7metricproptokilling}), from eq.
(\ref{WThetarelationinD=4}) one gets
  \begin{equation}
  W_{(2)}^\alpha{}_M = -2 D^\beta_M{}^N D^\alpha_{NP}
  \Theta^P_\beta \quad . \label{W2thetaconditioninD=4}
  \end{equation}
Eq. (\ref{2.16}) for $n=2$, which is the condition that the Jacobi
identity involving the 1-form, the 2-form and momentum is satisfied,
reads
  \begin{equation}
  S^{M\alpha}_A W_{(3)}^A{}_\beta = - W_{(2)}^\alpha{}_N
  D_\beta^{MN} + \Theta^M_\gamma f^{\gamma \alpha}{}_\beta \quad .
  \end{equation}
This has to be compatible with the conditions of eqs.
(\ref{gravitinoconditiononS}) and (\ref{otherconditiononS}) that $S$
satisfies. The first condition gives
  \begin{equation}
  2 \Theta^M_\alpha D_{\beta , M}{}^N D^\alpha_N{}^P +
  \Theta^P_\beta - X^{MN}{}_M D_{\beta , N}{}^P = 0 \quad ,
  \label{onlythetaconditioninD=4}
  \end{equation}
while the second is identically satisfied. If we then contract this
last equation with $D^\beta_P{}^Q$ we get
  \begin{equation}
  X^{MN}{}_M = \Theta^M_\alpha D^\alpha_M{}^N = 0 \quad , \label{gravitinoconditiononthetainD=4}
  \end{equation}
and plugging this into (\ref{onlythetaconditioninD=4}) and comparing
with (\ref{W2thetaconditioninD=4}) one obtains
  \begin{equation}
  W_{(2)}^\alpha{}_M = - \Theta^\alpha_M \quad .
  \end{equation}
Substituting this in eq. (\ref{WThetarelationinD=4}) gives
  \begin{equation}
  X^{(MNP)} = 0 \quad , \label{otherconditiononthetaD=4}
  \end{equation}
and contracting this with $D_{\alpha , NP}$ gives
  \begin{equation}
  \Theta^M_\alpha = -2 D_{\alpha , N}{}^P   \Theta^N_\beta D^\beta_P{}^M  \quad
  . \label{DDconditiononthetainfour}
  \end{equation}
The two conditions of eqs. (\ref{gravitinoconditiononthetainD=4})
and (\ref{DDconditiononthetainfour}) project the embedding tensor
$\Theta$ to belong to the ${\bf 912}$ of $E_7$. Therefore we have
shown that $E_{11}$ produces all the linear (or representation)
constraints on $\Theta^M_\alpha$. The Jacobi identities at the next
level then give
  \begin{equation}
  W_{(4)}^{\alpha\beta}{}_A = -2 \Theta_M^{[\alpha} S^{M \beta] }_A
  \quad .
  \end{equation}

The embedding tensor also satisfies quadratic constraints that
follow from the general analysis of section 2. In particular eq.
(\ref{2.28}), resulting from the Jacobi identity involving
$\Theta^M_\alpha R^\alpha$, $R^{a, M}$ and momentum, together with
eq. (\ref{2.27}) for $n=2$, resulting from the Jacobi identity
involving the 2-form and two momentum operators, and which reads in
this case
   \begin{equation}
   \Omega_{MN} \Theta^M_\alpha \Theta^N_\beta = 0 \quad ,
   \end{equation}
imply that the quantities $(X^M )^N{}_P$ are the generators of the
subgroup of $E_7$ that in gauged. This analysis thus exactly
reproduces all the constraints of \cite{dWSTD=4}. It is important to
stress that from $E_{11}$ all the constraints arise from imposing
the consistency of the deformed algebra.

To summarise, the Jacobi identities impose that the commutators of
the deformed $E_{11}$ generators with momentum are
   \begin{eqnarray}
   & & [ R^{a ,M} , P_b ] = - g \delta^a_b \Theta^M_\alpha R^\alpha
   \nonumber \\
   & & [ R^{a_1 a_2 , \alpha} , P_b ] = g \Theta^\alpha_M
   \delta^{[a_1}_b R^{a_2 ] , M} \nonumber \\
   & & [ R^{a_1 a_2 a_3 ,A} , P_b ] = - g S^A_{M\alpha} [
   \Theta^\alpha_N D_\beta^{MN} + \Theta^M_\gamma f^{\gamma
   \alpha}{}_\beta ] \delta^{[a_1}_b R^{a_2 a_3 ] , \beta} \nonumber
   \\
   & & [ R^{a_1 ...a_4 , \alpha \beta} , P_b ]  = 2 g
   \Theta_M^{[\alpha} S^{M \beta] }_A \delta^{[a_1}_b R^{a_2 ...a_4
   ], A} \quad .
   \end{eqnarray}
From these commutators, as well as the $E_{11}$ commutators of eq.
(\ref{E11algebrainfourdimensions}), and using the group element in
eq. (\ref{fourdimensionalgroupelement}), one determines the field
strengths and gauge transformations of the fields. The result is
   \begin{eqnarray}
   & & F_{a_1 a_2 , M} = 2 [\partial_{[a_1} A_{a_2 ] , M} - g
   \Theta^\alpha_M A_{a_1 a_2 , \alpha} + {g \over 2} A_{[a_1 , N}
   A_{a_2 ] , P} X^{NP}{}_M ] \nonumber \\
   & & F_{a_1 a_2 a_3 , \alpha} = 3 [\partial_{[a_1} A_{a_2 a_3 ] ,
   \alpha} + {1 \over 2} \partial_{[a_1} A_{a_2 , M} A_{a_3] ,N}
   D_\alpha^{MN} + g S^A_{M \beta} (\Theta_N^\beta D_\alpha ^{MN} +
   \Theta^M_\gamma f^{\gamma \beta}{}_\alpha ) A_{a_1 a_2 a_3 , A} \nonumber \\
   & & \qquad \quad -
   g \Theta_M^\beta D_\alpha ^{MN} A_{[ a_1 a_2 , \beta} A_{a_3 ],
   N} + {g \over 6} A_{[a_1 , M} A_{a_2 , N} A_{a_3 ] , P}
   X^{MN}{}_Q D_\alpha^{QP} ]\nonumber \\
   & & F_{a_1 ...a_4 , A} = 4[ \partial_{[a_1} A_{a_2 a_3 a_4 ] , A} -
   S^{M \alpha}_A  \partial_{[a_1} A_{a_2 a_3 , \alpha } A_{a_4 ] ,
   M} - {1 \over 6} D_\alpha^{MN} S^{P\alpha}_A  \partial_{[a_1}
   A_{a_2 , M} A_{a_3 , N} A_{a_4 ], P} \nonumber \\
   & & \qquad \quad -2 g \Theta_M^\alpha S^{M\beta}_A A_{a_1 ...a_4
   , \alpha \beta} - g S_{M \alpha}^B ( \Theta_N^\alpha D_\beta^{MN}
   + \Theta^M_\gamma f^{\gamma\alpha}{}_\beta ) S^{P\beta}_A A_{[a_1
   a_2 a_3 , B} A_{a_4 ] , P} \nonumber \\
   & & \quad \qquad - {g \over 2} \Theta_M^\alpha S^{M\alpha}_A
   A_{[a_1 a_2 , \alpha} A_{a_3 a_4 ] , \beta} + { g \over 2}
   \Theta_P^\alpha D_\beta^{MP} S^{N\beta}_A A_{[a_1 a_2 , \alpha}
   A_{a_3 , M} a_{a_4 ], N} \nonumber \\
   & & \quad \qquad - {g \over 24} X^{MN}{}_R D_\alpha^{PR}
   S^{Q\alpha}_A A_{[a_1 , M} A_{a_2 , N} A_{a_3 , P} A_{a_4 ] , Q}
   ]
   \quad ,
  \end{eqnarray}
transforming covariantly under
  \begin{eqnarray}
  & & \delta A_{a , M} = a_{a , M} - g X^{NP}{}_M \Lambda_N A_{a, P}
  \nonumber \\
  & & \delta A_{a_1 a_2 , \alpha } = a_{a_1 a_2, \alpha } + {1 \over 2} D_\alpha^{MN} a_{[a_1 ,
  M} A_{a_2 ] , N}
  - g
  \Theta^M_\beta \Lambda_M f^{\beta \gamma}{}_\alpha A_{a_1 a_2 ,
  \gamma}\nonumber \\
  & & \delta A_{a_1 a_2 a_3 , A} = a_{a_1 a_2 a_3 , A} +
  S^{M\alpha}_A a_{[a_1 , M} A_{a_2 a_3 ]  , \alpha} - {1 \over 6}
  S^{M \alpha}_A D_\alpha^{NP} A_{[a_1 , M} A_{a_2 , N} a_{a_3 ] ,
  P} \nonumber \\
  & & \quad \qquad - g \Theta^M_\alpha D^\alpha_A{}^B \Lambda_M A_{a_1 a_2 a_3 ,
  B} \nonumber \\
  & & \delta A_{a_1 ...a_4 , \alpha \beta} = \partial_{[a_1}
  \Lambda_{a_2 a_3 a_4 ], \alpha \beta} + {1 \over 2} a_{[ a_1 a_2 ,
  \alpha} A_{a_3 a_4 ] , \beta} + C^{MA}_{\alpha \beta} a_{[a_1 ,
  M} A_{a_2 a_3 a_4 ] , A} \nonumber \\
  & & \qquad \quad - {1 \over 24} D_\gamma^{PQ}
  C^{MA}_{\alpha \beta} S^{N \gamma}_A A_{[a_1 , M} A_{a_2 , N}
  A_{a_3 , P} a_{a_4 ] , Q} + {1 \over 4} D_{[\alpha}^{MN} A_{[a_1
  a_2 , \beta] } A_{a_3 , M} a_{a_4 ] , N} \nonumber \\
  & & \quad \qquad + 2 g \Theta^M_\delta
  f^{\delta \gamma}{}_{[\alpha} \Lambda_M A_{a_1 ...a_4 , \beta ]
  \gamma} \quad ,
  \end{eqnarray}
where $D^\alpha_A{}^B$ are the generators in the ${\bf 912}$ and the
parameters $a_{a , M}$, $a_{a_1 a_2 , \alpha}$ and $a_{a_1 a_2 a_3 ,
A}$ are defined in terms of the gauge parameters as
  \begin{eqnarray}
  & & a_{a , M} = \partial_a \Lambda_M + g \Theta_M^\alpha
  \Lambda_{a , \alpha} \nonumber \\
  & & a_{a_1 a_2 , \alpha} = \partial_{[a_1 } \Lambda_{a_2 ],
  \alpha} - g S^A_{M\beta} ( \Theta_N^\beta D_\alpha^{MN} +
  \Theta^M_\gamma f^{\gamma \beta}{}_\alpha ) \Lambda_{a_1 a_2 , A}
  \nonumber \\
  & & a_{a_1 a_2 a_3 , A} = \partial_{[a_1 } \Lambda_{a_2 a_3 ],
  A} + 2 g \Theta_M^\alpha S^{M\beta}_A \Lambda_{a_1 a_2 a_3 ,
  \alpha \beta} \quad .
  \end{eqnarray}
These are the field strengths and gauge transformations of any
gauged maximal supergravity theory in four dimensions.

\section{D=5}
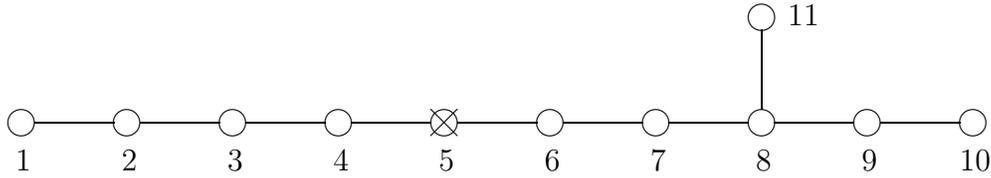
\begin{figure}[h]
\begin{center}
\begin{picture}(380,70)
\multiput(10,10)(40,0){6}{\circle{10}}
\multiput(250,10)(40,0){3}{\circle{10}} \put(370,10){\circle{10}}
\multiput(15,10)(40,0){9}{\line(1,0){30}} \put(290,50){\circle{10}}
\put(290,15){\line(0,1){30}} \put(8,-8){$1$} \put(48,-8){$2$}
\put(88,-8){$3$} \put(128,-8){$4$} \put(168,-8){$5$}
\put(208,-8){$6$} \put(248,-8){$7$} \put(288,-8){$8$}
\put(328,-8){$9$} \put(365,-8){$10$} \put(300,47){$11$}
\put(165,5){\line(1,1){10}} \put(165,15){\line(1,-1){10}}
\end{picture}
\caption{\sl The $E_{11}$ Dynkin diagram corresponding to
5-dimensional supergravity. The internal symmetry group is
$E_{6(+6)}$. \label{Dynkinfive}}
\end{center}
\end{figure}

We now consider the five-dimensional case. The bosonic sector of the
maximal massless supergravity theory in five dimensions
\cite{sugraD=5} contains 42 scalars parametrising the manifold
$E_{6(+6)}/USp(8)$, the metric and a 1-form in the ${\bf 27}$. This
theory arises from the decomposition of $E_{11}$ appropriate to five
dimensions whose Dynkin diagram is shown in fig. \ref{Dynkinfive}.
The form generators up to the 4-form included that occur in this
decomposition of $E_{11}$ with respect to $GL(5, \mathbb{R}) \otimes
E_6$ are \cite{fabiopeterE11origin}
  \begin{equation}
  R^\alpha  \ ({\bf 78}) \quad R^{a , M} \ ({\bf \overline{27}})
  \quad R^{ab}{}_M \ ({\bf 27})
  \quad R^{abc , \alpha} \ ({\bf 78}) \quad
  R^{abcd}{}_{MN} \ ({\bf \overline{351}}) \quad ,
  \label{listofgeninD=5}
  \end{equation}
where  $R^\alpha$, $\alpha = 1 ,\dots ,78$ are the $E_6$ generators,
and an upstairs $M$ index, $M= 1 ,\dots , 27$, corresponds to the
${\bf \overline{27}}$ representation of $E_6$, a downstairs  $M$
index to the ${\bf 27}$ of $E_6$ and the 4-forms are antisymmetric
in $MN$, thus belonging to the ${\bf \overline{351}}$. The
commutation relations for the $E_6$ generators is
  \begin{equation}
  [R^\alpha , R^\beta ]= f^{\alpha\beta}{}_{\gamma} R^\gamma \quad ,
  \end{equation}
where $f^{\alpha\beta}{}_{\gamma}$ are the structure constants of
$E_6$. The commutation relations of $R^\alpha$ with all the other
generators is determined by the $E_6$ representations that they
carry. This gives
  \begin{eqnarray}
  & & [R^\alpha , R^{a,M} ]= (D^\alpha )_N{}^M R^{a, N} \nonumber \\
  & & [R^\alpha , R^{ab}{}_M ]= -(D^\alpha )_M{}^N R^{ab}{}_N
  \nonumber \\
  & &  [R^\alpha , R^{abc ,\beta} ]= f^{\alpha\beta}{}_{\gamma}
  R^{abc,\gamma} \nonumber \\
  & & [R^\alpha , R^{abcd}{}_{MN} ]= -(D^\alpha )_M{}^P
  R^{abcd}{}_{PN}-(D^\alpha )_N{}^P  R^{abcd}{}_{MP}\quad ,
  \end{eqnarray}
where $(D^\alpha )_N{}^M $ obey
  \begin{equation}
  [D^\alpha , D^\beta ]_M{}^N = f^{\alpha\beta}{}_\gamma (D^\gamma )_M{}^N \quad
  .
  \end{equation}
The commutation relations of all the other generators are
  \begin{eqnarray}
  & & [R^{a,M} , R^{b,N} ]= d^{MNP} R^{ab}{}_P \nonumber \\
  & & [R^{a,N} , R^{bc}{}_M ]= g_{\alpha\beta} (D^\alpha )_M{}^N R^{abc, \beta}
  \nonumber \\
  & & [R^{ab}{}_M , R^{cd}{}_N ]= R^{abcd}{}_{MN} \nonumber \\
  & & [R^{a, P}, R^{bcd , \alpha} ]= S^{\alpha P, MN} R^{abcd}{}_{MN}
  \quad \label{E11algebrainfivedimensions}
  \end{eqnarray}
where $d^{MNP}$ is the completely symmetric invariant tensor of
$E_6$ and $g^{\alpha\beta}$ is defined by the relation
  \begin{equation}
  D^\alpha_M{}^N D^\beta_N{}^M = g^{\alpha \beta}
  \end{equation}
and is thus proportional to the Cartan-Killing metric of $E_6$, and
is the metric that is used to raise and lower indices in the adjoint
(we are using the $E_6$ conventions of \cite{dWSTgeneral}, that are
summarised in appendix A). Another useful identity is
  \begin{equation}
  f_{\alpha \beta\gamma} f^{\alpha \beta \delta} = - 4
  \delta^\delta_\gamma
  \quad ,
  \end{equation}
where $f^{\alpha \beta \gamma}$ are the structure constants of
$E_6$. $S^{\alpha P, MN}$ is also an invariant tensor, antisymmetric
with respect to $MN$, and the Jacobi identity between two 1-forms
and one 2-form gives
  \begin{equation}
  g_{\alpha \beta} D^\alpha_Q{}^{(P} S^{\beta R ) , MN} =  -{1 \over 2} \delta_{Q}^{[M} d^{N]PR}
  \quad . \label{gDS=deltad}
  \end{equation}
Using the fact that $d^{MNP}$ is completely antisymmetric one
derives from this the condition
  \begin{equation}
  g_{\alpha \beta} D^\alpha_M{}^N S^{\beta M , PQ} = 0 \quad .
  \label{gravitinoconditiononSinfivedimensions}
  \end{equation}
One can show that all the Jacobi identities involving the generators
are satisfied using the commutators listed above
\cite{fabiopeterE11origin}.

We now show that $S^{\alpha M, NP}$ is proportional to
$D^{\alpha}_Q{}^{[N} d^{P]MQ}$ and determine the coefficient of
proportionality. We introduce the invariant tensor $d_{MNP}$ in the
completely symmetric product of three ${\bf 27}$ indices, that
satisfies \cite{dWSTD=5}
  \begin{equation}
  d^{MNP} d_{MNQ} = \delta^P_Q \quad .
  \label{ddisdeltaE6relationinD=5}
  \end{equation}
Observe that the normalisation used here differs from the one used
in \cite{fabiopeterextendedspacetime}, where the same contraction
produced the delta function with a factor 5. This simply corresponds
to a rescaling of $d$ by $\sqrt{5}$. In appendix A we derive the
useful relation
  \begin{equation}
  g_{\alpha \beta} D^\alpha_M{}^N D^\beta_P{}^Q = {1 \over 6} \delta^N_P
  \delta_M^Q + {1 \over 18} \delta_M^N \delta_P^Q -{5 \over 3} d^{NQR} d_{MPR}
  \quad . \label{finalDDequationinD=5}
  \end{equation}
Using this relation one shows that eq. (\ref{gDS=deltad}) implies
  \begin{equation}
  S^{\alpha M , NP} = -3 D^\alpha_Q{}^{[N} d^{P] MQ} \quad
  .\label{S=Ddinfivedimensions}
  \end{equation}
Notice that this relation differs from the one in
\cite{fabiopeterextendedspacetime} because of the different
conventions used in that paper. In particular in
\cite{fabiopeterextendedspacetime} the generators were normalised in
such a way that the first coefficient in eq.
(\ref{finalDDequationinD=5}) was equal to 1. Contracting eq.
(\ref{gDS=deltad}) with $D^\gamma_P{}^Q$ and using eq.
(\ref{S=Ddinfivedimensions}) one finally gets
  \begin{equation}
  S^{\alpha M ,NP} + {3 \over 2} ( D^\alpha D_\beta )_Q{}^M S^{\beta
  Q , NP} = 0\quad . \label{S=-3over2DDS}
  \end{equation}
As we will describe in detail in appendix A, the conditions of eqs.
(\ref{gravitinoconditiononSinfivedimensions}) and
(\ref{S=-3over2DDS}) are the conditions that the indices $\alpha M$
in $S^{\alpha M ,NP}$ are in the ${\bf \overline{351}}$. Indeed,
given that the $NP$ indices are antisymmetric and thus form the
${\bf 351}$, the only way of building an invariant tensor from
tensoring this with the product ${\bf \overline{27} \otimes 78}$ is
to project this product on the ${\bf \overline{351}}$. Later in this
section we will derive from $E_{11}$ the same projection rules for
the embedding tensor.

From the group element
 \begin{equation}
  g = e^{x \cdot P} e^{A_{a_1
  ...a_4}^{MN} R^{a_1 ... a_4}_{MN} } e^{A_{a_1 a_2 a_3, \alpha}
  R^{a_1 a_2 a_3, \alpha}} e^{A_{a_1 a_2}^M R^{a_1 a_2}_M } e^{A_{a
  ,M} R^{a , M}} e^{\phi_\alpha R^\alpha} \quad ,
  \label{fivedimgroupelement}
  \end{equation}
one can compute the Maurer-Cartan form using the fact that the
generators commute with momentum in the massless theory. The
complete antisymmetrisation of the indices leads to the
gauge-invariant field-strengths of the massless theory obtained in
\cite{fabiopeterE11origin}.

We now consider the deformed case. This was analysed in detail in
\cite{fabiopeterogievetsky}, where it was shown that introducing the
Ogievetsky generators and deforming the algebra one obtains the
field strengths of all the fields and dual fields of the gauged
maximal five-dimensional supergravity which had been previously
obtained in \cite{fabiopeterextendedspacetime}. We now only
concentrate on the deformed $E_{11}$ generators, as we do in all
other cases in this paper, which is  all one needs to determine the
field strengths of the massive theory. This is completely consistent
if one simply assumes that the indices are antisymmetrised, and
indeed the completely antisymmetric part of the Ogievetsky
generators vanishes. As it is clear from the discussion in section
2, considering only the constraints coming from taking into account
the deformed $E_{11}$ generators is enough to determine the whole
deformed algebra. The following analysis thus determines all the
possible massive deformations of the algebra of eq.
(\ref{E11algebrainfivedimensions}).

The general analysis of section 2 can be applied to the
five-dimensional case making the identifications
  \begin{eqnarray}
  & & R^{a_1 , M_1 } \rightarrow R^{a_1 , M} \qquad \qquad \quad
  \Theta^{M_1}{}_\alpha \rightarrow \Theta^M_\alpha \nonumber \\
  & & R^{a_1 a_2 , M_2} \rightarrow R^{a_1 a_2}{}_M \qquad \quad \ \
  W^{M_2}{}_{N_1} \rightarrow W_{MN} \nonumber \\
  & & R^{a_1 a_2 a_3, M_3} \rightarrow R^{a_1 a_2 a_3 ,\alpha}
  \qquad \ W^{M_3}{}_{M_2} \rightarrow W_{(3)}^{\alpha M} \nonumber\\
  & & R^{a_1 ... a_4, M_4} \rightarrow R^{a_1 ... a_4}{}_{MN}
  \qquad W^{M_4}{}_{M_3} \rightarrow W_{(4) MN \alpha} \quad .
  \end{eqnarray}
The Jacobi identity among two 1-forms and momentum gives eq.
(\ref{2.19}), which in this case is
  \begin{equation}
  d^{MNQ} W_{QP} = 2 X^{(MN)}{}_P \quad ,
  \end{equation}
where as usual
  \begin{equation}
  X^{MN}{}_P = \Theta^M_\alpha D^\alpha_P{}^N \quad .
  \end{equation}
Contracting with $d_{MNR}$ one then gets
  \begin{equation}
  W_{RP} = 2d_{MN[R} \Theta^M_\alpha D^\alpha_{P]}{}^N - d_{RPN}
  X^{MN}{}_M \quad , \label{WsymmWantisymm}
  \end{equation}
where the first term is antisymmetric and the second is symmetric in
$RP$. The Jacobi identity involving the 1-form, the 2-form and
momentum gives eq. (\ref{2.16}) for $n=2$, which is
  \begin{equation}
  W_{NP} d^{PMQ} + X^{MQ}{}_N = - W_{(3)}^Q{}_{\alpha}
  D^\alpha_N{}^M \label{W2W3Thetainfivedimensions}
  \quad ,
  \end{equation}
while the Jacobi identity involving two 2-forms and momentum gives
  \begin{equation}
  W_{(4)}^\alpha{}_{MN} = W_{MP} D^\alpha_N{}^P - W_{NP}
  D^\alpha_M{}^P \label{W4=W2Dinfivedimensions}
  \end{equation}
and the Jacobi identity involving the 1-form, the 3-form and
momentum gives
  \begin{equation}
  W_{(4)}^\gamma{}_{NP} S_\alpha^{M ,NP} = \Theta^M_\beta
  f^\beta{}_{\alpha \gamma} - W_{(3)}^N{}_\alpha D_{\gamma , N}{}^M
  \quad . \label{jacobi13pinfivedimensions}
  \end{equation}
Substituting $W_{(4)}^\alpha{}_{MN}$ given in eq.
(\ref{W4=W2Dinfivedimensions}) in this last equation and contracting
$\alpha$ and $\gamma$ gives
  \begin{equation}
  2 W_{NQ} D_{\alpha , P}{}^Q S^{\alpha M , NP} = - W_{(3)}^N{}_\alpha D_{\alpha , N}{}^M
  \quad ,
  \end{equation}
and using eq. (\ref{W2W3Thetainfivedimensions}), as well as eqs.
(\ref{S=Ddinfivedimensions}) and (\ref{finalDDequationinD=5}), one
obtains
  \begin{equation}
  W_{MN} d^{MNP} = 0 \quad .
  \end{equation}
From eq. (\ref{W2W3Thetainfivedimensions}) one can deduce that this
implies
  \begin{equation}
  X^{MN}{}_M = D^\alpha_M{}^N \Theta^M_\alpha = 0 \quad ,
  \label{gravitinoconditionforthetainfived}
  \end{equation}
so that from eq. (\ref{WsymmWantisymm}) one gets that $W_{MN}$ must
be antisymmetric, that is it belongs to the ${\bf \overline{351}}$.
This also implies that
  \begin{equation}
  W_{(3)}^M{}_\alpha  = \Theta^M_\alpha \quad ,
  \end{equation}
and substituting everything in eq. (\ref{jacobi13pinfivedimensions})
one gets
  \begin{equation}
  f^{\alpha\beta}{}_\gamma \Theta^Q_\beta - D^\alpha_P{}^Q \Theta^P_\gamma = 2 D^\alpha_M{}^P W_{PN} g_{\beta
  \gamma} S^{\beta Q,MN} \quad ,
  \label{finaleqfivedimensionsfthetaDtheta}
  \end{equation}
where $W_{MN}$ is in the ${\bf \overline{351}}$. If $\Theta$ was not
along the ${\bf \overline{351}}$ this equation would be inconsistent
because it would imply the invariance of $\Theta$ under $E_6$.
Therefore the embedding tensor has to belong to the ${\bf
\overline{351}}$. To determine this more rigorously, we now show
that eq. (\ref{finaleqfivedimensionsfthetaDtheta}) leads to the
projection for $\Theta$ analogous to that in eq.
(\ref{S=-3over2DDS}). Contracting eq.
(\ref{finaleqfivedimensionsfthetaDtheta}) with $D^\alpha_Q{}^R$
gives
  \begin{equation}
  {26 \over 9} \Theta^R_\gamma + ( D_\gamma D^\beta )_Q{}^R
  \Theta^Q_\beta = {20 \over 9} W_{NP} S_\gamma^{R, NP} \quad ,
  \end{equation}
while contracting it with $f_{\alpha \gamma \delta}$ gives
  \begin{equation}
  4 \Theta^R_\gamma + ( D_\gamma D^\beta )_Q{}^R
  \Theta^Q_\beta = {10 \over 3} W_{NP} S_\gamma^{R, NP} \quad ,
  \end{equation}
and combining these two equations one gets
  \begin{equation}
  \Theta^R_\gamma + {3 \over 2}( D_\gamma D^\beta )_Q{}^R
  \Theta^Q_\beta = 0 \quad .
  \end{equation}
This equation, together with eq.
(\ref{gravitinoconditionforthetainfived}), projects the embedding
tensor on the ${\bf \overline{351}}$.

The embedding tensor also satisfies quadratic constraints, as
discussed in complete generality in section 2. The Jacobi identity
between $\Theta^M_\alpha {R}^\alpha$, ${R}^{a , N}$ and $P_b$ gives
   \begin{equation}
   \Theta^M_\alpha \Theta^N_\beta f^{\alpha \beta}{}_\gamma -
   \Theta^P_\gamma X^{MN}{}_P =0 \quad ,
   \end{equation}
while the Jacobi identity between the 2-form and two momentum
operators gives
  \begin{equation}
  \Theta^M_\alpha W_{MN} = 0 \quad .
  \end{equation}
Combining these two conditions one obtains the condition that the
embedding tensor is invariant under the gauge group, which is the
subgroup of $E_6$ generated by $\Theta^M_\alpha R^\alpha$.

To summarise, we have shown that the Jacobi identities constrain the
commutators of the deformed $E_{11}$ generators and momentum to be
  \begin{eqnarray}
  & &   [ {R}^{a , M} , P_b ] = - g \delta^a_b \Theta^M_\alpha
  {R}^\alpha \nonumber \\
  & &   [ {R}^{a_1 a_2}{}_M , P_b ] = -  g W_{MN} \delta^{[a_1}_b
  {R}^{ a_2 ] , N} \nonumber \\
  & &   [ {R}^{a_1 a_2 a_3}{}_\alpha , P_b ] = - g \Theta^M_\alpha
  \delta^{[a_1}_b {R}^{a_2 a_3 ]}{}_M \nonumber \\
  & &   [ {R}^{a_1 ... a_4}{}_{MN} , P_b ] = - 2 g W_{[M\vert P
  \vert } D^\alpha_{N]}{}^P \delta^{[ a_1}_b {R}^{a_2 a_3 a_4
  ]}{}_\alpha \quad .
  \end{eqnarray}
From this commutators,as well as the $E_{11}$ commutators of eq.
(\ref{E11algebrainfivedimensions}), and using the group element of
eq. (\ref{fivedimgroupelement}), we determine the field strengths of
the fields using the general results of section 2. The result is
  \begin{eqnarray}
  & & \! \! \! \! \! \! \!   F_{a_1 a_2 ,M} = 2 [\partial_{[a_1} A_{a_2 ] ,M} + {1 \over 2} g
  X^{[NP]}_M A_{[a_1 ,N} A_{a_2], P} -  g W_{MN}A_{a_1
  a_2}^N ]\nonumber \\
  & &  \! \! \! \! \! \! \!  F^{M}_{a_1 a_2 a_3} = 3 [\partial_{[a_1} A^{M}_{a_2 a_3 ]} + {1 \over 2} \partial_{[a_1} A_{a_2 ,N}
  A_{a_3 ] ,P} d^{MNP} - 2 g X^{(MN)}_P A_{[a_1 a_2}^P A_{a_3 ], N}
  \nonumber \\
  & &  +{1 \over 6} g X^{[NP]}_R d^{RQM} A_{[a_1 , N} A_{a_2 , P} A_{a_3 ] , Q} + g \Theta^M_\alpha A_{a_1 a_2
  a_3 }^\alpha ] \nonumber \\
  &  & \! \! \! \! \! \! \!
  F^{\alpha}_{a_1 \dots  a_4} = 4 [\partial_{[a_1} A^{\alpha}_{a_2 \dots a_4 ]} - {1 \over 6} \partial_{[a_1}
  A_{a_2 ,M} A_{a_3 ,N} A_{a_4 ] ,P} d^{MNQ} D^\alpha_Q{}^P - \partial_{[a_1} A^{M}_{a_2
  a_3} A_{a_4 ],N} D^\alpha_M{}^N \nonumber \\
  & &   +  g D^\alpha_M{}^P \Theta^M_\beta A_{[a_1 , P} A^\beta_{a_2 \dots a_4 ]} + 2 g  D^\alpha_M{}^P  W_{PN}
  A^{MN}_{a_1 \dots a_4} - {g  \over 2} D^\alpha_M{}^P W_{PN} A^M_{[a_1 a_2} A^N_{a_3 a_4
  ]} \nonumber \\
  & &   -  g  D^\alpha_M{}^P X^{(MR)}_Q A_{[a_1 , P} A_{a_2 , R} A_{a_3 a_4 ]}^Q - {1 \over 24} g X^{[MN]}_R d^{RPS}
  D^\alpha_S{}^Q A_{[a_1 , M} A_{a_2 , N} A_{a_3 , P} A_{a_4 ] , Q}
  ]
  \quad .
  \end{eqnarray}
These are the field-strengths of the five-dimensional gauged maximal
supergravity \cite{fabiopeterextendedspacetime}. One can also derive
the gauge transformations of the fields from the non-linear
realisation. The result is
  \begin{eqnarray}
  & &   \delta A_{a , M}= a_{a ,M} -g \Lambda_N X^{NP}{}_M A_{a ,P}  \nonumber \\
  & &   \delta A^M_{a_1 a_2}=  a^M_{a_1 a_2}  +{1\over 2} a_{[a_1 , N}  A_{a_2 ]P}
  d^{MNP} +g \Lambda_N X^{NM}{}_P A_{a_1a_2}^P  \nonumber
  \\
  & &   \delta A_{a_1a_2a_3}^\alpha=  a_{a_1 a_2 a_3 }^\alpha +  a_{[a_1 , M}
  A^{N}_{a_2 a_3 ]} D^\alpha_N{}^M + {1 \over 6}
  a_{[a_1 , M}   A_{a_2 , N}  A_{a_3 ], P} d^{MNQ}
  D^\alpha_Q{}^P \nonumber \\
  & & \quad
  - g \Lambda_M  \Theta^M_\beta f^{\alpha\beta}{}_\gamma A^\gamma_{a_1 a_2
  a_3}  \nonumber \\
  & & \delta A_{a_1 ...a_4}^{MN} = \partial_{[a_1} \Lambda_{a_2 a_3
  a_4 ]}^{MN} +{1 \over 2} a_{[a_1 a_2}^{[M} A_{a_3 a_4}^{N]} +
  a_{[a_1 , P} A_{a_2 a_3 a_4 ]}^\alpha g_{\alpha \beta} S^{\beta P,
  MN} \nonumber \\
  & & \quad - {1 \over 24} a_{[a_1 , P} A_{a_2 , Q} A_{a_3 , R} A_{a_4 ] ,
  S} d^{PQT} D^\alpha_T{}^R S^{\beta S, MN} g_{\alpha \beta} - {1
  \over 4} a_{ [a_1 , P} A_{a_2 , Q} A_{a_3 a_4 ]}^{[M} d^{N]PQ}
  \nonumber \\
  & & \quad - 2 g \Lambda_P X^{P[M}{}_Q A_{a_1 ...a_4 }^{N]Q}
   \quad ,
  \end{eqnarray}
where the parameters $a_{a , M}$, $a_{a_1 a_2}^M$ and $a_{a_1 a_2
a_3}^\alpha$ are defined in terms of the gauge parameters as
  \begin{eqnarray}
  & & a_{a , M} = \partial_a \Lambda_M + g W_{MN} \Lambda_a^N \nonumber \\
  & & a_{a_1 a_2}^M = \partial_{[a_1 } \Lambda_{a_2 ]}^M - g \Theta^M_\alpha \Lambda_{a_1 a_2}^\alpha
  \nonumber \\
  & & a_{a_1 a_2 a_3}^\alpha =  \partial_{[a_1 } \Lambda_{a_2 a_3
  ]}^\alpha -2 g W_{MP} D^\alpha_N{}^P \Lambda_{a_1 a_2 a_3}^{MN}
  \quad .
  \end{eqnarray}
We also compute the field strength of the 4-form up to a term
involving the 5-form, the result being
  \begin{eqnarray}
  & &   F^{MN}_{a_1 \dots  a_5} = 5 [\partial_{[a_1} A^{MN}_{a_2 \dots a_5
  ]}+  \partial_{[a_1} A^{\alpha}_{a_2 \dots a_4} A_{a_5 ] , P }
  S^{\beta P, MN} g_{\alpha \beta} \nonumber \\
  & & \quad - {1 \over 2} D_{\alpha P}{}^Q
  S^{\alpha R , MN} \partial_{[a_1} A_{a_2 a_3 }^P A_{a_4 , Q}
  A_{a_5 ] , R} + { 1 \over 2} \partial_{[a_1 } A_{a_2 a_3}^{[M}
  A_{a_4 a_5 ]}^{N]} \nonumber \\
  & & \quad -{1 \over 24} d^{PRQ} D_{\alpha R}{}^S
  S^{\alpha T, MN} \partial_{[a_1} A_{a_2 ,P} A_{a_3 ,Q} A_{a_4 , S}
  A_{a_5 ] , T} \nonumber \\
  & & \quad + 2 g W_{RQ} D_{\alpha S}{}^Q S^{\alpha P,MN} A_{[a_1
  ...a_4}^{RS} A_{a_5] , P} + g \Theta^{[M}_\alpha A_{[a_1 a_2
  a_3}^\alpha A_{a_4 a_5]}^{N]} \nonumber \\
  & & \quad - {g \over 2} \Theta^P_\alpha D_{\beta , P}{}^Q S^{\beta R, MN}
  A_{[a_1 a_2 a_3}^\alpha A_{a_4 , Q} A_{a_5 ] , R} + {g \over 2}
  W_{RQ} D_{\alpha ,S}{}^Q S^{\alpha P, MN} A_{[a_1 a_2 }^R A_{a_3
  a_4 }^S A_{a_5 ] , P} \nonumber \\
  & & \quad - {g \over 6} W_{TR} d^{RUS} D_{\alpha S}{}^P S^{\alpha
  Q, MN} A_{[a_1 a_2}^T A_{a_3 , U} A_{a_4 , P} A_{a_5 ] ,
  Q}\nonumber \\
  & & \quad - {g \over 5!} X^{PQ}{}_U d^{URV} D_{\beta , V}{}^S S^{
  \beta T, MN} A_{[ a_1 , P} A_{a_2 , Q} A_{a_3 , R} A_{a_4 , S}
  A_{a _5 ], T} ] \quad .
  \end{eqnarray}
In order to compute the complete field strength for the 4-form one
should consider the contribution of the 5-form generators in the
deformed algebra.

\section{D=6}
\begin{figure}[h]
\begin{center}
\begin{picture}(380,70)
\multiput(10,10)(40,0){6}{\circle{10}}
\multiput(250,10)(40,0){3}{\circle{10}} \put(370,10){\circle{10}}
\multiput(15,10)(40,0){9}{\line(1,0){30}} \put(290,50){\circle{10}}
\put(290,15){\line(0,1){30}} \put(8,-8){$1$} \put(48,-8){$2$}
\put(88,-8){$3$} \put(128,-8){$4$} \put(168,-8){$5$}
\put(208,-8){$6$} \put(248,-8){$7$} \put(288,-8){$8$}
\put(328,-8){$9$} \put(365,-8){$10$} \put(300,47){$11$}
\put(205,5){\line(1,1){10}} \put(205,15){\line(1,-1){10}}
\end{picture}
\caption{\sl The $E_{11}$ Dynkin diagram corresponding to
6-dimensional supergravity. The internal symmetry group is
$SO(5,5)$. \label{Dynkinsix}}
\end{center}
\end{figure}
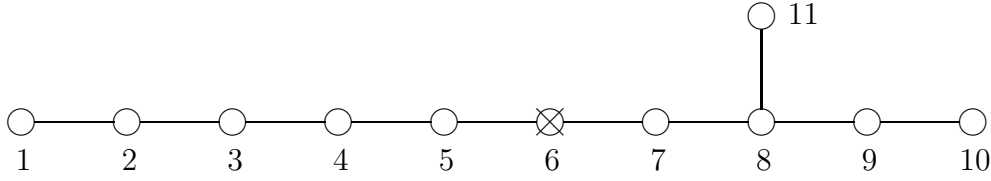

In this section we consider the six-dimensional case. The symmetry
of the massless maximal supergravity theory in 6 dimensions
\cite{taniisixdim} is $SO(5,5)$, and the bosonic sector of the
theory describes 25 scalars parametrising $SO(5,5)/[SO(5) \times
SO(5)]$, the metric, a 1-form in the ${\bf 16}$ and a 2-form in the
${\bf 10}$, whose field strength satisfies a self-duality condition.
From $E_{11}$ this theory arises after deleting node 6 in the
$E_{11}$ Dynkin diagram, as shown in fig. \ref{Dynkinsix}. From the
diagram it is manifest that the 1-forms belong to the spinor
representation.

The positive level $E_{11}$ generators with completely antisymmetric
spacetime indices that arise in six dimensions, without considering
the 6-forms, are
  \begin{eqnarray}
  & & R^{MN} \quad ({\bf {45}}) \qquad R^{a , \dot{\alpha}} \quad ({\bf
  \overline{16}} )
  \qquad R^{a_1 a_2, M} \quad ( {\bf 10} ) \qquad R^{a_1 a_2 a_3 , {\alpha}}
  \quad  ({\bf 16} )\nonumber \\
  & &  R^{a_1 a_2 a_3 a_4, MN} \quad  ({\bf 45} )
  \qquad R^{a_1 ... a_5, M {\alpha}} \quad  ( {\bf \overline{144}})
  \quad , \label{genlistD=6}
  \end{eqnarray}
where  $M = 1 ,\dots ,10$ is a vector index of $SO(5,5)$ and
$\alpha, \dot{\alpha} = 1, \dots , 16$ denote the two spinor
representations of $SO(5,5)$. The scalar generators $R^{MN}$ and the
4-form generators $R^{a_1 a_2 a_3 a_4, MN}$ are antisymmetric in
$MN$ and thus belong to the $\bf 45$ of $SO(5,5)$. Note that the
1-form generators belong to the ${\bf {\overline{16}}}$ of
$SO(5,5)$, which is denoted by the $\dot{\alpha}$ index, as they
belong to the representation complex conjugate to the one of the
vector fields. The 2-forms belong to the $\bf 10$, the 3-forms
belong to the $\bf 16$ and the 5-forms to the $\bf
{\overline{144}}$.

It is useful to list the conventions for the $SO(5,5)$ Gamma
matrices that we are using. In particular, we are using a Weyl
basis, so that the Gamma matrices have the form
  \begin{equation}
  \Gamma_{M ,A}{}^B =  \left( \begin{array}{cc}
  0  &  \Gamma_{M, \alpha}{}^{\dot{\beta}} \\
  \Gamma_{M, \dot{\alpha}}{}^\beta & 0 \end{array} \right) \quad ,
  \end{equation}
where $A=1, ..., 32$. They satisfy the Clifford algebra
  \begin{equation}
  \{ \Gamma_M , \Gamma_N \} = 2 \eta_{MN}
  \end{equation}
where $\eta_{MN}$ is the Minkowski metric. The charge conjugation
matrix is
  \begin{equation}
  C^{AB} =  \left( \begin{array}{cc}
  0 &   C^{\alpha \dot{\beta}}  \\
  C^{\dot{\alpha} \beta} & 0 \end{array} \right) \quad ,
  \end{equation}
which is antisymmetric and unitary, that is
  \begin{equation}
  C^{\alpha \dot{\beta}} =-  C^{\dot{\beta} \alpha}
  \end{equation}
and
  \begin{equation}
  C^\dagger_{\alpha \dot{\beta}} C^{\dot{\beta} \gamma} =
  \delta_\alpha^\gamma \qquad   C^\dagger_{\dot{\alpha}{\beta}}
  C^{{\beta} \dot{\gamma}} = \delta_{\dot{\alpha}}^{\dot{\gamma}}
  \quad ,
  \end{equation}
and satisfies the property
  \begin{equation}
  C \Gamma_M C^\dagger = - \Gamma_M^T \quad .
  \end{equation}
In this section we will make use of various Fierz identities, the
most relevant being
  \begin{equation}
  (C \Gamma_M )^{(\alpha \beta} \Gamma^M{}_{\dot{\alpha}}{}^{\gamma)}
  =0  \qquad \quad
  (C \Gamma_M )^{(\dot{\alpha}  \dot{\beta}} \Gamma^M{}_{{\alpha}}{}^{\dot{\gamma})}
  =0 \quad ,\label{famousfierzidentity}
  \end{equation}
which is the well-known identity of Gamma matrices in ten
dimensions. The 5-form generators satisfy the constraint
  \begin{equation}
  R^{a_1 ... a_5, M {\alpha}} \Gamma_{M {\alpha}}{}^{\dot{\alpha}} =
  0 \quad ,
  \end{equation}
which indeed restricts them in the ${\bf {\overline{144}}}$ of
$SO(5,5)$.

We now analyse the commutation relations. The $SO(5,5)$ algebra is
  \begin{equation}
  [ R^{MN} , R^{PQ} ] = \eta^{MP} R^{NQ} - \eta^{NP} R^{MQ} +
  \eta^{NQ} R^{MP} - \eta^{MQ} R^{NP}
  \end{equation}
while the commutators of the $SO(5,5)$ generators is
  \begin{eqnarray}
  & & [ R^{MN} , R^{a ,\dot{\alpha}} ] = - {1 \over 2}
  \Gamma^{MN}{}_{\dot{\beta}}{}^{\dot{\alpha}} R^{a , \dot{\beta}}
  \nonumber \\
  & & [ R^{MN} , R^{ab ,P} ] = \eta^{MP} R^{ab , N} - \eta^{NP} R^{ab ,
  M}
  \nonumber \\
  & &
  [ R^{MN} , R^{abc ,\alpha} ] = - {1 \over 2}
  \Gamma^{MN}{}_{\beta}{}^{\alpha} R^{abc , {\beta}}
  \end{eqnarray}
and similarly for the higher rank generators. The commutators of the
positive level generators of eq. (\ref{genlistD=6}) are
  \begin{eqnarray}
  & & [ R^{ a_1 ,\dot{\alpha}} , R^{a_2 , \dot{\beta}} ] = ( C \Gamma_M )^{
  \dot{\alpha} \dot{\beta}} R^{a_1 a_2 , M}
  \nonumber \\
  & & [ R^{ a_1 ,\dot{\alpha}} , R^{a_2 a_3 , M} ] =
  \Gamma^M{}_\alpha{}^{\dot{\alpha}}
  R^{a_1 a_2 a_3 , \alpha}
  \nonumber \\
  & &
  [ R^{ a_1 a_2 , M} , R^{a_3 a_4 , N} ] = R^{a_1 ... a_4  , MN}
  \nonumber \\
  & & [ R^{ a_1 ,\dot{\alpha}} , R^{a_2 a_3 a_4 , \alpha} ] = {1 \over 4}( C \Gamma_{MN} )^{
  \dot{\alpha} \alpha} R^{a_1 ... a_4  , MN}
  \nonumber \\
  & & [ R^{ a_1 ,\dot{\alpha}} , R^{a_2 ... a_5 , MN} ] = \Gamma^{[M}{}_\alpha{}^{\dot{\alpha}}
  R^{a_1 ... a_5  , N] \alpha}
  \nonumber \\
  & & [ R^{a_1 a_2 , M} , R^{a_3 a_4 a_5 , \alpha} ] = - {1 \over 2}
  R^{a_1 \dots a_5 ,  M \alpha} \quad .
  \end{eqnarray}
One can show that all the Jacobi identities are satisfied. From this
algebra, one can write down the group element
  \begin{eqnarray}
  g &=&  e^{x \cdot P} e^{A_{a_1
  ...a_5, M \alpha} R^{a_1 ... a_5, M \alpha}} e^{A_{a_1 ... a_4, MN}
  R^{a_1 ... a_4, MN}} e^{A_{a_1 a_2 a_3 , \alpha} R^{a_1 a_2 a_3 ,
  \alpha}} \nonumber \\
  & & e^{A_{a_1 a_2 , M} R^{a_1 a_2  , M}}
  e^{A_{a
  , \dot{\alpha}} R^{a , \dot{\alpha}}} e^{\phi_{MN} R^{MN}}
  \end{eqnarray}
and compute the Maurer-Cartan form. The field strengths of the
massless theory are then obtained antisymmetrising the spacetime
indices of the various terms in the Maurer-Cartan form, and the
field equations of the supergravity theory arise as duality
relations. In particular, the field-strength of the vector is dual
to the field-strength of the 3-form, while the field strength of the
4-form is dual to the derivative of the scalars. The 2-forms satisfy
self-duality conditions, while the field strength of the 5-form
vanishes in the massless theory. In deriving the field strengths of
the massless theory one takes the positive level $E_{11}$ generators
to commute with momentum. In the following we will consider the
deformation of the $E_{11}$ algebra which results from modifying the
commutation relations of the $E_{11}$ generators with momentum
compatibly with the Jacobi identities. As already shown in other
cases these deformations exactly coincide with all the possible
massive deformations of the corresponding supergravity theory.

We thus consider all the consistent deformations of the massless
algebra. Restricting the general analysis of section 2 to the
particular case of six dimensions, we write the most general
commutators of the first three positive level $E_{11}$ generators
with momentum as
  \begin{eqnarray}
  & &   [ R^{a , \dot{\alpha} }, P_b ] = - g \Theta^{\dot{\alpha},
  M N}
  R_{MN} \nonumber \\
  & & [ R^{a_1 a_2 , M} , P_b ] = - g W_{(2)}^M{}_{\dot{\alpha}}
  \delta^{[a_1}_b R^{a_2 ] , \dot{\alpha}} \nonumber \\
  & &
  [ R^{a_1 a_2 a_3 , \alpha} , P_b ] =  - g W_{(3)}^\alpha{}_M
  \delta^{[a_1 }_b R^{a_2 a_3 ], M} \quad ,
  \end{eqnarray}
where $\Theta$ is antisymmetric in $MN$. It turns out that the
Jacobi identities involving these operators are enough to restrict
the representation of $\Theta$ completely and to determine $W_{(2)}$
and $W_{(3)}$ uniquely in terms of $\Theta$. This is what we are
showing now. As done already in other sections for different
dimensions, we can assume that the upstairs spacetime indices are
all antisymmetrised when we compute the Jacobi identities. Indeed,
the terms which are not completely antisymmetric are cancelled by
deforming the $E_{11}$ commutation relations in terms of Og 1
operators. The details of this mechanism were shown in
\cite{fabiopeterogievetsky} for various examples. In this paper we
are only interested in the part of the algebra which is relevant for
the determination of the field strengths, and thus we do not need to
determine the part of the deformation which involves the Og
generators.

The Jacobi identity between two 1-forms and momentum gives the
relation
  \begin{equation}
  ( C \Gamma )^{\dot{\alpha} \dot{\beta}} W_{(2)}^M{}_{\dot{\gamma}}
  = - {1 \over 2} \Gamma_{MN , \dot{\gamma}}{}^{\dot{\alpha}}
  \Theta^{\dot{\beta} , MN} - {1 \over 2} \Gamma_{MN , \dot{\gamma}}{}^{\dot{\beta}}
  \Theta^{\dot{\alpha} , MN} \quad , \label{thetaw2constraintsixdimensions}
  \end{equation}
while the Jacobi identity between the 1-form, the 2-form and
momentum gives
  \begin{equation}
  \Gamma^M_\alpha{}^{\dot{\alpha}} W_{(3)}^{\alpha , N} = 2 \Theta^{
  \dot{\alpha}, MN} - W_{(2)}^M{}_{\dot{\beta}} ( C \Gamma^N
  )^{\dot{\alpha} \dot{\beta}} \label{w2w3andthetainsixdimensions}
  \end{equation}
The antisymmetry of $\Theta$ in $MN$  in the last equation implies
  \begin{equation}
  W_{(3)}^{\alpha , M} = C^{\alpha \dot{\alpha}}
  W_{(2)}^M{}_{\dot{\alpha}} \quad ,
  \end{equation}
as can be shown taking the part of eq.
(\ref{w2w3andthetainsixdimensions}) which is symmetric in $MN$, and
therefore does not contain $\Theta$, and suitably contracting with
Gamma matrices. Eq. (\ref{w2w3andthetainsixdimensions}) thus becomes
  \begin{equation}
  \Theta^{\dot{\alpha}, MN} = - (C \Gamma^{[M} )^{\dot{\alpha}
  \dot{\beta}} W_{(2)}^{N]}{}_{\dot{\beta}} \quad ,
  \end{equation}
and substituting this back in eq.
(\ref{thetaw2constraintsixdimensions}) gives
  \begin{equation}
  [ ( C \Gamma_M )^{\dot{\alpha} \dot{\beta} }
  \delta^{\dot{\delta}}_{\dot{\gamma}} + {1 \over 2} (C \Gamma^N
  )^{\dot{\beta} \dot{\delta}} \Gamma_{MN,
  \dot{\gamma}}{}^{\dot{\alpha}} + {1 \over 2} (C \Gamma^N
  )^{\dot{\alpha} \dot{\delta}} \Gamma_{MN,
  \dot{\gamma}}{}^{\dot{\beta}} ] W_{(2)}^M{}_{\dot{\delta}} = 0
  \quad .
  \end{equation}
Using the Fierz identity of eq. (\ref{famousfierzidentity}) one can
show that this equation implies
  \begin{equation}
  \Gamma_{M , \alpha}{}^{\dot{\alpha}} W_{(2)}^M{}_{\dot{\alpha}} =
  0 \quad .
  \end{equation}
This analysis shows that the most general deformation of the algebra
is encoded in the embedding tensor
  \begin{equation}
  \Theta^M_{\dot{\alpha}} = - W_{(2)}^M{}_{\dot{\alpha}}
  \end{equation}
which belongs to the $\bf {\overline{144}}$. This is exactly the
embedding tensor of the maximal supergravity theory in six
dimensions \cite{embeddingtensorD=6}, and this analysis shows again,
as in any dimension, that the linear (or representation) constraint
of the embedding tensor is completely encoded in the (deformed)
$E_{11}$ algebra.

One can determine the commutation relation of the 4-form and the
5-form with momentum requiring that all the Jacobi identities close.
The final result is
  \begin{eqnarray}
  & & [ R^{a , \dot{\alpha} }, P_b ] = - g ( C \Gamma^M )^{\dot{\alpha}
  \dot{\beta}} \Theta^N_{\dot{\beta}} \delta^a_b R_{MN}
  \nonumber \\
  & & [ R^{a_1 a_2 , M} , P_b ] = g \Theta^M_{\dot{\alpha}}
  \delta^{[a_1}_b R^{a_2 ] , \dot{\alpha}}
  \nonumber \\
  & & [ R^{a_1 a_2 a_3 , \alpha} , P_b ] = g C^{\alpha \dot{\alpha}}
  \Theta^M_{\dot{\alpha}} \delta^{[a_1 }_b R^{a_2 a_3 ]}{}_M
  \\
  & & [ R^{a_1 ... a_4 , MN} , P_b ] = - 2 g
  \Gamma^{[M}_\alpha{}^{\dot{\alpha}} \Theta^{N]}_{\dot{\alpha}}
  \delta^{[a_1}_b R^{a_2 a_3 a_4 ] , \alpha}
  \nonumber \\
  & & [ R^{a_1 ... a_5 , M\alpha} , P_b ] = -2 g C^{\alpha \dot{\alpha}} \Theta_{N \dot{\alpha}}
  \delta^{[a_1}_b R^{a_2 ...a_5 ] , MN} - {g \over 2} (C
  \Gamma_{NP})^{\dot{\alpha} \alpha} \Theta^M_{\dot{\alpha}}
  \delta^{[a_1}_b R^{a_2 ...a_5 ] , NP} \nonumber
  \end{eqnarray}
All the quadratic constraints on the embedding tensor result from
the Jacobi identities involving a positive level $E_{11}$ generator
and two momentum operators, as well as the Jacobi identities
involving a positive level $E_{11}$ generator, the momentum operator
and the scalar operator $R_{MN} \Theta^{M}_{\dot{\alpha}}$.

It is straightforward to compute the field strengths from the
algebra above, using the general results of section 2 and appendix
B. The field strength of the vectors is
  \begin{equation}
  F_{a_1 a_2 , \dot{\alpha}} = 2[ \partial_{[a_1} A_{a_2 ] ,
  \dot{\alpha}} - g \Theta^M_{\dot{\alpha}} A_{a_1 a_2, M} + {g
  \over 4 } \Theta^M_{\dot{\gamma}} ( C \Gamma^N )^{\dot{\beta}
  \dot{\gamma}} \Gamma_{MN, \dot{\alpha}}{}^{\dot{\delta}} A_{[a_1 , \dot{\beta} } A_{a_2 ] , \dot{\delta}} ] \quad
  ,
  \end{equation}
the field strength of the 2-form is
  \begin{eqnarray}
  & & F_{a_1 a_2 a_3 ,M} = 3 [ \partial_{[a_1} A_{a_2 a_3 , M}  + {1
  \over 2} ( C \Gamma_M )^{\dot{\alpha} \dot{\beta}}
  \partial_{[a_1}  A_{a_2, \dot{\alpha}} A_{a_3 ] , \dot{\beta}}
  -g C^{\alpha \dot{\alpha}} \Theta_{M \dot{\alpha}} A_{a_1 a_2 a_3
  , \alpha} \nonumber \\
  &  &\quad \qquad - g (C \Gamma_M )^{\dot{\alpha} \dot{\beta}}
  \Theta^N_{\dot{\beta}} A_{[a_1 a_2 , N} A_{a_3 ] , \dot{\alpha}} \nonumber \\
  & & \quad \qquad +
  { g \over 12 } ( C \Gamma_{MNP} )^{\dot{\beta} \dot{\gamma}} (C
  \Gamma^N )^{\dot{\alpha} \dot{\delta}} \Theta^P_{\dot{\delta}}
  A_{[a_1 , \dot{\alpha}} A_{a_2 , \dot{\beta}} A_{a_3 ] ,
  \dot{\gamma}} ] \quad ,
  \end{eqnarray}
the field strength of the 3-form is
  \begin{eqnarray}
  && F_{a_1 ...a_4 , \alpha} = 4 [ \partial_{[a_1} A_{a_2 ...a_4 ] ,
  \alpha} - \Gamma^M_\alpha{}^{\dot{\alpha}} \partial_{[a_1} A_{a_2
  a_3 ,M} A_{a_4 ] , \dot{\alpha}} \nonumber \\
  & & \quad \qquad - {1 \over 6} ( C \Gamma_M
  )^{\dot{\alpha}\dot{\beta}} \Gamma^M_\alpha{}^{\dot{\gamma}}
  \partial_{[a_1} A_{a_2 , \dot{\alpha}} A_{a_3 , \dot{\beta}}
  A_{a_4 ] , \dot{\gamma}} + 2 g \Gamma^M_\alpha{}^{\dot{\alpha}}
  \Theta^N_{\dot{\alpha}}A_{a_1 ...a_4 ,MN} \nonumber \\
  & & \quad \qquad + g C^{\beta \dot{\beta}} \Theta^M_{\dot{\beta}}
  \Gamma_{M, \alpha}{}^{\dot{\alpha}} A_{[a_1 a_2 a_3 , \beta}
  A_{a_4 ] , \dot{\alpha}} - {g \over 2} \Theta^M_{\dot{\alpha}}
  \Gamma^N_\alpha{}^{\dot{\alpha}} A_{[a_1 a_2 , M} A_{a_3 a_4 ] ,
  N} \nonumber \\
  & & \quad \qquad + { g \over 2} \Theta^M_{\dot{\gamma}} ( C
  \Gamma_N )^{\dot{\alpha} \dot{\gamma}}
  \Gamma^N_\alpha{}^{\dot{\beta}} A_{[a_1 a_2 , M} A_{a_3 ,
  \dot{\alpha}} A_{a_4 ] , \dot{\beta}}\nonumber \\
  & & \quad \qquad  - {g \over 48} ( C \Gamma^M )^{\dot{\alpha} \dot{\epsilon}}
  \Theta^N_{\dot{\epsilon}} ( C \Gamma_{MNP} )^{\dot{\beta}
  \dot{\gamma} } \Gamma^P_\alpha{}^{\dot{\delta}} A_{[a_1 ,
  \dot{\alpha}} A_{a_2 , \dot{\beta}} A_{a_3 , \dot{\gamma}} A_{a_4
  ] , \dot{\delta}} ]
  \end{eqnarray}
and the field strength of the 4-form is
  \begin{eqnarray}
  & & F_{a_1 ...a_5, MN} = 5 [ \partial_{[a_1} A_{a_2 ...a_5 ], MN} +
  {1 \over 4} ( C \Gamma_{MN})^{\dot{\alpha} \alpha} A_{[a_1,
  \dot{\alpha}} \partial_{a_2} A_{a_3 a_4 a_5 ], \alpha} - {1 \over
  2} A_{[a_1 a_2 , [M} \partial_{a_3} A_{a_4 a_5 ] , N]} \nonumber
  \\
  & & \quad \qquad + {1 \over 8} ( C \Gamma_{MNP})^{\dot{\alpha}
  \dot{\beta}} A_{[a_1 , \dot{\alpha}} A_{a_2 , \dot{\beta}}
  \partial_{a_3} A_{a_4 a_5 ] }^P \nonumber \\
  & & \quad \qquad + {1 \over 4 \cdot 4!} (C
  \Gamma_{MNP})^{\dot{\alpha} \dot{\beta}} (C \Gamma^P
  )^{\dot{\gamma} \dot{\delta}} A_{[a_1 , \dot{\alpha}} A_{a_2
  \dot{\beta}} A_{a_3 , \dot{\gamma}} \partial_{a_4 } A_{a_5 ] ,
  \dot{\delta}} \nonumber \\
  & & \quad \qquad + 2 g A_{a_1 ...a_5 , [M \alpha} C^{\alpha
  \dot{\alpha}} \Theta_{N] \dot{\alpha}} + {g \over 2} A_{a_1 ...a_5
  , P \alpha} (C \Gamma_{MN})^{\dot{\alpha} \alpha}
  \Theta^P_{\dot{\alpha}} \nonumber \\
  & & \quad \qquad + {g \over 2} (C \Gamma_{MN})^{\dot{\alpha}
  \alpha} \Gamma^P_\alpha{}^{\dot{\beta}} \Theta^Q_{\dot{\beta}}
  A_{[a_1 , \dot{\alpha}} A_{a_2 ...a_5 ], PQ} - g C^{\alpha
  \dot{\alpha}} \Theta_{[M \dot{\alpha}} A_{[ a_1 a_2 , N]} A_{a_3
  a_4 a_5 ] , \alpha} \nonumber \\
  & & \quad \qquad - {g \over 8} ( C \Gamma_{MNP} )^{\dot{\alpha}
  \dot{\beta}} C^{\alpha \dot{\gamma}} \Theta^P_{\dot{\gamma}}
  A_{[a_1 , \dot{\alpha}} A_{a_2 , \dot{\beta}} A_{a_3 a_4 a_5 ],
  \alpha} \nonumber \\
  & & \quad \qquad - {g \over 8} (C \Gamma_{MN})^{\dot{\alpha}
  \alpha} \Gamma^P_\alpha{}^{\dot{\beta}} \Theta^Q_{\dot{\beta}}
  A_{[a_1 , \dot{\alpha}} A_{a_2 a_3 , P} A_{a_4 a_5 ] , Q}
  \nonumber \\
  & & \quad \qquad - { g \over 4 \cdot 3!} ( C \Gamma_{MNQ}
  )^{\dot{\alpha} \dot{\beta}} ( C \Gamma^Q )^{\dot{\gamma}
  \dot{\delta}} \Theta^P_{\dot{\delta}} A_{[a_1 ,\dot{\alpha}}
  A_{a_2 , \dot{\beta}} A_{a_3 , \dot{\gamma}} A_{a_4 a_5 ] , P}
  \nonumber \\
  & & \quad \qquad + { g \over 8 \cdot 5!} ( C \Gamma_{MNR}
  )^{\dot{\alpha} \dot{\beta}} (C \Gamma^R{}_{PQ} )^{\dot{\gamma}
  \dot{\delta}} ( C \Gamma^P )^{\dot{\epsilon} \dot{\rho}}
  \Theta^Q_{\dot{\rho}} A_{[a_1 , \dot{\alpha}} A_{a_2 ,\dot{\beta}}
  A_{a_3 , \dot{\gamma}} A_{a_4 , \dot{\delta}} A_{a_5 ] ,
  \dot{\epsilon}} ] \quad .
  \end{eqnarray}
Using the general results of section 2 that are summarised in
appendix B we also determine the gauge transformations of the fields
under which the field strengths above transform covariantly. The
gauge transformation of the 1-form is
  \begin{equation}
  \delta A_{a, \dot{\alpha}} = a_{a , \dot{\alpha}} - {1 \over 2}
  a^{MN} \Gamma_{MN, \dot{\alpha}}{}^{\dot{\beta}} A_{a ,
  \dot{\beta}} \quad ,
  \end{equation}
the gauge transformation of the 2-form is
  \begin{equation}
  \delta A_{a_1 a_2 ,M} = a_{a_1 a_2 , M} - {1 \over 2} ( C \Gamma_M
  )^{\dot{\alpha} \dot{\beta} } A_{[a_1 ,\dot{\alpha} } a_{a_2 ] ,
  \dot{\beta}} + 2 a_M{}^N A_{a_1 a_2 ,N} \quad ,
  \end{equation}
the gauge transformation of the 3-form is
   \begin{eqnarray}
   & & \delta A_{a_1 a_2 a_3 , \alpha} = a_{a_1 a_2 a_3, \alpha} +
   \Gamma^M_\alpha{}^{\dot{\alpha}} A_{[a_1 a_2 , M} a_{a_3 ],
   \dot{\alpha}} - { 1 \over 3!} (C \Gamma_M )^{\dot{\beta}
   \dot{\gamma}} \Gamma^M_\alpha{}^{\dot{\alpha}} A_{[a_1 ,
   \dot{\alpha}} A_{a_2 , \dot{\beta}} a_{a_3 ] , \dot{\gamma}} \nonumber \\
   & & \quad \qquad  - {1
   \over 2 } a^{MN} \Gamma_{MN, \alpha}{}^\beta A_{a_1 a_2 a_3 ,
   \beta} \quad ,
   \end{eqnarray}
the gauge transformation of the 4-form is
   \begin{eqnarray}
   & & \delta A_{a_1 ...a_4 , MN} = a_{a_1 ...a_4 , MN} -{ 1 \over
   2} A_{[a_1 a_2, [M} a_{a_3 a_4 ] , N]} - {1 \over 4} (C
   \Gamma_{MN})^{\dot{\alpha} \alpha} A_{[a_1 a_2 a_3 , \alpha}
   a_{a_4 ], \dot{\alpha}} \nonumber \\
   & & \quad \qquad - {1 \over 4 \cdot 4!} ( C \Gamma_{MNP}
   )^{\dot{\alpha} \dot{\beta}} (C \Gamma^P )^{\dot{\gamma}
   \dot{\delta}} A_{[a_1 , \dot{\alpha} } A_{a_2 , \dot{\beta}}
   A_{a_3 , \dot{\gamma}} a_{a_4 ] , \dot{\delta}}  \nonumber \\
   & & \quad \qquad + { 1 \over 4} (
   C \Gamma_{[M})^{\dot{\alpha} \dot{\beta}}A_{[a_1 a_2 , N]} A_{a_3
   , \dot{\alpha}} a_{a_4 ] , \dot{\beta}}+ 4 a_{[M}{}^P A_{a_1 ...a_4 , \vert P \vert N]}
   \end{eqnarray}
and the gauge transformation of the 5-form is
   \begin{eqnarray}
    & & \delta A_{a_1 ...a_5 , M \alpha} = \partial_{[a_1} \Lambda_{a_2
   ...a_5 ], M\alpha} -{ 1 \over 2} A_{[a_1 a_2 a_3 ,\alpha} a_{a_4
   a_5 ] , M} - \Gamma^N_\alpha{}^{\dot{\alpha}} A_{[a_1 ...a_4, MN}
   a_{a_5 ], \dot{\alpha}} \nonumber \\
   & & \quad \qquad - {1 \over 4} \Gamma^N_\alpha{}^{\dot{\alpha}}
   A_{[a_1 a_2 , M} A_{a_3 a_4 , N} a_{a_5 ],\dot{\alpha}} + {1 \over 2 \cdot 3!}
   (C \Gamma_N )^{\dot{\beta}\dot{\gamma}} \Gamma^N_\alpha{}^{\dot{\alpha}}
   A_{[a_1 a_2 , M} A_{a_3 , \dot{\alpha}} A_{a_4 , \dot{\beta}}
   a_{a_5 ] , \dot{\gamma}}
   \nonumber \\
   & & \quad \qquad + { 1
   \over 4 \cdot 5 !} (C \Gamma_{MNP})^{\dot{\beta}\dot{\alpha}}
   \Gamma^N_\alpha{}^{\dot{\alpha}} (C \Gamma^P
   )^{\dot{\delta}\dot{\epsilon}} A_{[a_1 , \dot{\alpha}} A_{a_2 ,
   \dot{\beta}} A_{a_3 , \dot{\gamma}} A_{a_4 , \dot{\delta}} a_{a_5
   ] , \dot{\epsilon}} \nonumber \\
   & & \quad \qquad -{1 \over 2} a^{NP} \Gamma_{NP, \alpha}{}^\beta
   A_{a_1 ...a_5 , M \beta} + 2 a_M{}^N A_{a_1 ...a_5 , N \alpha}
   \quad ,
   \end{eqnarray}
where the parameters $a$ are given in terms of the gauge parameters
$\Lambda$ according to eqs. (\ref{massiveidentification}) and
(\ref{thisisthegaugetransf}), which in the six-dimensional case are
  \begin{eqnarray}
  & & a_{MN} = - g \Lambda_{\dot{\alpha}} (C \Gamma_{[M}
  )^{\dot{\alpha} \dot{\beta}} \Theta_{N] \dot{\beta}} \nonumber \\
  & & a_{a , \dot{\alpha}} = \partial_a \Lambda_{\dot{\alpha}} + g
  \Theta^M_{\dot{\alpha}} \Lambda_{a, M} \nonumber \\
  & & a_{a_1 a_2 , M} = \partial_{[a_1} \Lambda_{a_2 ], M} + g
  C^{\alpha \dot{\alpha}} \Theta_{M \dot{\alpha}} \Lambda_{a_1 a_2 ,
  \alpha} \nonumber \\
  & & a_{a_1 a_2 a_3, \alpha} = \partial_{[a_1}
  \Lambda_{a_2 a_3 ], \alpha} -2 g \Gamma^M_\alpha{}^{\dot{\alpha}}
  \Theta^N_{\dot{\alpha}} \Lambda_{a_1 a_2 a_3 , MN} \nonumber \\
  & & a_{a_1 ...a_4 , MN} = \partial_{[a_1} \Lambda_{a_2 a_3 a_4 ]
  ,MN} -( 2 g C^{\alpha \dot{\alpha}} \delta^P_{[M}
  \Theta_{N]\dot{\alpha}} +{g \over 2} (C \Gamma_{MN})^{\dot{\alpha}
  \alpha} \Theta^P_{\dot{\alpha}} ) \Lambda_{a_1 ...a_4, P\alpha}
  \ .
  \end{eqnarray}
One can easily determine also the field strength of the 5-form (up
to the term containing the 6-form potential) using the formulae in
appendix B.

\section{D=7}
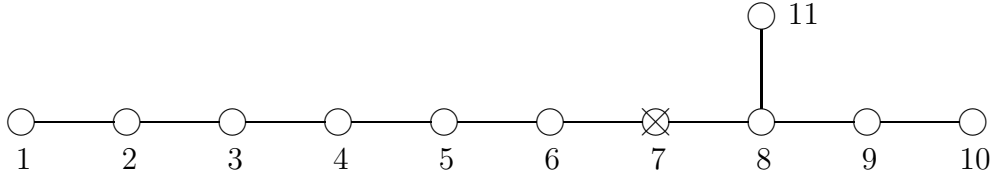
\begin{figure}[h]
\begin{center}
\begin{picture}(380,70)
\multiput(10,10)(40,0){6}{\circle{10}}
\multiput(250,10)(40,0){3}{\circle{10}} \put(370,10){\circle{10}}
\multiput(15,10)(40,0){9}{\line(1,0){30}} \put(290,50){\circle{10}}
\put(290,15){\line(0,1){30}} \put(8,-8){$1$} \put(48,-8){$2$}
\put(88,-8){$3$} \put(128,-8){$4$} \put(168,-8){$5$}
\put(208,-8){$6$} \put(248,-8){$7$} \put(288,-8){$8$}
\put(328,-8){$9$} \put(365,-8){$10$} \put(300,47){$11$}
\put(245,5){\line(1,1){10}} \put(245,15){\line(1,-1){10}}
\end{picture}
\caption{\sl The $E_{11}$ Dynkin diagram corresponding to
7-dimensional supergravity. The internal symmetry group is
$SL(5,\mathbb{R})$. \label{Dynkinseven}}
\end{center}
\end{figure}

The multiplet describing massless maximal supergravity theory in 7
dimensions \cite{maxsugrasevendim} has a bosonic sector containing
14 scalars parametrising $SL(5,\mathbb{R})/SO(5)$, the metric, a
1-form in the ${\bf \overline{10}}$ and a 2-form in the ${\bf 5}$ of
$SL(5,\mathbb{R})$. This theory results from $E_{11}$ after deletion
of node 7, as shown in the Dynkin diagram of fig. \ref{Dynkinseven}.
One can see from the diagram that the 1-forms carry two
antisymmetric indices of $SL(5,\mathbb{R})$.

The positive-level $E_{11}$ generators with completely antisymmetric
spacetime indices and up to the 6-form included are
  \begin{eqnarray}
  & & R^M{}_N  \ \ ({\bf 24}) \quad R^{a , MN} \ \ ({\bf 10}) \qquad R^{a_1 a_2}{}_M \ \ ({\bf \overline{5}})
  \quad R^{a_1 a_2 a_3 , M} \ \ ({\bf 5})
  \qquad  R^{a_1 a_2 a_3 a_4}{}_{MN} \ \ ({\bf \overline{10}}) \nonumber \\
  && R^{a_1 ... a_5, M}{}_N \ \ ({\bf 24})
  \qquad R^{a_1 ...a_6 }{}_{MN,P} \ \ ({\bf 40 }) \qquad R^{a_1 ...a_6,
  M N} \ \ ({\bf 15 }) \label{poslevgenD=7}
  \quad ,
  \end{eqnarray}
where $M= 1, ..., 5$. The scalar generators and the 5-forms are in
the adjoint of $SL(5,\mathbb{R})$ and thus satisfy $R^M{}_M = R^{a_1
...a_5 , M}{}_M = 0$. The 1-form and the 4-form are antisymmetric in
$MN$. The 6-form $ R^{a_1 ...a_6 }{}_{MN,P}$ is antisymmetric in
$MN$ and satisfies $ R^{a_1 ...a_6 }{}_{[ MN,P] } =0$, which
corresponds to the ${\bf 40 }$ of $SL(5,\mathbb{R})$. Finally, the
6-form $R^{a_1 ...a_6, M N}$ is symmetric in $MN$, corresponding to
the ${\bf 15}$ of $SL(5,\mathbb{R})$.

The scalars generate $SL(5, \mathbb{R})$,
  \begin{equation}
  [ R^M{}_N , R^P{}_Q ] = \delta^P_N R^M{}_Q - \delta^M_Q R^P{}_N
  \quad ,
  \end{equation}
while the commutators of the other generators with the scalars are
  \begin{eqnarray}
  & & [ R^M{}_N , R^{a , PQ} ]= \delta^P_N R^{a , MQ} + \delta^Q_N
  R^{a, PM} - {2 \over 5} \delta^M_N R^{a, PQ}
  \nonumber \\
  & & [ R^M{}_N , R^{a b}{}_P ] = - \delta^M_P R^{ab}{}_N + {1 \over 5}
  \delta^M_N R^{ab}{}_P
  \end{eqnarray}
and similarly for the higher level generators.

The commutators of the positive level generators in eq.
(\ref{poslevgenD=7}) are
  \begin{eqnarray}
  & & [ R^{a_1 , MN} , R^{a_2 , PQ} ] = \epsilon^{MNPQR} R^{a_1
  a_2}{ }_R
  \nonumber \\
  & & [ R^{a_1 , MN} , R^{a_2 a_3}{}_{ P} ] = \delta^{[M}_P R^{a_1 a_2
  a_3 , N]}
  \nonumber \\
  & & [ R^{a_1 a_3}{}_M , R^{a_3 a_4}{}_N ] = R^{a_1 ...a_4 }{}_{MN}
  \nonumber \\
  & &
  [ R^{a_1 , MN} , R^{a_2 a_3 a_4, P}] = \epsilon^{MNPQR} R^{a_1
  ...a_4}{}_{QR}
  \nonumber \\
  & & [ R^{a_1 a_2}{}_M , R^{a_3 a_4 a_5 , N} ] = R^{a_1 ...a_5, N}{}_M
  \nonumber \\
  & & [ R^{a_1 , MN} , R^{a_2 ...a_5}{}_{PQ} ]= -2 \delta^{[M}_{[P}
  R^{a_1 ...a_5 , N]}{}_{Q]}
  \nonumber \\
  & & [R^{a_1 a_2 a_3, M} , R^{a_4 a_5 a_6, N} ]= R^{a_1 ...a_6 ,MN}
  \nonumber \\
  & & [ R^{a_1 a_2}{}_M , R^{a_3 ...a_6}{}_{NP} ] = R^{a_1
  ...a_6}{}_{NP,M}
  \nonumber \\
  & & [ R^{a_1 , MN} , R^{a_2 ...a_6, Q}{}_P ] = \epsilon^{MNQRS}
  R^{a_1 ...a_6}{}_{RS,P} + \delta^{[M}_P R^{a_1 ...a_6, N]Q}\quad .
  \end{eqnarray}
To prove that all the Jacobi identities close one makes use of the
identity
  \begin{equation}
  \epsilon^{M_1 ..M_5} \epsilon_{N_1 ...N_5} = 5! \delta^{[M_1
  ...M_5]}_{[N_1 ...N_5 ]} \quad .
  \end{equation}

If one considers the group element
  \begin{eqnarray}
  g &=&  e^{x \cdot P} e^{A_{a_1
  ...a_6, M N} R^{a_1 ... a_6, MN }} e^{A_{a_1
  ...a_6}^{MN,P} R^{a_1 ... a_6}{}_{MN,P}} e^{A_{a_1
  ...a_5, M}{}^N R^{a_1 ... a_5, M}{}_N} e^{A_{a_1 ... a_4}^{ MN}
  R^{a_1 ... a_4}{}_{MN}} \nonumber \\
  & &e^{A_{a_1 a_2 a_3 , M} R^{a_1 a_2 a_3 ,
  M}}  e^{A_{a_1 a_2}^{M} R^{a_1 a_2}_{  M}}
  e^{A_{a
  , M N} R^{a , MN}} e^{\phi_{M}{}_N R^{M}{}_N}
  \end{eqnarray}
and compute the Maurer-Cartan form using the fact that the positive
level generators commute with momentum, the field strengths of the
massless theory are obtained antisymmetrising the spacetime indices
of the various terms in the Maurer-Cartan form. Requiring that the
field strengths satisfy duality relations, that is the
field-strength of the vector is dual to the field-strength of the
4-form, the field strength of the 2-form is dual to the field
strength of the 3-form and the field strength of the 6-form is dual
to the derivative of the scalars, one recovers the field equations
of the massless supergravity theory. We now consider the
deformations of the $E_{11}$ algebra resulting from modifying the
commutation relations of the $E_{11}$ generators with momentum
compatibly with the Jacobi identities. In this way we will derive
all the gauged supergravities in seven dimensions.

Following the results of section 2, the most general deformation
that we can write down is
  \begin{eqnarray}
  & &   [ R^{a , MN }, P_b ] = - g \Theta^{MN, P}{}_Q
  R^Q{}_P \nonumber \\
  & & [ R^{a_1 a_2}{}_{M} , P_b ] = - g W^{(2)}_{M,NP}
  \delta^{[a_1}_b R^{a_2 ] , NP} \nonumber \\
  & &
  [ R^{a_1 a_2 a_3 , M} , P_b ] =  - g W_{(3)}^{M,N}
  \delta^{[a_1 }_b R^{a_2 a_3 ]}{}_N \quad ,
  \end{eqnarray}
where $\Theta^{MN,P}{}_Q$ is antisymmetric in $MN$ and
$\Theta^{MN,P}{}_P =0$, and $W^{(2)}_{M,NP}$ is antisymmetric in
$NP$. For simplicity we have only considered the $E_{11}$ generators
up to level 4. Indeed it turns out that the constraints that the
Jacobi identities impose on $\Theta$, $W_{(2)}$ and $W_{(3)}$ are
enough to restrict the representations of $\Theta$ uniquely.

The Jacobi identity involving two 1-forms and the momentum operator
gives
  \begin{equation}
  \Theta^{PQ,[M}{}_{[S} \delta^{N]}_{T]} +
  \Theta^{MN,[P}{}_{[S} \delta^{Q]}_{T]} = {1 \over 2}
  \epsilon^{MNPQR} W^{(2)}_{R,ST}
  \quad ,
  \end{equation}
which can be solved for $W^{(2)}$ in terms of $\Theta$ giving
  \begin{equation}
  W^{(2)}_{R,ST} = \epsilon_{MTPQR} \Theta^{PQ,M}{}_S - \epsilon_{MSPQR}
  \Theta^{PQ,M}{}_T \quad . \label{W2Thetasevendimensions}
  \end{equation}
The Jacobi identity involving the 1-form, the 2-form and the
momentum operator gives
  \begin{equation}
  W^{(2)}_{P, RS} \epsilon^{MNQRS} + \Theta^{MN,Q}{}_P = - {1 \over
  2} W_{(3)}^{N,Q} \delta^M_P + {1 \over 2} W_{(3)}^{M,Q} \delta^N_P
  \quad .\label{W3W2Thetasecondjacobisevendimensions}
  \end{equation}

We can analyse the solutions of eqs. (\ref{W2Thetasevendimensions})
and (\ref{W3W2Thetasecondjacobisevendimensions}) for different
representations of $\Theta$. If we take $\Theta^{MN,P}{}_Q$ such
that $\Theta^{[MN,P]}{}_Q =0$, then eq.
(\ref{W2Thetasevendimensions}) implies
  \begin{equation}
  W^{(2)}_{M,NP} =0 \quad .
  \end{equation}
Substituting this in eq.
(\ref{W3W2Thetasecondjacobisevendimensions}) and using
$\Theta^{[MN,P]}{}_Q =0$ and $\Theta^{MN,P}{}_P =0$ one obtains that
$W_{(3)}^{M,N}$ is symmetric in $MN$. One thus obtains the embedding
tensor $\Theta^{MN}= W_{(3)}^{MN}$ in the $\bf 15$ of
$SL(5,\mathbb{R})$ and we will then show that the inclusion of the
forms of rank higher that 3 is also compatible with this
deformation. If we instead take $\Theta^{MN,P}{}_Q$ to be completely
antisymmetric in $MNP$, then we can write
  \begin{equation}
  \Theta^{MN,P}{}_Q = \epsilon^{MNPRS} \Theta_{RS, Q} \quad ,
  \end{equation}
and the condition $\Theta^{MN,P}{}_P =0$ implies $\Theta_{[MN,Q]}
=0$. Therefore the embedding tensor $\Theta_{MN,P}$ belongs to the
${\bf 40}$ of $SL(5,\mathbb{R})$. Eq. (\ref{W2Thetasevendimensions})
then gives
  \begin{equation}
  W^{(2)}_{M,NP} = - \Theta_{NP,M} \quad ,
  \end{equation}
and eq. (\ref{W3W2Thetasecondjacobisevendimensions}) gives
  \begin{equation}
  W_{(3)}^{M,N} =0 \quad .
  \end{equation}
Also in this case we will show that one can consistently include in
the algebra the higher rank form generators. We now proceed with the
analysis of the algebra and the derivation of the field strengths
and gauge transformations for the two different cases corresponding
to an embedding tensor in the  ${\bf 15}$ and in the  ${\bf 40}$.
One can show that a linear combination of these two deformations is
not allowed because of the quadratic constraints.

\subsection{Embedding tensor in the {\bf 15} of $SL(5,\mathbb{R})$}
We now determine the commutation relations or all the generators in
eq. (\ref{poslevgenD=7}) with the momentum operator requiring the
closure of the Jacobi identities. In the case of the embedding
tensor $\Theta^{MN}$ in the ${\bf 15}$ we get
  \begin{eqnarray}
  & & [ R^{a , MN} , P_b ] = - g \Theta^{[M \vert P \vert } \delta^a_b
  R^{N]}{}_P
  \nonumber \\
  & & [ R^{a_1 a_2 }{}_M , P_b ] =0
  \nonumber \\
  & & [ R^{a_1 a_2 a_3 , M} , P_b ] = - g \Theta^{MN} \delta^{[ a_1}_b
  R^{a_2 a_3 ]}{}_N
  \nonumber \\
  & & [ R^{a_1 a_2 a_3 a_4 }{}_{MN} , P_b ] = 0
  \nonumber \\
  & & [ R^{a_1 ...a_5 , M}{}_N , P_b ] = g \Theta^{MP}\delta^{[a_1}_b
  R^{a_2 ... a_5 ]}{}_{PN}
  \nonumber \\
  & & [ R^{a_1 ... a_6}{}_{MN , P} , P_b ] =0
  \nonumber \\
  & & [ R^{a_1 ... a_6, MN } , P_b ] = -2g \Theta^{P (M}
  \delta^{[a_1}_b R^{a_2 ...a_6 , N)}{}_P \quad .
  \end{eqnarray}

From the algebra above one computes the field strengths using the
general results of section 2 and appendix B. The field strength of
the vectors is
  \begin{equation}
  F_{a_1 a_2, MN} = 2[\partial_{[a_1} A_{a_2 ] , MN} + g \Theta^{PQ}
  A_{[a_1 , PM} A_{a_2 ] , QN} ] \quad ,
  \end{equation}
the field strength of the 2-form is
  \begin{eqnarray}
  & & F_{a_1 a_2 a_3 }^M = 3 [ \partial_{[a_1} A_{a_2 a_3]}^M + {1
  \over 2} \epsilon^{MNPQR} \partial_{[a_1} A_{a_2 , NP } A_{a_3 ] ,
  QR} + g \Theta^{MN} A_{a_1 a_2 a_3 , N} \nonumber \\
  & & \quad\qquad + {g \over 6} \Theta^{NQ}
  \epsilon^{MPRST} A_{[a_1, NP} A_{a_2 , QR} A_{a_3 ], ST} ] \quad ,
  \end{eqnarray}
the field strength of the 3-form is
  \begin{eqnarray}
  & & F_{a_1 ...a_4 , M} = 4[ \partial_{[a_1} A_{a_2 ...a_4 ], M} -
  \partial_{[a_1} A_{a_2 a_3}^N A_{a_4 ] ,NM} -{1 \over 6}
  \partial_{[a_1} A_{a_2, NP} A_{a_3 , QR} A_{a_4 ] , SM}
  \epsilon^{SNPQR} \nonumber \\
  & & \quad \qquad - g \Theta^{NP} A_{[a_1 a_2 a_3, N} A_{a_4 ], PM}
  - { g \over 12} \Theta^{NQ} \epsilon^{PRSTU} A_{[a_1 , NP} A_{a_2
  , QR} A_{a_3 , ST} A_{a_4 ] , UM} ]
  \end{eqnarray}
and the field strength of the 4-form is
  \begin{eqnarray}
  & & F_{a_1 ...a_5}^{MN} = 5[ \partial_{[a_1} A_{a_2 ...a_5 ]}^{MN}
  + \epsilon^{PQRMN} A_{[a_1, PQ} \partial_{a_2} A_{a_3 ...a_5 ], R} -
  {1 \over 2} A_{[a_1 a_2}^{[M} \partial_{a_3} A_{a_4 a_5 ]}^{N]}
  \nonumber \\
  & & \quad \qquad + {1 \over 2} \epsilon^{PQSMN} A_{[a_1 , PQ}
  A_{a_2, RS} \partial_{a_3} A_{a_4 a_5 ]}^R \nonumber \\
  & & \quad \qquad + {1 \over 4!}
  \epsilon^{TUVWR} \epsilon^{PQSMN} A_{[a_1, PQ} A_{a_2, RS} A_{a_3
  , TU} \partial_{a_4} A_{a_5 ], VW} \nonumber \\
  & & \quad \qquad - g \Theta^{P[M} A_{a_1 ...a_5}^{N]} -g
  \Theta^{P[M} A_{[a_1 a_2}^{N]} A_{a_3 a_4 a_5], P} +{g \over 2}
  \epsilon^{PQSMN} \Theta^{RT} A_{[a_1, PQ} A_{a_2, RS} A_{a_3 a_4
  a_5 ],T} \nonumber \\
  & & \quad \qquad - {g \over 2 \cdot 5 !} \epsilon^{PQSMN}
  \epsilon^{TUYWR} \Theta^{VX} A_{[a_1 , PQ} A_{a_2, RS } A_{a_3,
  TU} A_{a_4 , VW} A_{a_5 ] , XY}] \quad .
  \end{eqnarray}
These field strengths transform covariantly under the gauge
transformations
  \begin{eqnarray}
  & & \delta A_{a, MN} = a_{a, MN} + 2 a_{[M}{}^P A_{a, \vert P
  \vert N]} \nonumber \\
  & & \delta A_{a_1 a_2}^M = a_{a_1 a_2}^M - {1 \over 2}
  \epsilon^{MNPQR} A_{[a_1, NP} a_{a_2 ], QR} - a_N{}^M A_{a_1
  a_2}^N \nonumber \\
  & & \delta A_{a_1 a_2 a_3 , M} = a_{a_1 a_2 a_3 ,M} + A_{[a_1
  a_2}^N a_{a_3], NM} +{1 \over 3!} \epsilon^{NQRST} A_{[a_1, MN}
  A_{a_2, QR} a_{a_3 ] , ST} \nonumber \\
  & & \quad \qquad + a_M{}^N A_{a_1 a_2 a_3 , N} \nonumber \\
  & & \delta A_{a_1 ...a_4}^{MN} = a_{a_1 ...a_4}^{MN} -{1 \over 2}
  A _{[a_1 a_2}^{[M} a_{a_3 a_4]}^{N]} - \epsilon^{MNPQR} A_{[a_1
  a_2 a_3 , P} a_{a_4] , QR} \nonumber \\
  & & \quad \qquad - { 1 \over 4!} \epsilon^{TUVWR} \epsilon^{PQSMN} A_{[a_1, PQ}
  A_{a_2 , RS} A_{a_3, TU } a_{a_4 ], VW}\nonumber \\
  & & \quad \qquad +{1 \over 4} \epsilon^{QRST[M} A_{[a_1 a_2}^{N]}
  A_{a_3 , QR} a_{a_4 ] , ST} \nonumber \\
  & & \delta A_{a_1 ...a_5 , M}{}^N = \partial_{[a_1} \Lambda_{a_2
  ...a_5], M}{}^N - A_{[a_1 a_2 a_3 ,M} a_{a_4 a_5 ]}^N -2 A_{[a_1
  ...a_4}^{PN} a_{a_5 ] , PM}\nonumber \\
  & & \quad \qquad + {1 \over 2} A_{[a_1 a_2}^N A_{a_3 a_4}^Q
  a_{a_5 ], QM} + {2 \over 5!} \epsilon^{VWXYT} \epsilon^{RSUPN}
  A_{[a_1 , PM} A_{a_2 ,RS} A_{a_3,TU} A_{a_4, VW} a_{a_5 ] , XY}
  \nonumber \\
  &  &\quad \qquad -{1 \over 3!} \epsilon^{STUVQ} A_{[a_1 a_2}^N
  A_{a_3 , QM} A_{a_4 , ST} a_{a_5 ] ,UV}  -2 a_P{}^{[M} A_{a_1 ...a_5}^{\vert P\vert N]}\quad
  , \label{gaugetransfsind=7}
  \end{eqnarray}
where the parameters $a$ are given in terms of the gauge parameters
as
  \begin{eqnarray}
  & & a_M{}^N = g \Lambda_{MP} \Theta^{PN}  \nonumber \\
  & & a_{a, MN} =\partial_{a} \Lambda_{MN} \nonumber \\
  & & a_{a_1 a_2}^M = \partial_{[a_1} \Lambda_{a_2 ]}^M - g
  \Theta^{MN} \Lambda_{a_1 a_2, N} \nonumber \\
  & & a_{a_1 a_2 a_3 , M} = \partial_{[a_1 } \Lambda_{a_2 a_3 ], M}
  \nonumber \\
  & & a_{a_1 ...a_4}^{MN} = \partial_{[a_1} \Lambda_{a_2 ..a_4 ]}^{MN}
  + g \Theta^{P[M} \Lambda_{a_1 ...a_4 . P}{}^{N]} \quad .
  \label{identificationparameters15}
  \end{eqnarray}
One can easily determine the field strengths and the gauge
transformations of the fields of higher rank using the general
results on section 2 which are explicitly expanded in appendix B.

\subsection{Embedding tensor in the {\bf 40} of $SL(5,\mathbb{R})$}
In the case of the embedding tensor $\Theta_{MN,P}$, which belongs
to the ${\bf 40}$, one gets
  \begin{eqnarray}
  & & [ R^{a , MN} , P_b ] = - g \epsilon^{MNPQR} \Theta_{PQ , S} \delta^a_b
  R^{S}{}_R
  \nonumber \\
  & & [ R^{a_1 a_2 }{}_M , P_b ] = g \Theta_{NP , M} \delta^{[a_1}_b
  R^{a_2 ] , NP}
  \nonumber \\
  & & [ R^{a_1 a_2 a_3 , M} , P_b ] = 0
  \nonumber \\
  & & [ R^{a_1 a_2 a_3 a_4 }{}_{MN} , P_b ] =  - g \Theta_{ MN , P}
  \delta^{[a_1}_b R^{a_2 a_3 a_4 ] , P}
  \nonumber \\
  & & [ R^{a_1 ...a_5 , M}{}_N , P_b ] = g \epsilon^{MPQRS} \Theta_{ PQ , N} \delta^{[a_1}_b
  R^{a_2 ... a_5 ]}{}_{RS}
  \nonumber \\
  & & [ R^{a_1 ... a_6}{}_{MN , P} , P_b ] = - g \Theta_{ NP , Q}
  \delta^{[a_1}_b R^{a_2 ...a_6 ] , Q}{}_M -g \Theta_{NQ , M}
  \delta^{[a_1}_b R^{a_2 ...a_6 ] , Q}{}_P \nonumber \\
  & & \qquad  \quad +g \Theta_{PQ , M}
  \delta^{[a_1}_b R^{a_2 ...a_6 ] , Q}{}_N
  \nonumber \\
  & & [ R^{a_1 ... a_6, MN} , P_b ] = 0 \quad .
  \end{eqnarray}

For this deformation the field strength of the vector is
  \begin{equation}
  F_{a_1 a_2 , MN} = 2[ \partial_{[a_1} A_{a_2 ] , MN} - g \Theta_{MN,
  P} A_{a_1 a_2}^P + g \epsilon^{PQRTU} \Theta_{TU, [M} A_{[a_1 ,
  \vert PQ} A_{a_2 ] , R \vert N]} ]\quad ,
  \end{equation}
the field strength of the 2-form is
  \begin{eqnarray}
  & & F_{a_1 a_2 a_3 }^M = 3 [ \partial_{[a_1} A_{a_2 a_3]}^M + {1
  \over 2} \epsilon^{MNPQR} \partial_{[a_1} A_{a_2 , NP } A_{a_3 ] ,
  QR} - g \Theta_{ST, P} \epsilon^{MSTQR} A_{[a_1 a_2}^P A_{a_3 ],
  QR} \nonumber \\
  & & \quad \qquad + { g \over 3} \epsilon^{PQVWR} \Theta_{VW,N}
  \epsilon^{MNSTU} A_{[a_1 , PQ} A_{a_2 , RS} A_{a_3 ] , TU} ] \quad
  ,
  \end{eqnarray}
the field strength of the 3-form is
  \begin{eqnarray}
  & & F_{a_1 ...a_4 , M} = 4[ \partial_{[a_1} A_{a_2 ...a_4 ], M} -
  \partial_{[a_1} A_{a_2 a_3}^N A_{a_4 ] ,NM} -{1 \over 6}
  \partial_{[a_1} A_{a_2, NP} A_{a_3 , QR} A_{a_4 ] , SM}
  \epsilon^{SNPQR} \nonumber \\
  & & \quad \qquad + g \Theta_{NP, M} A_{a_1 ...a_4 }^{NP} -{g \over
  2} \Theta_{PM, N} A_{[a_1 a_2}^N A_{a_3 a_4 ]}^P + {g \over 2}
  \Theta_{TU,N} \epsilon^{TUPQR} A_{[a_1 a_2}^N A_{a_3 , PQ} A_{a_4
  ], RM}\nonumber \\
  & & \quad \qquad  - {g \over 12} \epsilon^{NPWXQ} \Theta_{WX, Z}
  \epsilon^{UZRST} A_{[a_1 , NP} A_{a_2 , QR} A_{a_3 , ST} A_{a_4 ],
  UM} ]
  \end{eqnarray}
and the field strength of the 4-form is
  \begin{eqnarray}
  & & F_{a_1 ...a_5}^{MN} = 5[ \partial_{[a_1} A_{a_2 ...a_5 ]}^{MN}
  + \epsilon^{PQRMN} A_{[a_1, PQ} \partial_{a_2} A_{a_3 ...a_5 ], R} -
  {1 \over 2} A_{[a_1 a_2}^{[M} \partial_{a_3} A_{a_4 a_5 ]}^{N]}
  \nonumber \\
  & & \quad \qquad + {1 \over 2} \epsilon^{PQSMN} A_{[a_1 , PQ}
  A_{a_2, RS} \partial_{a_3} A_{a_4 a_5 ]}^R \nonumber \\
  & & \quad \qquad + {1 \over 4!}
  \epsilon^{TUVWR} \epsilon^{PQSMN} A_{[a_1, PQ} A_{a_2, RS} A_{a_3
  , TU} \partial_{a_4} A_{a_5 ], VW} \nonumber \\
  & & \quad \qquad - g \epsilon^{MNPQS} \Theta_{PQ,T} A_{a_1 ...a_5,
  S}{}^T + g \epsilon^{MNPQT} \Theta_{RS,T} A_{[a_1 , PQ} A_{a_2
  ...a_5 ] }^{RS} \nonumber \\
  & & \quad\qquad - {g \over 2} \epsilon^{MNPQT} \Theta_{RT,S}
  A_{[a_1 , PQ} A_{a_2 a_3}^R A_{a_4 a_5 ] }^S \nonumber \\
  & & \quad \qquad - {g \over 3!} \epsilon^{MNPQS} \epsilon^{TUWXR}
  \Theta_{WX,V} A_{[a_1 , PQ} A_{a_2 ,RS} A_{a_3 , TU } A_{a_4 a_5 ]
  }^V \nonumber \\
  & & \quad \qquad -{ 2 g \over 5!} \epsilon^{MNPQS}
  \epsilon^{TUDWR} \epsilon^{XYABV} \Theta_{AB,D} A_{[a_1 , PQ}
  A_{a_2, RS} A_{a_3 , TU} A_{a_4 , VW} A_{a_5 ] , XY} ] \quad .
  \end{eqnarray}

The field strengths transform covariantly under the gauge
transformations of eq. (\ref{gaugetransfsind=7}) where the
parameters $a$ are given as
  \begin{eqnarray}
  & & a_M{}^N = - g \Lambda_{PQ} \epsilon^{PQRSN} \Theta_{RS, M}
  \nonumber \\
  & & a_{a, MN} = \partial_{a} \Lambda_{MN} + g \Theta_{MN,P}
  \Lambda_a^P \nonumber \\
  & & a_{a_1 a_2}^M = \partial_{[a_1} \Lambda_{a_2 ]}^M
  \nonumber \\
  & & a_{a_1 a_2 a_3, M} = \partial_{[a_1} \Lambda_{a_2 a_3 ],M} -g
  \Theta_{NP,M} \Lambda_{a_1 a_2 a_3}^{NP} \nonumber \\
  & & a_{a_1 ...a_4}^{MN} = \partial_{[a_1} \Lambda_{a_2 a_3 a_4
  ]}^{MN} + g \epsilon^{PQRMN} \Theta_{QR,S} \Lambda_{a_1 ...a_4 ,
  P}{}^S \quad .
  \end{eqnarray}

The field strengths and the gauge transformations of the higher rank
fields can also easily been determined from the above algebra.

\section{D=8}
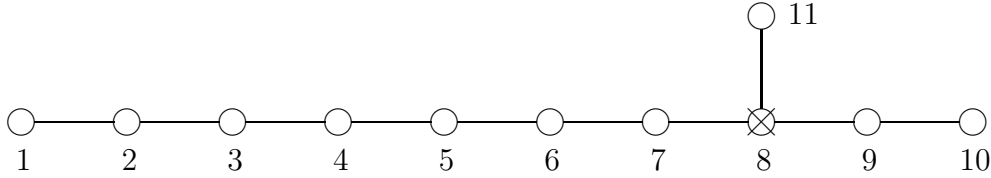
\begin{figure}[h]
\begin{center}
\begin{picture}(380,70)
\multiput(10,10)(40,0){6}{\circle{10}}
\multiput(250,10)(40,0){3}{\circle{10}} \put(370,10){\circle{10}}
\multiput(15,10)(40,0){9}{\line(1,0){30}} \put(290,50){\circle{10}}
\put(290,15){\line(0,1){30}} \put(8,-8){$1$} \put(48,-8){$2$}
\put(88,-8){$3$} \put(128,-8){$4$} \put(168,-8){$5$}
\put(208,-8){$6$} \put(248,-8){$7$} \put(288,-8){$8$}
\put(328,-8){$9$} \put(365,-8){$10$} \put(300,47){$11$}
\put(285,5){\line(1,1){10}} \put(285,15){\line(1,-1){10}}
\end{picture}
\caption{\sl The $E_{11}$ Dynkin diagram corresponding to
8-dimensional supergravity. The internal symmetry group is
$SL(3,\mathbb{R}) \times SL(2,\mathbb{R})$. \label{Dynkineight}}
\end{center}
\end{figure}

The bosonic sector of maximal massless eight-dimensional
supergravity \cite{salamsezginD=8} contains seven scalars
parametrising the manifold $SL(3,\mathbb{R})/SO(3) \times
SL(2,\mathbb{R})/SO(2)$, the metric, a vector in the ${\bf
(\overline{3},2)}$ of the internal symmetry group $SL(3,\mathbb{R})
\times SL(2,\mathbb{R})$, a 2-form in ${\bf (3,1)}$ and an
$SL(2,\mathbb{R})$ doublet of 3-forms which satisfy self-duality
conditions. The $E_{11}$ Dynkin diagram corresponding to this theory
is shown in fig. \ref{Dynkineight}.

The positive-level $E_{11}$ generators with completely antisymmetric
spacetime indices and up to the 6-form included, together with their
$SL(3,\mathbb{R}) \times SL(2,\mathbb{R})$ representations, are
  \begin{eqnarray}
  & & R^i \ \ {\bf (1,3)} \quad R^M{}_N \ \ {\bf (8,1)}
  \quad R^{a , M \alpha} \ \ {\bf (3,2)}
  \quad R^{a_1 a_2}{}_M \  \ {\bf (\overline{3},1)} \quad R^{a_1 a_2 a_3 , \alpha}
  \ \ {\bf (1,2)} \nonumber \\
  & &  R^{a_1 a_2 a_3 a_4, M} \ \ {\bf (3,1)}
  \quad R^{a_1 ... a_5, \alpha}{}_M \ \ {\bf (\overline{3},2)}
  \quad R^{a_1 ...a_6 , i} \ \ {\bf (1,3)}\qquad R^{a_1 ...a_6, M}{}_N \ \ {\bf (8,1)}
  \nonumber \\
  & &  R^{a_1
  ... a_7 , \alpha}{}_{MN} \ \ {\bf (\overline{6},2)} \qquad R^{a_1 ...a_7 , M \alpha}
  \ \ {\bf (3,2)} \quad . \label{eightdimlistofgen}
  \end{eqnarray}
Here the index $i=1,2,3$ and $\alpha =1,2$ denote the adjoint and
the fundamental of $SL(2,\mathbb{R})$ respectively, while the
upstairs index $M=1,2,3$ denotes the fundamental of
$SL(3,\mathbb{R})$. The scalars $R^i$, $i=1,2,3$, are the
$SL(2,\mathbb{R})$ generators, while the scalars $R^M{}_N$ are the
$SL(3,\mathbb{R})$ generators and thus satisfy the constraint
$R^M{}_M=0$. The 6-form $R^{a_1 ...a_6, M}{}_N$ also satisfies the
constraint $R^{a_1 ...a_6, M}{}_M =0$, while the 7-form $R^{a_1
  ... a_7 , \alpha}{}_{MN}$ is symmetric in $MN$.

The algebra of the scalars is
  \begin{eqnarray}
  & & [R^i , R^j ] = f^{ij}{}_k R^k
  \nonumber \\
  & & [ R^M{}_N , R^P{}_Q ] = \delta^P_N R^M{}_Q - \delta^M_Q
  R^P{}_N \quad ,
  \end{eqnarray}
while the other commutation relations with the scalar generators are
  \begin{eqnarray}
  & & [ R^i , R^{a, M\alpha} ] = D^i_\beta{}^\alpha R^{a, M \beta}
  \nonumber \\
  & & [ R^M{}_N , R^{a , P\alpha} ] = \delta^P_N R^{a , M \alpha} - {1
  \over 3} \delta^M_N R^{a , P \alpha}
  \end{eqnarray}
and similarly for the higher rank forms. Here $D^i_\beta{}^\alpha$
are the generators of $SL(2, \mathbb{R})$ satisfying
  \begin{equation}
  [ D^i , D^j ]_\beta{}^\alpha = f^{ij}{}_k D^k_\beta{}^\alpha
  \end{equation}
and $f^{ij}{}_k$ are the structure constants of $SL(2,\mathbb{R})$.
In terms of Pauli matrices, a choice of $D^i_\beta{}^\alpha$ is
  \begin{equation}
  D_1 = {\sigma_1 \over 2} \qquad D_2 = {i \sigma_2 \over 2} \qquad D_3 = {\sigma_3 \over
  2} \quad .
  \end{equation}
We raise and lower $SL(2,\mathbb{R})$ indices using the
antisymmetric metric $\epsilon^{\alpha\beta}$, that is, for a
generic doublet $V^\alpha$,
  \begin{equation}
  V^\alpha = \epsilon^{\alpha \beta} V_\beta \qquad \quad V_\alpha =
  V^\beta \epsilon_{\beta \alpha} \quad , \label{sl2raiseandlower}
  \end{equation}
which implies
  \begin{equation}
  \epsilon^{\alpha \beta} \epsilon_{\beta \gamma} = -
  \delta^\alpha_\gamma \quad .
  \end{equation}
The generators
  \begin{equation}
  D^{i, \alpha \beta} = \epsilon^{\alpha \gamma} D^i_\gamma{}^\beta
  \end{equation}
are symmetric in $\alpha\beta$. Useful identities relating the
$SL(2, \mathbb{R})$ generators are
  \begin{equation}
  D_i^{\alpha \beta} D^{i , \gamma \delta }= - {1 \over 4} [
  \epsilon^{\alpha \gamma} \epsilon^{\beta \delta} +
  \epsilon^{\alpha \delta } \epsilon^{\beta \gamma} ]
  \end{equation}
and
  \begin{equation}
  D^i_\beta{}^\gamma D^j_\gamma{}^\alpha + D^j_\beta{}^\gamma
  D^i_\gamma{}^\alpha = {1 \over 2} g^{ij} \delta^\alpha_\beta \quad
  ,\label{symmetricproductsl2}
  \end{equation}
where $g^{ij}$ is the $SL(2, \mathbb{R})$ Killing metric.

The non-vanishing commutation relations involving all the non-scalar
generators of eq. (\ref{eightdimlistofgen}) are
  \begin{eqnarray}
  & & [ R^{a_1 , M\alpha} , R^{a_2, N \beta} ] = \epsilon^{\alpha \beta}
  \epsilon^{MNP} R^{a_1 a_2}{}_P
  \nonumber \\
  & & [ R^{a_1 , M\alpha} , R^{a_2 a_3 }{}_N ]= \delta^M_N R^{a_1 a_2 a_3
  , \alpha}
  \nonumber \\ & &
  [ R^{a_1 a_2}{}_M , R^{a_3 a_4 }{}_N ]= \epsilon_{MNP} R^{a_1 a_2
  a_3 a_4, P}
  \nonumber \\ & &
  [ R^{a_1 , M\alpha} , R^{a_2 a_3 a_4 , \beta} ]  = -\epsilon^{\alpha \beta} R^{a_1 a_2 a_3
  a_4 , M}
  \nonumber \\ & &
  [ R^{a_1 , M\alpha} , R^{a_2 ... a_5 , N} ]  = \epsilon^{MNP} R^{a_1 ...  a_5 ,
  \alpha}{}_P
 \nonumber \\ & &
  [ R^{a_1 a_2}{}_M , R^{a_3 a_4 a_5 , \alpha} ]= R^{a_1 ... a_5,
  \alpha}{}_M
\nonumber \\ & &
  [ R^{a_1 , M\alpha} , R^{a_2 ... a_6 , \beta}{}_N ]  = \epsilon^{\alpha \beta} R^{a_1 ...  a_6 ,
  M}{}_N + D_i^{\alpha \beta} \delta^M_N R^{a_1 ...a_6 , i}
\nonumber \\ & &
  [ R^{a_1 a_2}{}_M , R^{a_3 ... a_6 , N} ]= - R^{a_1 ... a_6, N }{}_M
\nonumber \\ & &
  [ R^{a_1 a_2 a_3 , \alpha} , R^{a_4 a_5 a_6 , \beta} ] =
  D_i^{\alpha \beta} R^{a_1 ...a_6, i}
\nonumber \\ & &
  [ R^{a_1 , M\alpha} , R^{a_2 ... a_7 , i} ]  = D^i_\beta{}^\alpha R^{a_1 ...a_7 , M \beta}
\nonumber \\ & &
  [ R^{a_1 , M\alpha} , R^{a_2 ... a_7 , N}{}_P ]  = {3 \over 8}
  \delta^M_P R^{a_1 ...a_7 , N \alpha} - {1 \over 8}   \delta^N_P R^{a_1 ...a_7 , M \alpha}
  + \epsilon^{MNQ} R^{a_1 ...a_7 , \alpha}{}_{PQ}
\nonumber \\ & &
  [ R^{a_1 a_2}{}_M , R^{a_3 ... a_7 , \alpha}{}_N  ]= {1 \over 8} \epsilon_{MNP}
   R^{a_1 ... a_7, P \alpha } - R^{a_1 ...a_7 , \alpha}{}_{MN}
\nonumber \\ & &
  [ R^{a_1 a_2 a_3 , \alpha} , R^{a_4 a_5 a_6 a_7, M} ] =
  - {1 \over 4} R^{a_1 ...a_7, M \alpha} \quad .
  \label{masslessE11algebraD=8}
  \end{eqnarray}
One can show that all Jacobi identities are satisfied. This requires
the use of the identities of eqs.
(\ref{sl2raiseandlower})-(\ref{symmetricproductsl2}), as well as the
identities
  \begin{equation}
  \epsilon^{M_1 M_2 M_3} \epsilon_{N_1 N_2 N_3} = 6 \delta^{[M_1
  M_2 M_3]}_{[N_1 N_2 N_3]}
  \end{equation}
and
  \begin{equation}
  \epsilon^{\alpha \beta} \epsilon_{\gamma \delta }V^\gamma W^\delta
  = V^\alpha W^\beta - V^\beta W^\alpha \quad ,
  \end{equation}
where in the last equation $V$ and $W$ are two generic $SL(2,
\mathbb{R})$ doublets.

The derivation of the field strengths of all the fields and dual
fields of massless maximal supergravity follows exactly the same
steps as in the other cases. One considers the group element
  \begin{equation}
  g =  e^{x \cdot P} e^{A_{a_1
  ...a_7, M \alpha} R^{a_1 ... a_7, M\alpha }}... e^{A_{a_1 a_2}^{M} R^{a_1 a_2}_{ M}}
  e^{A_{a
  , M \alpha} R^{a , M\alpha}} e^{\phi_{M}{}_N R^{M}{}_N} e^{\phi_i
  R^i} \quad , \label{groupelementinD=8}
  \end{equation}
and computes the Maurer-Cartan form using the fact that in the
massless case the positive level generators commute with momentum.
In this way one derives the field strengths of the massless theory
antisymmetrising the spacetime indices of the various terms in the
Maurer-Cartan form. The field equations are then obtained imposing
duality conditions for the various field strengths. We now consider
all the consistent deformations of the $E_{11}$ algebra resulting
from modifying the commutation relations of the $E_{11}$ generators
with momentum compatibly with the Jacobi identities. In this way we
will derive all the gauged supergravities in eight dimensions.

The representation of the embedding tensor is contained in the
tensor product of the representation of the 1-form generator and of
the scalar generators. In eight dimensions this leads to
  \begin{equation}
  {\bf
  ( 3,2) \otimes [   (1,3) \oplus (8,1)]= (3 ,2 ) \oplus (3, 2)
  \oplus (3,4) \oplus (\overline{6},2) \oplus (15,2) }\quad .
  \end{equation}
We now show that only including an embedding tensor in the ${\bf
(\overline{6},2)}$ or in one of the two ${\bf (3,2)}$
representations leads to a consistent deformation of the algebra.

We first show that the deformations in the ${\bf (15,2)}$ and in the
${\bf (3,4)}$ are ruled out. The first case corresponds to the
embedding tensor $\Theta^{MN ,\alpha}_P$, symmetric in $MN$ and
satisfying the traceless condition $\Theta^{MN ,\alpha}_N =0$. We
want to write down the commutator
  \begin{equation}
  [ R^{a , M\alpha} , P_b ] = - g \Theta^{MN , \alpha}_P \delta^a_b
  R^P{}_N \quad ,
  \end{equation}

but the Jacobi identity between $R^{a,M\alpha}$, $R^{b , N\beta}$
and $P_c$ shows that this is ruled out because of symmetry
arguments. Analogously, in the ${\bf (3,4)}$ case we would write
  \begin{equation}
  [  R^{a , M\alpha} , P_b ] = - g  \Theta^{M , \alpha \beta \gamma}
  D_{i , \beta \gamma} \delta^a_b R^i \quad ,
  \end{equation}
where $\Theta^{M , \alpha \beta \gamma}$ is completely symmetric in
$\alpha \beta \gamma$, but again the Jacobi identity between
$R^{a,M\alpha}$, $R^{b , N\beta}$ and $P_c$ rules this out.

We now consider the two ${\bf(3,2)}$ deformations. These lead to
  \begin{equation}
  [ R^{a , M\alpha} , P_b ] = - g \delta^a_b [ a \Theta^{N \alpha}
  R^M{}_N + b \Theta^{M \beta} D_{i, \beta}{}^\alpha R^i ]
  \end{equation}
where the parameters $a$ and $b$ are in principle arbitrary, and we
now determine the constraints on these parameters that come from the
Jacobi identities. The Jacobi identity between $R^{a,M\alpha}$,
$R^{b , N\beta}$ and $P_c$ gives
  \begin{equation}
  b = - {8 \over 3} a \quad ,
  \end{equation}
and we can fix the parameter $a$ to 1. Therefore, only one of the
two ${\bf(3,2)}$ deformations can lead to a consistent algebra. In
the remaining of this section we show that this deformation is
indeed consistent, and we also show that the embedding tensor in the
${\bf (\overline{6},2)}$ leads to a consistent deformation. We do
this determining the commutation relations of all the generators in
eq. (\ref{eightdimlistofgen}) with momentum consistently with the
Jacobi identities.

\subsection{Embedding tensor in the ${\bf (3,2)}$ of $SL(3,\mathbb{R}) \times
SL(2,\mathbb{R})$} We first consider the deformation in the ${\bf
(3,2)}$, that is the embedding tensor $\Theta^{M \alpha}$. The
result is
  \begin{eqnarray}
  & & [ R^{a, M\alpha} , P_b ] =- g \delta^a_b [ \Theta^{N \alpha}
  R^M{}_N - {8 \over 3} \Theta^{M \beta} D_{i , \beta}{}^\alpha R^i
  ] \nonumber \\
  & & [ R^{a_1 a_2}{}_M , P_b ] = - {1 \over 3} g  \epsilon_{MNP}
  \epsilon_{\alpha \beta} \Theta^{N \alpha} \delta^{[a_1}_b R^{a_2 ]
  , P \beta} \nonumber \\
  & & [ R^{a_1 a_2 a_3 ,\alpha} , P_b ] = {2 \over 3} g \Theta^{M
  \alpha} \delta^{[a_1}_b R^{a_2 a_3 ] }{}_M \nonumber \\
  & & [ R^{a_1 ... a_4 , M} , P_b ] = {2 \over 3} g \epsilon_{\alpha
  \beta } \Theta^{M \alpha} \delta^{[a_1}_b R^{a_2 a_3 a_4 ] ,
  \beta} \nonumber \\
  & & [ R^{a_1 ...a_5 , \alpha}{}_M , P_b ] = {1 \over 3} g
  \epsilon_{MNP} \Theta^{N \alpha} \delta^{[a_1}_b R^{a_2 ...a_5 ] ,
  P} \nonumber \\
  & & [ R^{a_1 ...a_6, i} , P_b ] = - {8 \over 3} g D^i_{\alpha
  \beta} \Theta^{M \alpha} \delta^{[a_1}_b R^{a_2 ...a_6 ] ,
  \beta}{}_M \nonumber \\
  & & [ R^{a_1 ...a_6 , N}{}_M , P_b ] = -g \epsilon_{\alpha \beta}
  [ \Theta^{N \alpha} \delta_M^P - {1 \over 3} \Theta^{P \alpha}
  \delta^N_M ] \delta^{[a_1}_b R^{a_2 ...a_6 ] , \beta}{}_P
  \nonumber \\
  & & [ R^{a_1 ...a_7 , M\alpha} , P_b ] = - {8 \over 3} g \Theta^{M
  \beta} D_{i , \beta}{}^\alpha \delta^{[a_1}_b R^{a_2 ...a_ 7 ] , i} + { 8 \over 3}
  g \Theta^{N \alpha} \delta^{[a_1}_b R^{a_2 ... a_7 ] , M}{}_N
  \nonumber \\
  & & [ R^{a_1 ...a_7 , \alpha}{}_{MN} , P_b ] =0 \quad .
  \end{eqnarray}
From the algebra above as well as the algebra in eq.
(\ref{masslessE11algebraD=8}) and using the group element of eq.
(\ref{groupelementinD=8}) one can compute the field strengths and
the gauge transformations of the fields following the general
analysis of section 2. The field strength of the 1-form is
  \begin{eqnarray}
  & & F_{a_1 a_2 , M \alpha} = 2[ \partial_{[ a_1} A_{a_2 ] , M\alpha}
  +{g \over 3} \epsilon_{MNP} \epsilon_{\alpha \beta} \Theta^{N
  \beta} A_{a_1 a_2}^P -{5g \over 6} \Theta^{N \beta} A_{[a_1 ,
  N\alpha} A_{a_2 ], M\beta}\nonumber \\
  & & \quad \qquad +{g \over 6} \Theta^{N \beta} A_{[a_1 ,M\alpha}
  A_{a_2 ], N\beta} - {g \over 3} \Theta^N_\alpha A_{[a_1 ,M}^\beta
  A_{a_2 ], N\beta} ] \quad ,
  \end{eqnarray}
the field strength of the 2-form is
  \begin{eqnarray}
  & & F_{a_1 a_2 a_3}^M = 3[ \partial_{[a_1}A_{a_2 a_3]}^M +{1 \over 2 }
  \epsilon^{\alpha \beta} \epsilon^{MNP} A_{[a_1 ,N\alpha}
  \partial_{a_2}A_{a_3 ] , P\beta} - {2 g \over 3} \Theta^{M\alpha}
  A_{a_1 a_2 a_3 ,\alpha} \nonumber \\
  & & \quad \qquad + {g \over 3} \Theta^{N\alpha} A_{[a_1 ,N\alpha}
  A_{a_2 a_3 ]}^M -{g \over 3} \Theta^{M\alpha} A_{[a_1 ,N\alpha}
  A_{a_2 a_3]}^N \nonumber \\
  & & \quad \qquad + {7 g \over 3 \cdot 3!} \Theta^{P\beta}
  \epsilon^{MNQ} A_{[a_1 ,N}^\alpha A_{a_2 ,P \alpha} A_{a_3
  ],Q\beta} ]
  \end{eqnarray}
and the field strength of the 3-form is
  \begin{eqnarray}
  & & F_{a_1 ...a_4 , \alpha} = 4 [ \partial_{[a_1}A_{a_2 a_3 a_4 ]
  , \alpha} +A_{[a_1 , M\alpha} \partial_{a_2} A_{a_3 a_4 ]}^M +{1
  \over 3!} \epsilon^{\beta\gamma} \epsilon^{MNP} A_{[a_1 , M\alpha}
  A_{a_2 , N\beta} \partial_{a_3} A_{a_4 ] , P \gamma}\nonumber \\
  & & \qquad \quad +{2 g \over 3} \epsilon_{\alpha \beta} \Theta^{M
  \beta} A_{a_1 ...a_4 , M} - {2 g \over 3} \Theta^{M \beta} A_{[a_1
  , M\alpha} A_{a_2 a_3 a_4 ] , \beta} + {g \over 6} \Theta^{N\beta}
  A_{[a_1 , M \alpha } A_{a_2 , N\beta} A_{a_3 a_4]}^M\nonumber \\
  & & \quad \qquad - {g \over 6} \Theta^{M\beta}A_{[a_1 , M \alpha}
  A_{a_2 , N\beta} A_{a_3 a_4 ]}^N \nonumber \\
  & & \quad \qquad - {7 g \over 3 \cdot 4 !}
  \Theta^{P\delta} \epsilon^{\beta\gamma} \epsilon^{MNQ} A_{[a_1 ,
  M\alpha} A_{a_2 ,N\beta} A_{a_3 , P \gamma} A_{a_4] , Q\delta} ]
  \quad .
  \end{eqnarray}
These field strengths transform covariantly under the gauge
transformations
  \begin{eqnarray}
  & & \delta A_{a, M\alpha } = a_{a, M\alpha} + a_M{}^N
  A_{a,N\alpha} -{1 \over 3} a_N{}^N A_{a , M\alpha} + a_i
  D^i_\alpha{}^\beta A_{a , M \beta} \nonumber \\
  & & \delta A_{a_1 a_2}^M = a_{a_1 a_2}^M -{1 \over 2}
  \epsilon^{\alpha \beta} \epsilon^{MNP} A_{[a_1 , N\alpha } a_{a_2]
  , P\beta} - a_N{}^M A_{a_1 a_2}^N + {1 \over 3 } a_N{}^N A_{a_1 a_2}^M \nonumber \\
  & & \delta A_{a_1 a_2 a_3 , \alpha} = a_{a_1 a_2 a_3,\alpha} +
  A_{[a_1 a_2}^M a_{a_3 ], M\alpha} -{1 \over 3!} \epsilon^{MNP }
  \epsilon^{\beta\gamma} A_{[a_1 , M\alpha} A_{a_2 N \beta} a_{a_3 ]
  ,P\gamma} + a_i D^i_\alpha{}^\beta A_{a_1 a_2 a_3 ,
  \beta}\nonumber \\
  & & \delta A_{a_1 ...a_4, M} = \partial_{[a_1 } \Lambda_{a_2 a_3
  a_4 ], M} -{1 \over 2} \epsilon_{MNP} A_{[a_1 a_2}^N a_{a_3
  a_4]}^P - \epsilon^{\alpha \beta} A_{[a_1 a_2 a_3,\alpha} a_{a_4 ]
  ,\beta}\nonumber \\
  & & \quad \qquad +{1 \over 4!} \epsilon^{\alpha \beta}
  \epsilon^{\gamma \delta} \epsilon^{NPQ} A_{[a_1 , M \alpha} A_{a_2
  , N \beta} A_{a_3 , P \gamma} A_{a_4 ] , Q\delta} -{1 \over 4}
  \epsilon^{\alpha\beta} A_{[a_1 a_2 }^N A_{a_3 , M\alpha} a_{a_4 ]
  , N\beta} \nonumber \\
  & &\quad \qquad  +{1 \over 4} \epsilon^{\alpha\beta} A_{[a_1 a_2 }^N A_{a_3 ,
  N\alpha} a_{a_4 ] , M \beta} + a_M{}^N A_{a_1 ...a_4 ,N} -{1 \over
  3} a_N{}^N A_{a_1 ...a_4 , M} \quad , \label{gaugetransfsind=8}
  \end{eqnarray}
where the parameters $a$ are given in terms of the gauge parameters
$\Lambda$ as
   \begin{eqnarray}
   & & a_M{}^N = - g  \Lambda_{M\alpha} \Theta^{N\alpha} \nonumber
   \\
   & & a^i = {8\over 3}g \Theta^{M\beta} D^i_\alpha{}^\beta
   \Lambda_{M \alpha} \nonumber \\
   & & a_{a, M\alpha } = \partial_a \Lambda_{M\alpha} + {g \over 3}
   \epsilon_{MNP } \epsilon_{\alpha\beta} \Lambda_a^N
   \Theta^{P\beta} \nonumber \\
   & & a_{a_1 a_2}^M = \partial_{[a_1} \Lambda_{a_2]}^M +{2 \over 3}
   g \Theta^{M\alpha} \Lambda_{a_1 a_2, \alpha} \nonumber \\
   & & a_{a_1 a_2 a_3 ,\alpha} = \partial_{[a_1}\Lambda_{a_2 a_3 ],
   \alpha} -{2 \over 3} g \epsilon_{\alpha \beta} \Theta^{M\beta}
   \Lambda_{a_1 a_2 a_3 , M} \quad .
   \end{eqnarray}
Using the formulae given in this paper, the reader can easily
determine the field strengths and gauge transformations for the
remaining fields.

\subsection{Embedding tensor in the ${\bf (\overline{6},2)}$ of $SL(3,\mathbb{R}) \times
SL(2,\mathbb{R})$}

The deformation ${\bf (\overline{6},2 ) }$, corresponding to the
embedding tensor $\Theta_{MN}^\alpha$ symmetric in $MN$, leads to
the commutation relations
  \begin{eqnarray}
  & & [ R^{a, M\alpha} , P_b ] = - g \epsilon^{MNP} \Theta_{NQ}^\alpha
  \delta^a_b R^Q{}_P \nonumber \\
  & & [ R^{a_1 a_2}{}_M , P_b ] = g
  \epsilon_{\alpha \beta} \Theta^{\alpha}_{MN} \delta^{[a_1}_b R^{a_2 ]
  , N \beta} \nonumber \\
  & & [ R^{a_1 a_2 a_3 ,\alpha} , P_b ] = 0 \nonumber \\
  & & [ R^{a_1 ... a_4 , M} , P_b ] = 0 \nonumber \\
  & & [ R^{a_1 ...a_5 , \alpha}{}_M , P_b ] = g
  \Theta^{\alpha}_{MN} \delta^{[a_1}_b R^{a_2 ...a_5 ] ,
  N} \nonumber \\
  & & [ R^{a_1 ...a_6, i} , P_b ] =0 \nonumber \\
  & & [ R^{a_1 ...a_6 , N}{}_M , P_b ]=  g \epsilon_{\alpha \beta}
  \epsilon^{NPQ}
  \Theta^{\alpha}_{MP} \delta^{[a_1}_b R^{a_2 ...a_6 ] ,
  \beta}{}_Q
  \nonumber \\
  & & [ R^{a_1 ...a_7 , M\alpha} , P_b ] = 0
  \nonumber \\
  & & [ R^{a_1 ...a_7 , \alpha}{}_{MN} , P_b ] = 2 g
  \Theta^\alpha_{P (M} \delta^{[a_1}_b R^{a_2 ...a_7 ] ,P}{}_{N)} +
  g \Theta^\beta_{MN} D_{i, \beta}{}^\alpha \delta^{[a_1}_b R^{a_2
  ...a_7 ] , i} \quad .
  \end{eqnarray}
From the algebra above as well as the algebra in eq.
(\ref{masslessE11algebraD=8}) and using the group element of eq.
(\ref{groupelementinD=8}) one can compute the field strengths and
the gauge transformations of the fields corresponding to this
deformation.  We obtain
  \begin{equation}
  F_{a_1 a_2 , M \alpha} = 2[ \partial_{[ a_1} A_{a_2 ] , M\alpha} +
  g \epsilon_{\alpha \beta} \Theta^\beta_{MN} A_{a_1 a_2}^N -{g
  \over 2} \epsilon^{NPQ} \Theta^\beta_{QM} A_{[a_1 ,N\alpha} A_{a_2
  ], P \beta} ]
  \end{equation}
for the field strength of the 1-form,
  \begin{eqnarray}
  & & F_{a_1 a_2 a_3}^M = 3[ \partial_{[a_1}A_{a_2 a_3]}^M +{1 \over 2 }
  \epsilon^{\alpha \beta} \epsilon^{MNP} A_{[a_1 ,N\alpha}
  \partial_{a_2}A_{a_3 ] , P\beta} - g \epsilon^{MNQ}
  \Theta^\alpha_{PQ} A_{[a_1 ,N \alpha} A_{a_2 a_3 ]}^P \nonumber \\
  & & \quad \qquad -{g \over 3!} \epsilon^{\alpha \beta}\epsilon^{MNT}\epsilon^{PQR}
  \Theta^\gamma_{RT} A_{[a_1 ,N \alpha} A_{a_2 , P\beta} A_{a_3],
  Q\gamma}]
  \end{eqnarray}
for the field strength of the 2-form,
  \begin{eqnarray}
  & & F_{a_1 ...a_4 , \alpha} = 4 [ \partial_{[a_1}A_{a_2 a_3 a_4 ]
  , \alpha} +A_{[a_1 , M\alpha} \partial_{a_2} A_{a_3 a_4 ]}^M +{1
  \over 3!} \epsilon^{\beta\gamma} \epsilon^{MNP} A_{[a_1 , M\alpha}
  A_{a_2 , N\beta} \partial_{a_3} A_{a_4 ] , P \gamma}\nonumber \\
  & & \quad \qquad +{g \over 2} \epsilon_{\alpha \beta}
  \Theta^\beta_{MN} A_{[a_1 a_2}^M A_{a_3 a_4 ] }^N -{g \over 2}
  \epsilon^{MNQ} \Theta^\beta_{PQ} A_{[a_1 , M\alpha} A_{a_2 ,
  N\beta}A_{a_3 a_4 ] }^P \nonumber \\
  & & \quad \qquad -{g \over 4!} \epsilon^{MNT} \epsilon^{\beta
  \gamma} \epsilon^{PQR} \Theta^\delta_{RT} A_{[a_1 , M \alpha}
  A_{a_2 , N\beta} A_{a_3, P \gamma} A_{a_4 ] , Q \delta} ]
  \end{eqnarray}
for the field strength of the 3-form. These field strengths
transform covariantly under the gauge transformations determined
from eq. (\ref{gaugetransfsind=8}) once one expresses the parameters
$a$ in terms of the gauge parameters as
  \begin{eqnarray}
  & & a_M{}^N = g \epsilon^{NPQ} \Theta^\alpha_{MP}
  \Lambda_{Q\alpha}  \nonumber \\
  & & a^i = 0 \nonumber \\
  & & a_{a, M\alpha} = \partial_a \Lambda_{M\alpha} - g
  \epsilon_{\alpha \beta} \Theta^\beta_{MN} \Lambda_a^N \nonumber \\
  & & a_{a_1 a_2}^M = \partial_{[a_1} \Lambda_{a_2]}^M \nonumber \\
  & & a_{a_1 a_2 a_3 ,\alpha} = \partial_{[a_1} \Lambda_{a_2 a_3 ]
  ,\alpha} \quad .
  \end{eqnarray}

Also for this deformation one can determine the field strengths and
the gauge transformations of the higher rank fields using the
formulae in section 2 and appendix B.

\section{D=9}
\begin{figure}[h]
\begin{center}
\begin{picture}(380,70)
\multiput(10,10)(40,0){6}{\circle{10}}
\multiput(250,10)(40,0){3}{\circle{10}} \put(370,10){\circle{10}}
\multiput(15,10)(40,0){9}{\line(1,0){30}} \put(290,50){\circle{10}}
\put(290,15){\line(0,1){30}} \put(8,-8){$1$} \put(48,-8){$2$}
\put(88,-8){$3$} \put(128,-8){$4$} \put(168,-8){$5$}
\put(208,-8){$6$} \put(248,-8){$7$} \put(288,-8){$8$}
\put(328,-8){$9$} \put(365,-8){$10$} \put(300,47){$11$}
\put(325,5){\line(1,1){10}} \put(325,15){\line(1,-1){10}}
\put(285,45){\line(1,1){10}} \put(285,55){\line(1,-1){10}}
\end{picture}
\caption{\sl The $E_{11}$ Dynkin diagram corresponding to
9-dimensional supergravity. The non-abelian part of the internal
symmetry group is $SL(2,\mathbb{R})$. \label{Dynkinnine}}
\end{center}
\end{figure}
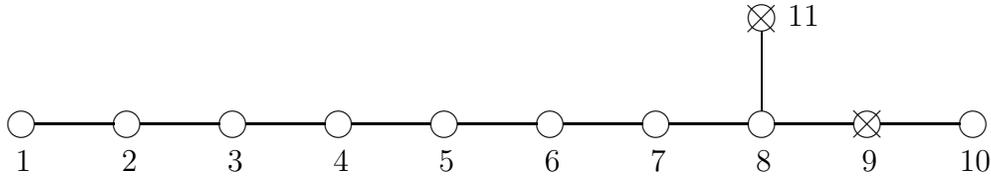

The three scalars of maximal massless nine-dimensional supergravity
parametrise the manifold $\mathbb{R}^+ \times
SL(2,\mathbb{R})/SO(2)$. The theory also contains the metric, a
doublet and a singlet of vectors, a doublet of 2-forms and a 3-form.
The decomposition of $E_{11}$ appropriate to the nine-dimensional
theory is shown in fig. \ref{Dynkinnine}. The form generators of
rank less that 8 that result are associated to the fields of the
supergravity theory and their duals.
The $E_{11}$ algebra also contains 8 and 9 forms, as well as
generators with mixed symmetry. The generators with completely
antisymmetric indices, not including the 9-forms, are
  \begin{eqnarray}
  & & R \quad R^i \quad R^{a} \quad R^{a , \alpha} \quad R^{a_1 a_2, \alpha}
  \quad R^{a_1 a_2 a_3 }
  \quad  R^{a_1 a_2 a_3 a_4} \quad R^{a_1 ... a_5, \alpha}
  \nonumber \\
  & &  R^{a_1 ...a_6 } \quad R^{a_1 ...a_6, \alpha} \qquad R^{a_1
  ... a_7} \qquad R^{a_1 ...a_7 , i } \quad R^{a_1 ...a_8 ,\alpha}
  \quad R^{a_1 ...a_8 , i}
  \quad , \label{listofgeneratorsninedim}
  \end{eqnarray}
where the $SL(2,\mathbb{R})$ conventions are as in the previous
section.

We now list all the non-vanishing commutators involving the
operators in eq. (\ref{listofgeneratorsninedim}). The scalars
satisfy
  \begin{equation}
  [ R^i , R^j ] = f^{ij}{}_k R^k \quad ,
  \label{ninedimscalaralgebra}
  \end{equation}
while all the commutators producing the 1-forms are
  \begin{eqnarray}
  & & [ R, R^a ] = - R^a \qquad \quad [ R , R^{a , \alpha} ] = R^{a
  ,\alpha} \nonumber \\
  & & [ R^i , R^{a , \alpha } ] = D^i_\beta{}^\alpha R^{a , \beta}
  \quad .
  \end{eqnarray}
The 2-form occurs in
  \begin{equation}
  [ R^i , R^{a_1 a_2 , \alpha} ] = D^i_\beta{}^\alpha R^{a_1 a_2 ,
  \beta} \quad \qquad [ R^{a_1} , R^{a_2 , \alpha} ] = - R^{a_1 a_2
  , \alpha} \quad ,
  \end{equation}
the 3-form in
  \begin{equation}
  [ R , R^{a_1 a_2 a_3 } ] = R^{a_1 a_2 a_3} \quad \qquad
  [ R^{a_1 , \alpha} , R^{a_2 a_3, \beta }]= \epsilon^{\alpha \beta} R^{a_1 a_2 a_3 }
  \end{equation}
and the 4-form in
  \begin{equation}
  [ R^{a_1 a_2, \alpha} , R^{a_3 a_4 , \beta} ]= \epsilon^{\alpha \beta} R^{a_1 a_2
  a_3 a_4}
  \quad \qquad
  [ R^{a_1} , R^{a_2 a_3 a_4 } ]  = -R^{a_1 a_2 a_3  a_4 } \quad .
  \end{equation}
The 5-form results from the commutators
  \begin{eqnarray}
  & & [ R , R^{a_1 ... a_5 , \alpha} ]= R^{a_1 ... a_5 , \alpha}
  \quad \qquad [ R^i , R^{a_1 ... a_5 , \alpha} ]= D^i_\beta{}^\alpha R^{a_1 ... a_5 ,
  \beta} \nonumber \\
  && [ R^{a_1 , \alpha} , R^{a_2 ... a_5 } ]  = R^{a_1 ...  a_5 ,
  \alpha} \quad \qquad
  [ R^{a_1 a_2, \alpha } , R^{a_3 a_4 a_5 } ]= R^{a_1 ... a_5,
  \alpha} \quad ,
  \end{eqnarray}
the 6-forms from the commutators
  \begin{eqnarray}
  & & [ R , R^{a_1 ...a_6 } ] = 2 R^{a_1 ...a_6} \quad \qquad
  [ R^i , R^{a_1 ... a_6 , \alpha} ]= D^i_\beta{}^\alpha R^{a_1 ... a_6 ,
  \beta} \nonumber \\
  & &   [ R^{a_1 , \alpha} , R^{a_2 ... a_6 , \beta} ]  = \epsilon^{\alpha \beta} R^{a_1 ...
  a_6} \quad \qquad   [ R^{a_1 a_2 a_3 } , R^{a_4 a_5 a_6 } ] =
  R^{a_1 ...a_6} \nonumber \\
  & &   [ R^{a_1 } , R^{a_2 ... a_6 , \alpha} ]  = - R^{a_1 ...
  a_6 , \alpha} \quad \qquad   [ R^{a_1 a_2 , \alpha} , R^{a_3 ... a_6 } ]=  R^{a_1 ... a_6, \alpha
  }
  \end{eqnarray}
and the 7-forms from the commutators
  \begin{eqnarray}
  & & [ R , R^{a_1 ...a_7 } ] = R^{a_1 ...a_7} \quad \qquad [ R , R^{a_1 ...a_7 , i} ] =
  R^{a_1 ...a_7 , i}\nonumber \\
  & & [ R^i , R^{a_1 ...a_7, j} ] = f^{ij}{}_k R^{a_1 ...a_7 , k}
  \quad \qquad
  [ R^{a_1 } , R^{a_2 ... a_7} ]  = R^{a_1 ...
  a_7} \nonumber \\
  & &
  [ R^{a_1 , \alpha} , R^{a_2 ... a_7 , \beta} ]  = D_i^{\alpha \beta}  R^{a_1 ...a_7 , i}
  + {3 \over 4} \epsilon^{\alpha \beta} R^{a_1 ...a_7}
  \nonumber \\
  & & [ R^{a_1 a_2 , \alpha} , R^{a_3 ... a_7 , \beta}  ]= D_i^{\alpha \beta}  R^{a_1 ...a_7 , i}
  - {1 \over 4} \epsilon^{\alpha \beta} R^{a_1 ...a_7}
  \nonumber \\
  & & [ R^{a_1 a_2 a_3 } , R^{a_4 a_5 a_6 a_7} ] =
  {1 \over 2} R^{a_1 ...a_7} \quad .
  \end{eqnarray}
Finally, the commutators giving rise to the 8-forms are
  \begin{eqnarray}
  & & [ R , R^{a_1 ...a_8 , \alpha} ] = 2 R^{a_1 ...a_8, \alpha}
  \quad \qquad [ R^i , R^{a_1 ...a_8, \alpha} ] = D^i_\beta{}^\alpha
  R^{a_1 ...a_8 , \beta} \nonumber \\
  & & [ R^i , R^{a_1 ...a_8, j} ] = f^{ij}{}_k R^{a_1 ...a_8 , k}
  \quad \qquad
  [ R^{a_1} , R^{a_2 ...a_8 , i}] = - R^{a_1 ...a_8 , i}
  \nonumber \\
  & & [ R^{a_1 , \alpha} , R^{a_2 ...a_8}] =  R^{a_1 ...a_8 , \alpha}
  \quad \qquad
  [ R^{a_1 , \alpha} , R^{a_2 ...a_8 , i}] = 3 D^i_\beta{}^\alpha R^{a_1 ...a_8 , \beta}
  \nonumber \\
  & & [ R^{a_1 a_2 , \alpha} , R^{a_3 ...a_8 , \beta}] = D_i^{\alpha \beta} R^{a_1 ...a_8 , i}
  \quad \qquad [ R^{a_1 a_2 , \alpha} , R^{a_3 ...a_8}] = - R^{a_1 ...a_8 , \alpha}
    \nonumber \\
  & &
  [ R^{a_1 a_2 a_3} , R^{a_3 ...a_8 , \alpha}] = - {1 \over 2} R^{a_1 ...a_8 ,
  \alpha} \quad . \label{eightformalgebraninedim}
  \end{eqnarray}
One can check that all Jacobi identities are satisfied.

The algebra of eqs.
(\ref{ninedimscalaralgebra})-(\ref{eightformalgebraninedim})
determines the fields strengths of all the forms of massless maximal
supergravity in nine-dimensions, with the exception of the 9-forms
that require the taking into account the 9-form generators. The
Maurer-Cartan form that results from the group element
  \begin{equation}
  g = e^{x \cdot P} e^{A_{a_1 ...a_8, \alpha} R^{a_1 ...a_8 ,
  \alpha} } ... e^{A_{a_1 a_2, \alpha}R^{a_1 a_2 ,\alpha}} e^{A_{a
  ,\alpha} R^{a ,\alpha}} e^{A_a R^a} e^{\phi_i R^i } e^{\phi R}
  \label{groupelementinnine}
  \end{equation}
produces indeed these field-strengths once all the spacetime indices
are antisymmetrised. In this derivation of the massless theory, one
imposes that all the generators in eq.
(\ref{listofgeneratorsninedim}) commute with momentum. We now show
that, exactly as in all the other cases discussed in this paper, the
field strengths of all the fields of the gauged maximal
supergravities in nine dimensions result from a deformed algebra,
called $\tilde{E}_{11,9}^{local}$, in which the generators in eq.
(\ref{listofgeneratorsninedim}) have non-trivial commutation
relations with the momentum operator compatibly with the Jacobi
identities.

As usual, from the $E_{11}$ perspective the commutator of the 1-form
generators with momentum give rise to the scalar generators
contracted with the embedding tensor. Therefore, in this
nine-dimensional case the embedding tensor in contained in the
$SL(2,\mathbb{R})$ tensor product
  \begin{equation}
  {\bf ( 1 \oplus 2 ) \otimes (1 \oplus 3 ) = 4 \oplus 3 \oplus 2
  \oplus 2 \oplus 1 }\quad .
  \end{equation}
The singlet $\Theta$ would correspond to the commutator
  \begin{equation}
  [ R^a , P_b ] =-g \Theta \delta^a_b R \quad ,
  \end{equation}
which is ruled out because of the Jacobi identity involving $R^a$,
$R^b$ and $P_c$. Similarly, the quadruplet $\Theta^{\alpha \beta
\gamma}$ would lead to the commutator
  \begin{equation}
  [ R^{a ,\alpha} , P_b ] = - g \Theta^{\alpha \beta \gamma} D_{i ,
  \beta \gamma} \delta^a_b R^i \quad ,
  \end{equation}
which is ruled out because of the Jacobi identity involving $R^{a,
\alpha}$, $R^{b, \beta}$ and $P_c$. The two doublet deformations
lead to the commutator
  \begin{equation}
  [ R^{a ,\alpha} , P_b ] = -g \delta^a_b [ a \Theta^\alpha R + b
  \Theta^\beta D_{i , \beta}{}^\alpha R^i ] \quad ,
  \label{1formmomentumdoubletdeformationD=9}
  \end{equation}
and the Jacobi identities impose the condition
  \begin{equation}
  b = - 4 a \quad . \label{D=9valueofparameters}
  \end{equation}
This leads to one possible doublet deformation, and we fix the
parameter $a$ to 1. To summarise, the only representations of the
embedding tensor which are not ruled out are the triplet and one of
the two doublets. We now show that both these embedding tensors lead
to a consistent algebra. In \cite{allgaugingsD=9} all the possible
gauged maximal supergravities in nine dimensions were constructed.
They indeed correspond to an embedding tensor in either the triplet
or the doublet of $SL(2,\mathbb{R})$.

\subsection{Embedding tensor in the 3 of $SL(2,\mathbb{R})$}
We first consider the deformation in the triplet, which corresponds
to the embedding tensor $\Theta_i$. This case as been considered in
\cite{fabiopeterogievetsky} up to the 5-forms. In that paper the
deformation parameter was denoted with $m_i$, and our conventions
here are such that $m_i = -g \Theta_i$. Deforming from the
commutator of $R^a$ with momentum, all the other commutators are
determined by requiring that the Jacobi identities close. The final
result is
  \begin{eqnarray}
  & & [R^a , P_b ] = - g \delta^a_b \Theta_i R^i \nonumber \\
  & & [ R^{a ,\alpha} , P_b ] = 0 \nonumber \\
  & & [ R^{a_1 a_2 , \alpha} , P_b ] = g \Theta_i D^i_\beta{}^\alpha
  \delta^{[a_1}_b R^{a_2 ] , \beta} \nonumber \\
  & & [ R^{a_1 a_2 a_3} , P_b ] =0 \nonumber \\
  & & [ R^{a_1 ...a_4} , P_b ] =0 \nonumber \\
  & & [ R^{a_1 ... a_5 , \alpha} , P_b ] =0 \nonumber \\
  & & [ R^{a_1 ...a_6 } , P_b ] =0 \nonumber \\
  & & [ R^{a_1 ...a_6 , \alpha} , P_b ] =g \Theta_i
  D^i_\beta{}^\alpha \delta^{[a_1}_b R^{a_2 ... a_6] , \beta}
  \nonumber \\
  & & [ R^{a_1 ...a_7 } , P_b ]=0 \nonumber \\
  & & [ R^{a_1 ... a_7 , i } , P_b ] = - g \Theta^i \delta^{[a_1}_b
  R^{a_2 ... a_7]} \nonumber \\
  & & [ R^{a_1 ...a_8 ,\alpha} , P_b ] =0 \nonumber \\
  & & [ R^{a_1 ...a_8 , i } , P_b ] = -g \Theta^i \delta^{[a_1}_b
  R^{a_2 ...a_8 ]} - g f^{ij}{}_k \Theta_j \delta^{[a_1}_b R^{a_2
  ...a_8 ] , k} \quad .
  \end{eqnarray}
From these commutation relations and the massless algebra of eqs.
(\ref{ninedimscalaralgebra})-(\ref{eightformalgebraninedim}), using
the group element in eq. (\ref{groupelementinnine}), one determines
the field strengths and the gauge transformations on the fields. The
field strengths of the 1-forms are
  \begin{eqnarray}
  & & F_{a_1 a_2 } = 2 \partial_{[a_1} A_{a_2]} \nonumber \\
  & & F_{a_1 a_2 , \alpha} = 2 [ \partial_{[a_1} A_{a_2] , \alpha} -
  g \Theta_i D^i_\alpha{}^\beta A_{a_1 a_2 ,\beta} ]
  \quad ,
  \end{eqnarray}
the field-strength of the 2-form is
  \begin{equation}
  F_{a_1 a_2 a_3 ,\alpha} = 3[ \partial_{[a_1 } A_{a_2 a_3 ] ,
  \alpha} - A_{[a_1} \partial_{a_2} A_{a_3 ], \alpha} + g \Theta_i
  D^i_\alpha{}^\beta A_{[a_1 } A_{a_2 a_3 ] ,\beta} ]
  \quad ,
  \end{equation}
the field strength of the 3-form is
  \begin{equation}
  F_{a_1 ...a_4} = 4 [ \partial_{[a_1} A_{a_2 ...a_4 ]} +
  \epsilon^{\alpha\beta} A_{[a_1} \partial_{a_2} A_{a_3 a_4] ,
  \beta} +{g \over 2} \Theta^i D_i^{\alpha \beta} A_{[a_1 a_2
  ,\alpha} A_{a_3 a_4 ] ,\beta} ]
  \quad,
  \end{equation}
the field strength of the 4-form is
  \begin{eqnarray}
  & & F_{a_1 ... a_5} = 5 [ \partial_{[a_1} A_{a_2 ...a_5 ]} -
  A_{[a_1} \partial_{a_2} A_{a_3 a_4 a_5 ]} - {1\over 2}
  \epsilon^{\alpha \beta} A_{[a_1 a_2, \alpha } \partial_{a_3}
  A_{a_4 a_5 ] , \beta} \nonumber \\
  & & \quad \qquad - \epsilon^{\alpha \beta} A_{[a_1} A_{a_2
  ,\alpha} \partial_{a_3} A_{a_4 a_5 ] ,\beta} -{g \over 2} \Theta^i
  D_i^{\alpha \beta} A_{[a_1} A_{a_2 a_3 , \alpha} A_{a_4 a_5 ]
  ,\beta}]
  \end{eqnarray}
and the field strength of the 5-from is
  \begin{eqnarray}
  & & F_{a_1 ... a_6, \alpha} = 6 [ \partial_{[a_1} A_{a_2 ...a_6 ],
  \alpha}+ A_{[a_1, \alpha} \partial_{a_2} A_{a_3 ...a_6 ]} - A_{ [ a_1
  a_2 ,\alpha} \partial_{a_3} A_{a_4 a_5 a_6]} \nonumber \\
  & & \quad \qquad -{1 \over 2} \epsilon^{\beta \gamma} A_{[a_1 ,
  \alpha} A_{a_2 a_3 , \beta} \partial_{a_4} A_{a_5 a_6 ] ,\gamma}  -
  {g \over 3!} \Theta^i D_i^{\beta \gamma} A_{[a_1 a_2 , \alpha }
  A_{a_3 a_4 , \beta} A_{a_5 a_6 ] ,\gamma} \nonumber \\
  & & \quad \qquad - g \Theta_i D^i_\alpha{}^\beta A_{a_1 ...a_6 ,\beta}] \quad .
  \end{eqnarray}
These field strengths transform covariantly under the gauge
transformations
  \begin{eqnarray}
  & & \delta A_a = a_a - a A_a \nonumber \\
  & & \delta A_{a , \alpha} = a_{a ,\alpha} - a A_{a,\alpha} + a_i
  D^i_\alpha{}^\beta A_{a ,\beta} \nonumber \\
  & & \delta A_{a_1 a_2 ,\alpha } = a_{a_1 a_2 , \alpha} + A_{[a_1
  ,\alpha} a_{a_2 ]} + a_i D^i_\alpha{}^\beta A_{a_1 a_2 ,\beta}
  \nonumber \\
  & & \delta A_{a_1 a_2 a_3 } = a_{a_1 a_2 a_3} - \epsilon^{\alpha
  \beta} A_{[a_1 a_2 ,\alpha} a_{a_3 ],\beta} + {1 \over 2}
  \epsilon^{\alpha \beta} A_{[a_1 ,\alpha} A_{a_2 ,\beta} a_{a_3]} +
  a A_{a_1 a_2 a_3} \nonumber \\
  & & \delta A_{a_1 ...a_4} = a_{a_1 ...a_4} -{1 \over 2}
  \epsilon^{\alpha \beta} A_{[a_1 a_2 ,\alpha } a_{a_3 a_4 ],\beta}
  + A_{[a_1 a_2 a_3} a_{a_4 ]} +{1 \over 2} \epsilon^{\alpha \beta}
  A_{[a_1 a_2 ,\alpha} A_{a_3 ,\beta} a_{a_4 ]} \nonumber \\
  & & \delta A_{a_1 ...a_5 ,\alpha} = a_{a_1 ...a_5 ,\alpha } +
  A_{[a_1 a_2 a_3 } a_{a_4 a_5 ],\alpha} + A_{[a_1 ...a_4 } a_{a_5
  ],\alpha} -{1 \over 2} \epsilon^{\beta \gamma} A_{[a_1 a_2,
  \alpha} A_{a_3 a_4 ,\beta } a_{a_5 ],\gamma} \nonumber \\
  & & \quad \qquad + {1 \over 2} \epsilon^{\beta \gamma} A_{[a_1 a_2
  ,\alpha} A_{a_3 ,\beta} A_{a_4 ,\gamma} a_{a_5]} + a A_{a_1 ...a_5
  ,\alpha} + a_i D^i_\alpha{}^\beta A_{a_1 ...a_5 , \beta} \nonumber
  \\
  & & \delta A_{a_1 ...a_6} = \partial_{[a_1 } \Lambda_{a_2 ...a_6
  ]} -{1 \over 2} A_{[a_1 a_2 a_3 } a_{a_4 a_5 a_6]} +
  \epsilon^{\alpha \beta} A_{[a_1 ... a_5 , \alpha} a_{a_6 ],\beta}
  -{1 \over 2} \epsilon^{\alpha \beta} A_{[a_1 a_2 a_3 } A_{a_4 a_5
  , \alpha} a_{a_6],\beta}\nonumber \\
  & & \quad \qquad +{1 \over 4} A_{[a_1 a_2 a_3 } A_{a_4 ,\alpha}
  A_{a_5 ,\beta} a_{a_6 ] } + 2 a A_{a_1 ...a_6} \nonumber \\
  & & \delta A_{a_1 ...a_6 ,\alpha} = \partial_{[a_1 } \Lambda_{a_2 ...a_6
  ] , \alpha } + A_{[ a_1 ...a_4 } a_{a_5 a_6 ] ,\alpha} -{1 \over
  3! } \epsilon^{\beta \gamma} A_{[a_1 a_3 ,\alpha} A_{a_3 a_4
  ,\beta} a_{a_5 a_6 ], \gamma} + A_{[a_1 ...a_5 ,\alpha} a_{a_6 ]}
  \nonumber \\
  &  &\quad \qquad - {1 \over 2} \epsilon^{\beta \gamma} A_{[a_1 a_2
  ,\alpha} A_{a_3 a_4 ,\beta} A_{a_5 ,\gamma} a_{a_6 ]} + a_i
  D^i_\alpha{}^\beta A_{a_1 ..a_6 , \beta} \quad ,
  \label{gaugetransfsinnine}
  \end{eqnarray}
where the parameters $a$ are given in terms of the gauge parameters
$\Lambda$ as
  \begin{eqnarray}
  & & a  = 0 \nonumber \\
  & & a^i = -g \Lambda \Theta^i \nonumber \\
  & & a_a = \partial_a \Lambda \nonumber \\
  & & a_{a ,\alpha} = \partial_a \Lambda_\alpha + g \Theta_i
  D^i_\alpha{}^\beta \Lambda_{ a,\beta} \nonumber \\
  & & a_{a_1 a_2 , \alpha } = \partial_{[a_1} \Lambda_{a_2 ]
  ,\alpha} \nonumber \\
  & & a_{a_1 a_2 a_3} = \partial_{[a_1 } \Lambda_{a_2 a_3 ] }
  \nonumber \\
  & & a_{a_1 ...a_4} = \partial_{[a_1} \Lambda_{a_2 a_3 a_4]}
  \nonumber \\
  & & a_{a_1 ...a_5 ,\alpha} = \partial_{[a_1} \Lambda_{a_2 ...a_5] ,\alpha} + g
  \Theta_i D^i_\alpha{}^\beta \Lambda_{a_1 ...a_5 ,\beta} \quad .
  \end{eqnarray}
The reader can easily evaluate the remaining field strengths and
gauge transformations.

\subsection{Embedding tensor in the 2 of $SL(2,\mathbb{R})$}
We now consider the doublet deformation, corresponding to the
embedding tensor $\Theta^\alpha$. We start from the commutator
between the 1-form and momentum as in eq.
(\ref{1formmomentumdoubletdeformationD=9}) with the parameters as in
eq. (\ref{D=9valueofparameters}) with $a=1$. Imposing the closure of
the Jacobi identities gives
  \begin{eqnarray}
  & & [R^a , P_b ] = 0 \nonumber \\
  & & [ R^{a ,\alpha} , P_b ] = - g \delta^a_b [ \Theta^\alpha R - 4 \Theta^\beta D_{i ,
  \beta}{}^\alpha  R^i ] \nonumber \\
  & & [ R^{a_1 a_2 , \alpha} , P_b ] = g \Theta^\alpha
  \delta^{[a_1}_b R^{a_2 ]} \nonumber \\
  & & [ R^{a_1 a_2 a_3} , P_b ] = -g \epsilon_{\alpha \beta} \Theta^\alpha
  \delta^{[a_1}_b R^{a_2 a_3 ] , \beta} \nonumber \\
  & & [ R^{a_1 ...a_4} , P_b ] =0 \nonumber \\
  & & [ R^{a_1 ... a_5 , \alpha} , P_b ] =0 \nonumber \\
  & & [ R^{a_1 ...a_6 } , P_b ] = -2 g \epsilon_{\alpha \beta} \Theta^\alpha
  \delta^{[a_1}_b R^{a_2 ...a_6 ] , \beta} \nonumber \\
  & & [ R^{a_1 ...a_6 , \alpha} , P_b ] =0
  \nonumber \\
  & & [ R^{a_1 ...a_7 } , P_b ]= -2 g \epsilon_{\alpha \beta} \Theta^\alpha
  \delta^{[a_1}_b R^{a_2 ...a_7 ] , \beta}  \nonumber \\
  & & [ R^{a_1 ... a_7 , i } , P_b ] = - 2 g \Theta^\alpha D^i_{ \alpha \beta} \delta^{[a_1}_b
  R^{a_2 ... a_7] , \beta} \nonumber \\
  & & [ R^{a_1 ...a_8 ,\alpha} , P_b ] = {1 \over 2} g \Theta^\alpha \delta^{[a_1}_b R^{a_2 ...a_8 ]} - 2 g
  D_{i, \beta}{}^\alpha \Theta^\beta \delta^{[a_1}_b R^{a_2 ...a_8 ], i}
   \nonumber \\
  & & [ R^{a_1 ...a_8 , i } , P_b ] = 0 \quad .
  \end{eqnarray}
From these commutation relations and the massless algebra of eqs.
(\ref{ninedimscalaralgebra})-(\ref{eightformalgebraninedim}), using
the group element in eq. (\ref{groupelementinnine}), one determines
the field strengths and the gauge transformations on the fields. The
field strengths of the 1-forms are
  \begin{eqnarray}
  & & F_{a_1 a_2 } = 2 [ \partial_{[a_1} A_{a_2]} - g \Theta^\alpha
  A_{a_1 a_2 , \alpha} + g \Theta^\alpha A_{[a_1} A_{a_2 ] ,\alpha}] \nonumber \\
  & & F_{a_1 a_2 , \alpha} = 2 [ \partial_{[a_1} A_{a_2] , \alpha} -
  g \Theta^\beta A_{[a_1 ,\alpha } A_{a_2 ] , \beta} - {g \over 2}
  \Theta_\alpha A_{[a_1}^\beta A_{a_2] , \beta} ]
  \quad ,
  \end{eqnarray}
the field-strength of the 2-form is
  \begin{eqnarray}
  & & F_{a_1 a_2 a_3 ,\alpha} = 3[ \partial_{[a_1 } A_{a_2 a_3 ] ,
  \alpha} - A_{[a_1} \partial_{a_2} A_{a_3 ], \alpha} - g
  \epsilon_{\alpha \beta} \Theta^\beta A_{a_1 a_2 a_3}  + g \Theta^\beta A_{[a_1 ,\alpha} A_{a_2 a_3 ] ,\beta}\nonumber \\
  & & \quad \qquad + g
  \Theta^\beta A_{[a_1} A_{a_2 , \alpha} A_{a_3 ],\beta} +{g \over
  2} \Theta_\alpha A_{[a_1} A_{a_2}^\beta A_{a_3 ] , \beta} ]
  \quad ,
  \end{eqnarray}
the field strength of the 3-form is
  \begin{eqnarray}
  & & F_{a_1 ...a_4} = 4 [ \partial_{[a_1} A_{a_2 ...a_4 ]} +
  \epsilon^{\alpha\beta} A_{[a_1} \partial_{a_2} A_{a_3 a_4] ,
  \beta} + g \Theta^\alpha A_{[a_1 ,\alpha} A_{a_2 a_3 a_4]} \nonumber \\
  & & \quad \qquad - {g
  \over 2} \Theta^\alpha A_{[a_1}^\beta A_{a_2 ,\beta} A_{a_3 a_4 ],
  \alpha} ]
  \quad,
  \end{eqnarray}
the field strength of the 4-form is
  \begin{eqnarray}
  & & F_{a_1 ... a_5} = 5 [ \partial_{[a_1} A_{a_2 ...a_5 ]} -
  A_{[a_1} \partial_{a_2} A_{a_3 a_4 a_5 ]} - {1\over 2}
  \epsilon^{\alpha \beta} A_{[a_1 a_2, \alpha } \partial_{a_3}
  A_{a_4 a_5 ] , \beta} \nonumber \\
  & & \quad \qquad - \epsilon^{\alpha \beta} A_{[a_1} A_{a_2
  ,\alpha} \partial_{a_3} A_{a_4 a_5 ] ,\beta} -g \Theta^\alpha
  A_{[a_1 a_2 , \alpha} A_{a_3 a_4 a_5 ]} \nonumber \\
  & & \quad \qquad - g \Theta^\alpha A_{[a_1} A_{a_2 ,\alpha} A_{a_3
  a_4 a_5]} -{g \over 2} \epsilon^{\alpha \beta} \Theta^\gamma
  A_{[a_1} A_{a_2 ,\alpha} A_{a_3, \beta} A_{a_4 a_5 ],\gamma}]
  \end{eqnarray}
and the field strength of the 5-from is
  \begin{eqnarray}
  & & F_{a_1 ... a_6, \alpha} = 6 [ \partial_{[a_1} A_{a_2 ...a_6 ],
  \alpha}+ A_{[a_1, \alpha} \partial_{a_2} A_{a_3 ...a_6 ]} - A_{ [ a_1
  a_2 ,\alpha} \partial_{a_3} A_{a_4 a_5 a_6]} \nonumber \\
  & & \quad \qquad -{1 \over 2} \epsilon^{\beta \gamma} A_{[a_1 ,
  \alpha} A_{a_2 a_3 , \beta} \partial_{a_4} A_{a_5 a_6 ] ,\gamma} -2 g \epsilon_{\alpha \beta} \Theta^\beta
  A_{a_1 ...a_6} \nonumber \\
  & & \quad \qquad -g \Theta^\beta A_{[a_1 , \alpha} A_{a_2 a_3 ,\beta} A_{a_4 a_5 a_6 ]} ]\quad .
  \end{eqnarray}
These field strengths transform covariantly under the gauge
transformations in eq. (\ref{gaugetransfsinnine}), with the
parameters $a$ given in terms of the gauge parameters $\Lambda$ as
  \begin{eqnarray}
  & & a  = -g \Theta^\alpha \Lambda_\alpha \nonumber \\
  & & a^i = 4 g \Theta^\beta D^i_\beta{}^\alpha  \Lambda_\alpha \nonumber \\
  & & a_a = \partial_a \Lambda + g \Theta^\alpha \Lambda_{a, \alpha} \nonumber \\
  & & a_{a ,\alpha} = \partial_a \Lambda_\alpha  \nonumber \\
  & & a_{a_1 a_2 , \alpha } = \partial_{[a_1} \Lambda_{a_2 ]
  ,\alpha} + g \epsilon_{\alpha \beta} \Theta^\beta \Lambda_{a_1 a_2 }  \nonumber \\
  & & a_{a_1 a_2 a_3} = \partial_{[a_1 } \Lambda_{a_2 a_3 ] }
  \nonumber \\
  & & a_{a_1 ...a_4} = \partial_{[a_1} \Lambda_{a_2 a_3 a_4]}
  \nonumber \\
  & & a_{a_1 ...a_5 ,\alpha } = \partial_{[a_1} \Lambda_{a_2 ...a_5], \alpha } + 2 g
  \epsilon_{\alpha \beta} \Theta^\beta  \Lambda_{a_1 ...a_5 } \quad .
  \end{eqnarray}
The reader can easily evaluate the remaining field strengths and
gauge transformations corresponding to this deformation.

\section{Form field equations and duality conditions}
In this section we write down the equations of motion for the form
fields taking into account that we have fields and their duals. Such
equations have been studied on an ad-hoc basis previously beginning
with \cite{cremmerjulialupope}. However, our discussions will be in
the context of $E_{11}$ and in particular the representations and
hierarchy of form fields it predicts \cite{fabiopeterE11origin} and
as is given in the table of \cite{fabiopeterE11origin} that is table
\ref{alwaysthesametable} of this paper.
\begin{table}
\begin{tiny}
\begin{center}
\begin{tabular}{|c|c||c|c|c|c|c|c|c|c|c|c|}
\hline \rule[-1mm]{0mm}{6mm}
D & G & 1-forms & 2-forms & 3-forms & 4-forms & 5-forms & 6-forms & 7-forms & 8-forms & 9-forms & 10-forms\\
\hline \rule[-1mm]{0mm}{6mm} \multirow{2}{*}{10A} &
\multirow{2}{*}{$\mathbb{R}^+$} & \multirow{2}{*}{${\bf 1}$} &
\multirow{2}{*}{${\bf 1}$} & \multirow{2}{*}{${\bf 1}$} &  &
\multirow{2}{*}{${\bf 1}$} & \multirow{2}{*}{${\bf 1}$} &
\multirow{2}{*}{${\bf 1}$} & \multirow{2}{*}{${\bf 1}$} &
\multirow{2}{*}{${\bf 1}$} & ${\bf 1}$ \\
& & & & & & & & & & & ${\bf 1}$ \\
\hline \rule[-1mm]{0mm}{6mm} \multirow{2}{*}{10B} &
\multirow{2}{*}{$SL(2,\mathbb{R})$} & & \multirow{2}{*}{${\bf 2}$} &
&
\multirow{2}{*}{${\bf 1}$} & & \multirow{2}{*}{${\bf 2}$} & & \multirow{2}{*}{${\bf 3}$} & & ${\bf 4}$ \\
& & & & & & & & & & & ${\bf 2}$ \\
\cline{1-12} \rule[-1mm]{0mm}{6mm} \multirow{3}{*}{ 9} &
\multirow{3}{*}{ $SL(2,\mathbb{R})\times \mathbb{R}^+$} & ${\bf 2}$
& \multirow{3}{*}{${\bf 2 }$} & \multirow{3}{*}{ ${\bf 1}$} &
\multirow{3}{*}{ ${\bf 1}$} & \multirow{3}{*}{ ${\bf
2 }$} & ${\bf 2}$ & ${\bf 3}$ & ${\bf 3}$ & ${\bf 4}$  \\
& & & & & & & & & & ${\bf 2}$ \\
& & ${\bf 1}$ & & & & & ${\bf 1}$ & ${\bf 1}$ & ${\bf 2}$ & ${\bf 2 }$  \\
\cline{1-11} \rule[-1mm]{0mm}{6mm} \multirow{4}{*}{8} &
\multirow{4}{*}{$SL(3,\mathbb{R}) \times SL(2,\mathbb{R})$} &
\multirow{4}{*}{${\bf (\overline{3}, 2)}$} & \multirow{4}{*}{${\bf
(3,1) }$} & \multirow{4}{*}{${\bf (1,2)}$} & \multirow{4}{*}{${\bf
(\overline{3},1)}$} & \multirow{4}{*}{${\bf (3,2) }$} &  &  & ${\bf (15,1)}$  \\
& & & & & & & ${\bf (8,1)}$ & ${\bf (6,2)}$ & ${\bf (3,3)}$  \\
& & & & & & & ${\bf (1,3)}$ & ${\bf (\overline{3},2)}$ & ${\bf (3,1)}$  \\
& & & & & & &  &  & ${\bf (3,1)}$ \\
 \cline{1-10} \rule[-1mm]{0mm}{6mm} \multirow{3}{*}{7} & \multirow{3}{*}{$SL(5,\mathbb{R})$} & \multirow{3}{*}{${\bf
\overline{10}}$} & \multirow{3}{*}{${\bf 5 }$} &
\multirow{3}{*}{${\bf \overline{5}}$} & \multirow{3}{*}{${\bf 10}$}
&
\multirow{3}{*}{${\bf 24 }$} & ${\bf \overline{40}}$ & ${\bf 70}$  \\
& & & & & & & & ${\bf  45}$  \\
& & & & & & & ${\bf \overline{15}}$ & ${\bf 5}$  \\
 \cline{1-9} \rule[-1mm]{0mm}{6mm}\multirow{3}{*}{6} & \multirow{3}{*}{$SO(5,5)$} & \multirow{3}{*}{${\bf
16}$} & \multirow{3}{*}{${\bf 10 }$} & \multirow{3}{*}{${\bf
\overline{16} }$} & \multirow{3}{*}{${\bf 45}$} &
\multirow{3}{*}{${\bf 144 }$} & ${\bf 320}$  \\
& & & & & & & ${\bf \overline{126}}$ \\
& & & & & & & ${\bf 10}$ \\
\cline{1-8} \rule[-1mm]{0mm}{6mm} \multirow{2}{*}{5} &
\multirow{2}{*}{$E_{6(+6)}$} & \multirow{2}{*}{${\bf 27}$} &
\multirow{2}{*}{${\bf \overline{27} }$} & \multirow{2}{*}{${\bf 78
}$} & \multirow{2}{*}{${\bf 351}$} &
${\bf \overline{1728}}$  \\
& & & & & &  ${\bf \overline{27}}$  \\
 \cline{1-7} \rule[-1mm]{0mm}{6mm} \multirow{2}{*}{4} & \multirow{2}{*}{$E_{7(+7)}$} & \multirow{2}{*}{${\bf
56}$} & \multirow{2}{*}{${\bf 133 }$} & \multirow{2}{*}{${\bf 912 }$} & ${\bf 8645}$ \\
 & & & & & ${\bf 133}$ \\
 \cline{1-6} \rule[-1mm]{0mm}{6mm}
 \multirow{3}{*}{3} &  \multirow{2}{*}{$E_{8(+8)}$} &  \multirow{2}{*}{${\bf 248}$} & ${\bf 3875}$
 & {\bf 147250}  \\
 & & & & {\bf 3875} \\
 & & & ${\bf 1}$ & {\bf 248} \\
 \cline{1-5}
\end{tabular}
\end{center}
\end{tiny}
\caption{\sl Table giving the representations of the symmetry group
$G$ of all the forms fields of maximal supergravities in any
dimension \cite{fabiopeterE11origin}. The 3-forms in three
dimensions were determined in \cite{ericembeddingtensor}. It is
important to observe that these are the representations of the
fields, which are the contragredient of the representations of the
corresponding generators, which have been considered in this paper.
\label{alwaysthesametable}}
\end{table}

If we assume that the form field equations are first order in
space-time derivatives they can only be  duality relations between
the field strengths obtained in this paper.  Let us first consider
gauge fields whose field strengths have a rank that is not half that
of the dimensions of space-time, that is those that do not obey some
kind of generalised self duality condition. Examining the table
\ref{alwaysthesametable} of the representations of $G$ of the form
fields we find that for every gauge field of rank $n $ for
$n<{1\over 2} D$, with a field strength $F_{n+1}$ of rank $n+1$,
that belongs to a representation ${\bf R_n}$ there is a dual gauge
field of rank $D-n-2$ with a field strength $F_{D-n-1}$ of rank
$D-n-1$ which is in that representation ${\bf R_{D-n-2}}$ which
turns out to be the conjugate representation, i.e. ${\bf
R_{D-n-2}=\overline{R}_{n}}$. The field strengths that occur in the
Cartan forms transform under $G$ with a non-linear action that is
only a transformation under the Cartan involution invariant subgroup
$I(G)$ rather than the above mentioned linear representations of
$G$. This is due to the scalar factors mentioned above which convert
the linear representation into the non-linear representation in the
well know manner. Thus demanding invariant field equations reduces
to finding those invariant under only $I(G)$ transformations.
Examining all such gauge fields for $D\le 7$, we find that under the
decomposition of their representations of $G$ from $G$ to $I(G)$ we
find one irreducible real representation of $I(G)$. Hence the gauge
fields and their duals belong to the same representation of $I(G)$.
For example in seven dimensions the two forms belong to the ${\bf
5}$ of $SL(5,\mathbb{R})$ while their dual gauge fields, the three
forms, belong to the ${\bf \overline{5}}$ of $SL(5,\mathbb{R})$. The
Cartan involution invariant subgroup is $I(SL(5, \mathbb{R}))=SO(5)$
and these two gauge fields both  belong to the real ${\bf 5}$
representation of this group.

The field equations for all such gauge fields can only be of the
form
 \begin{equation}
 F_{n+1}=\star F_{D-n-1} \label{firstsec10}
 \quad ,
 \end{equation}
where $\star$ is the space-time dual, since they each belong to the
same irreducible representation of $I(G)$. For dimensions $D\ge 8$
the gauge fields belong to representations of $G$ that decompose
into at most two distinct irreducible real representations of $I(G)$
and their dual gauge fields belong to precisely the same
representations of $I(G)$. Then the duality condition consists of as
many equations as there are representations of $I(G)$,   which are
of the form of eq. (\ref{firstsec10}) and they relate  the gauge
field and its dual in the same representation of $I(G)$.
\par
The scalars are a non-linear realisation of $G$ and obey duality
relations with the rank $D-2$ forms which are in the adjoint
representation. Under the decomposition from $G$ to $I(G)$ the
adjoint representation of the latter breaks into the adjoint of
$I(G)$  and the ``coset'' part. Only the latter enters into the
duality condition with the coset part of the Cartan form formed from
the scalars. The scalar equation results from the curl of these
duality relations. Such curl reproduces the field strengths of the
$D-1$ form fields, which are dual to the embedding tensor, and this
gives rise to the scalar potential. In general there is more than
one gauge covariant quantity that one can construct contracting the
scalars with the embedding tensor, and the method we have presented
in this paper of determining all the gauge covariant quantities of
the theory does not determine their relative coefficient, and
therefore does not determine the exact form of the scalar potential.
\par
For odd dimensions space-times there are clearly no generalised self
duality conditions. However, the cases when $D=4m$ and $D=4m+2$, for
integer $m$  are different due to the fact that $\star\star=-1$  and
$\star\star=+1$ respectively when acting on a ${D \over 2}$-form.
Let us begin with the latter case, that is dimensions ten and six.
As is well known in ten dimensions we have a four form gauge field
that is a singlet of $SL(2,\mathbb{R})$ and obeys a self duality
condition of the form $F_5=\star F_5$. In six dimensions that two
forms belong to the ${\bf 10}$ of SO(5,5) which decomposes into the
reducible representation ${\bf (5,1)\oplus (1,5)}$ of $SO(5)\otimes
SO(5)=I(SO(5,5))$. The  duality condition which is invariant under
the $SO(5)\otimes SO (5)$ transformations of the field strength can
only be of the form
  \begin{equation}
  \left( \begin{array}{cc}
  F_3 \\ F'_3 \end{array}\right)
  =\star
   \left( \begin{array}{cc} F_3 \\ - F'_3
  \end{array}\right)
  \quad ,
  \end{equation}
where $F_3$ and $ F'_3$ belong to the $(5,1)$ and $ (1,5)$
representations of $SO(5)\otimes SO(5)$ respectively. The minus sign
is required as there must be the same number of self-dual and
anti-self-dual forms as the resulting theory describes 5 tensors
that do not satisfy self-duality conditions.
\par
Let us now consider the case of $D=4m+2$ that is dimensions eight
and four. In this case the forms belong to an irreducible
representation of $G$ that breaks into two irreducible
representations of $I(G)$ which are related by complex conjugation.
In eight dimensions  the three forms belong to the ${\bf(1,2)}$
representation of $SL(3,\mathbb{R})\otimes SL(2,\mathbb{R})$ which
breaks into ${\bf (1,1^+)}$ and ${\bf (1,1^-)}$ representations of
$I(SL(3,\mathbb{R})\otimes SL(2,\mathbb{R}))= SO(3)\otimes SO(2)$.
As such the unique invariant field equation is of the form
  \begin{equation}
  \left( \begin{array}{cc}
  F_4 \\ F^*_4 \end{array}\right)
  = i \star
   \left( \begin{array}{cc}
  F_4 \\ - F^*_4
  \end{array}\right)
  \quad ,
  \end{equation}
where $F_4$ and $ F^*_4$ belong to the $(1,1^+)$ and $(1,1^-)$
representations of $ SO(3)\otimes SO(2)$. The $i$ found in this
equation is due to the fact that $\star\star=-1$ in this dimension
and the minus sign then results from the consistency with respect to
complex conjugation. In four dimensions the one forms belong to the
${\bf 56}$ dimensional representation of $E_7$ which decompose into
the ${\bf 28\oplus \overline{28}}$ of representations
$I(E_7)=SU(8)$. The self duality condition can only be of the form
  \begin{equation}
  \left( \begin{array}{cc}
  F_2\\ F^*_2 \end{array}\right)
  = i \star
   \left( \begin{array}{cc}
   F_2\\ - F^*_2
  \end{array}\right)
  \quad ,
  \end{equation}
where $F_2$ and $ F^*_2$ are the ${\bf 28\oplus \overline{28}}$
representations of $SU(8)$.
\par
These equations of motion are the correct equations although we have
not derived these duality relations as following from $E_{11}$ in
this paper. This remains a future project. There is a certain
freedom to rescale the form fields by constants which is reflected
that these duality relations can have constants that are not
explicitly shown above. These constants are fixed once one also
writes down the field equation for gravity as this involves the
stress tensor. It is impressive to see the way the representations
of the form fields, dictated by $E_{11}$,  cooperate with the
demands that the form field equations be duality conditions.

The duality relations discussed in this section have a crucial role
in determining the closure of the supersymmetry algebra. Indeed,
these relations are first order in derivatives, and the closure of
the supersymmetry algebra on fields and dual fields, as well as on
$D-1$ and $D$ forms, only occurs provided that they are satisfied.
In \cite{fabioericIIB} and \cite{fabioericIIA} it was shown that the
supersymmetry algebra of IIB and IIA respectively close on all the
fields and dual fields provided that the duality relations are
satisfied. The supersymmetry algebra also fixes the $D-1$ and $D$
forms that one can include, the result being exactly in agreement
with the predictions of $E_{11}$ \cite{axeligorpeter} (subsequently,
it was shown in \cite{peter10formsE11} that also the detailed
coefficients of the gauge algebra of the IIB theory are reproduced
by $E_{11}$). In particular the IIA algebra describes both the
massive and massless field equations, as the field strength of the
9-form can be set equal to the Romans cosmological constant
\cite{Romansfrom9form} or to zero respectively. More recently, the
closure of the supersymmetry algebra on higher rank forms was shown
for the case of gauged maximal supergravities in five dimensions
\cite{fabiopeterextendedspacetime} and in three dimensions
\cite{hierarchy2}.

\section{Conclusions}
It was previously  found
\cite{fabiopeterE11origin,ericembeddingtensor} that the maximal
gauged supergravity theories were classified by $E_{11}$. In
particular the forms of rank $D-1$ in the $D$ dimensional
supergravity theory, which lead to a cosmological constants, are in
the contragredient representation of the internal symmetry group of
the tensor which was known to label all such theories. Although this
discovery was kinematical in nature it demonstrated that $E_{11}$
provided, for the first time,  a unifying scheme within which to
consider all such theories. In this paper we show that $E_{11}$, by
the steps described in this paper, leads to all the field strengths
of the maximal gauged supergravity theories. The embedding tensor
arises as the tensor that uniquely determines the deformation of the
$E_{11}$ algebra from which the gauged supergravities arise as
non-linear realisations and we show that it is in the same
representation as the $D-1$ form generators. We have analysed each
dimension from three to nine, and these results, together with the
ten-dimensional deformation corresponding to the Romans theory
analysed in \cite{fabiopeterogievetsky}, give the field strengths of
all possible massive maximal supergravities in any dimension.

If one assumes, as is the case,  that the dynamics of the form
fields are first order in space-time derivatives then they must be
given by duality relations  on the field strengths calculated in
this paper. As a result, for all the fields apart from the scalars,
the dynamics of the bosonic sector is then determined up to a few
constants that multiply the field strengths. In the absence of the
gravity equation that contains the stress tensor one can fix these
constants by field redefinitions, hence in this sense the dynamics
is determined in the absence of gravity (in the case of the scalar
equation in general there is more than one gauge covariant quantity
that one can construct contracting the scalars with the embedding
tensor, and the procedure presented in this paper does not determine
their relative coefficient, and therefore does not determine the
exact form of the scalar potential). Thus most results on the
maximal gauged supergravity theories, including those that have been
derived over many years, can be found in a very quick, efficient and
unified manner from $E_{11}$.
\par
In the $E_{11}$ formulation of the maximal gauged supergravities
theories the field content in a given dimension is the same although
the actual physical degree of freedom in any given gauged
supergravity theory may differ. In particular the number of $D-1$
forms is the same and so a given maximal gauged supergravity theory
has a knowledge of all the other possible gauged supergravity
theories in the same dimension. This is analogous to having various
different theories and then discovering that there is a potential
from which they can all be derived as different minima.
\par
In this paper we have used a deformation of an algebra $\tilde
E_{11}$ containing the $E_{11}$ algebra, the usual space-time
translation and  the Ogievetsky generators. However, in reference
\cite{fabiopeterogievetsky} a detailed study of the nine dimensional
gauged supergravities was carried out and it was found that these
theories arose from the full $E_{11,10B}^{local}$ including the
parts associated with ten dimensions. However, these gauged
supergravities could be constructed from only a subalgebra of
$\tilde{E}_{11,9}^{local}$ which appeared to be a deformed
$E^{local}_{11,9}$ algebra as a result of the complicated field
redefinitions of the generators and the generators that were dropped
as they played no role in the dynamics. As such in these theories
one is dealing with a subalgebra of $\tilde{E}_{11,9}^{local}$ which
only appears as a deformation as a consequence of the way the
calculation is carried out. It would be of interest to see if this
is a general phenomenon.
\par
As discussed in the introduction, the original gauged supergravities
were derived by adding a deformations to the massless theory and
using supersymmetry to find the complete the theory. In such an
approach one did not use fields that were in representations of the
internal symmetry group $G$. Later gauged maximal supergravities
were constructed using fields that were representations of $G$, but
the theory also contained an embedding tensor that labelled the
theories and broke the internal symmetry group $G$. In this way of
proceeding  the fields always occurred together with this tensor in
just such a way that  the full $G$ representations of the fields was
not present into the equations of motion.  However, in the last few
years the bosonic sector of certain gauged supergravities have been
constructed \cite{hierarchy1} by taking the physical degrees of
freedom to be described by  fields that are representations of the
internal symmetry and demanding that these unconstrained fields
carry a gauge algebra extending the gauging of part of the local
internal symmetry. In carrying out this programme these authors have
found a hierarchy of fields of increasing rank \cite{hierarchy2}.
However, these are just those found previously in the $E_{11}$
approach \cite{fabiopeterE11origin}. It is obvious that this
procedure is just a bottom up way of discovering the form sector of
$E_{11}$ and it is not necessary to speculate about the mysterious
degrees of freedom of M theory that such a process may have
uncovered.
\par
While there can be no doubt of the calculational efficiency of the
approach of this paper it leaves open a number of more conceptual
questions. For example, what mathematical  object do the Ogievetsky
generators belong to. Also even though one uses  only the positive
and zero level  part of the $E_{11}$  algebra in the deformed
algebra, this algebra is only defined from the full $E_{11}$
algebra. As a result many properties of, and deductions from, the
full $E_{11}$ algebra are imported into the calculation. This would
at best seem unnatural. In a previous paper it was proposed to
derive the gauged supergravity theories, and explicitly the five
dimensional gauged supergravities, from a non-linear realisation of
$E_{11}\otimes_s l_1$  where $l_1$ is the fundamental representation
associated to the first node of $E_{11}$ and leads to a generalised
space-time. This approach has the advantage that it is conceptually
well defined from the mathematical viewpoint,  but it is less clear
how some of the physical aspects of the gauged supergravity theories
emerge in a natural way. These include  how the local gauge
transformations arise and how the slice of generalised space-time
that is active  arises from the full generalised space-time. Thus
one has a dilemma how to include space-time, local gauge
transformations within $E_{11}$. In this context we might mention
the interesting work of reference \cite{christian} that concerns the
role of diffeomorphism symmetries in the context of the non-linear
realisation relevant to supergravity theories. We hope to report
elsewhere on progress in this area.

\vskip 2cm

\section*{Acknowledgments}
This work is supported by the PPARC rolling grant PP/C5071745/1, the
EU Marie Curie research training network grant MRTN-CT-2004-512194
and the STFC rolling grant ST/G000/395/1.

\vskip 2cm
\begin{appendix}
\section{Group theory conventions and projectors}
In this appendix we first review some of the group-theoretic
techniques that have been used in this paper, and we then discuss
the $E_{11}$ derivation of various representation projectors,
fucusing in particular on the cases of $E_7$ and $E_6$ which have
been discussed in sections 4 and 5. These projectors arise in
$E_{11}$ as conditions on the structure constants that contract the
$D-1$ form generators, and the fact that the consistency of the
algebra imposes that the embedding tensor must satisfy the same
projection conditions proves that the embedding tensor and the
$D-1$-form generators must belong to the same representation. At the
end of this appendix we then show that these projectors are
precisely the ones that result from a purely group theoretic
analysis based on the representations of the internal symmetry
group.

The Cartan-Killing metric $\kappa^{\alpha \beta}$ is defined as
  \begin{equation}
  C_{\rm Adj}\kappa^{\alpha \beta} = f^{\alpha \gamma}{}_\epsilon f^{\beta
  \epsilon}{}_\gamma \quad , \label{cartankillingappendix}
  \end{equation}
where $C_{\rm Adj}$ is the quadratic Casimir in the adjoint
representation. Denoting with ${\bf D_\Lambda}$ the fundamental
representation, one then defines the quadratic Casimir in the
fundamental representation $C_\Lambda$ from the relation
  \begin{equation}
  C_\Lambda \delta^N_M = \kappa_{\alpha \beta} D^\alpha_M{}^P
  D^\beta_P{}^N  \quad. \label{casimirappendix}
  \end{equation}
When not otherwise specified in the paper, we use to raise and lower
indices in the adjoint representation the metric
  \begin{equation}
  g^{\alpha \beta} = {\rm Tr } (D^\alpha D^\beta ) =
  D^\alpha_M{}^N D^\beta_N{}^M  \label{metricappendix}
  \quad ,
  \end{equation}
where the trace is in the fundamental, {\it i.e.} lowest
dimensional, representation. This metric differs from the
Cartan-Killing metric of eq. (\ref{cartankillingappendix}) by a
constant, and indeed from eq. (\ref{casimirappendix}) one finds
  \begin{equation}
  \kappa^{\alpha \beta} = {d \over C_\Lambda d_\Lambda} g^{\alpha
  \beta} \quad , \label{maurercartanversusothermetricappendix}
  \end{equation}
where $d$ is the dimension of the adjoint and $d_\Lambda$ is the
dimension of the fundamental representation. Substituting the
inverse of eq. (\ref{maurercartanversusothermetricappendix}) in eq.
(\ref{casimirappendix}) one also derives
  \begin{equation}
  g_{\alpha \beta}  D^\alpha_M{}^P
  D^\beta_P{}^N  = {d \over d_\Lambda } \delta^N_M \quad ,
  \label{gDDisdoverdlambdaappendix}
  \end{equation}
while substituting it in eq. (\ref{cartankillingappendix}) one gets
  \begin{equation}
  f^{\alpha\beta \gamma} f_{\alpha \beta \delta} = - { d \over d_\Lambda}
  {C_{\rm Adj}
  \over C_\Lambda } \delta^\gamma_\delta \quad ,
  \label{ff=deltacasimirappendix}
  \end{equation}
as follows from raising the indices using the metric in eq.
(\ref{metricappendix}). The ratio  ${C_{\rm Adj} \over C_\Lambda}$
is given by the relation
  \begin{equation}
  {C_{\rm Adj}
  \over C_\Lambda } = {d_\Lambda \over d} {g^\vee \over
  \tilde{I}_\Lambda} \quad ,
  \end{equation}
where $g^\vee$ is the dual Coxeter number and $\tilde{I}_\Lambda$ is
the Dynkin index of the fundamental representation.

In this paper we have shown that the deformed $E_{11}$ algebras
resulting from suitably modifying the commutation relations of the
$E_{11}$ generators with momentum are entirely classified by the
tensor $\Theta^{M_1}_\alpha$ arising in eq.
(\ref{definitionofthetainalldimensions}), where $M_1$ denotes the
representation of the 1-form $E_{11}$ generator $R^{a_1 , M_1}$ in a
given dimension. As we have shown, this tensor is identified with
the embedding tensor. Denoting with ${\bf R_1}$ this representation,
and with ${\bf R_0}$ the adjoint, the embedding tensor is contained
in the tensor product ${\bf R_1 \otimes R_0}$. Since the $D-2$-form
generators also belong to the adjoint, that is ${\bf R_{D-2} =
R_0}$, this tensor product is the same as the tensor product ${\bf
R_1 \otimes R_{D-2}}$, which occurs in the commutator between the
1-form and the $D-2$-form. This commutator gives rise to the
$D-1$-forms, that belong to a representation ${\bf R_{D-1}}$ inside
${\bf R_1 \otimes R_{D-2}}$. However, ${\bf R_1 \otimes R_{D-2}}$
contains at least three irreducible representations, and the
$E_{11}$ derivation of the projection conditions on ${\bf R_{D-1}}$
plays an important role in this paper. In particular, it is
responsible for the demonstration that the $D-1$-form generators and
the embedding tensor belong to the same representation. This was
proven in detail for any dimension in this paper.

In table \ref{tablewithembeddingtensorinvariousdim} we list all the
irreducible representations arising in the tensor product ${\bf R_1
\otimes R_{0}}$ in any dimension, underlying the ones to which the
embedding tensor and the $D-1$ forms actually belong. As can be
noticed from the table, in four, five and six dimensions ${\bf R_1
\otimes R_{0}}$ generates three representations. These cases are
those in which the 1-form generator belongs to the fundamental
representation ${\bf D_\Lambda}$. Therefore, in four, five and six
dimensions the embedding tensor $\Theta^{M_1}_\alpha$ is contained
in the tensor product ${\bf D_\Lambda \otimes R_0}$. In the
following we will in general denote the adjoint representation  by
${\bf Adj}$. It is a property of any simple group with the exception
of $E_8$ that the tensor product ${\bf D_\Lambda \otimes Adj}$
always gives
  \begin{equation}
  {\bf D_\Lambda \otimes Adj} = {\bf D_\Lambda \oplus D_1 \oplus D_2
  } \quad ,\label{dlambdad1d2appendix}
  \end{equation}
where ${\bf D_1}$ and ${\bf D_2}$ are two other representations, and
we take the dimension of ${\bf D_1}$ to be lower than the dimension
of ${\bf D_2}$. As can be seen from the table, in four, five and six
dimensions the embedding tensor belongs to ${\bf D_1}$. Using the
fact that this is also the representation of the $D-1$-form, we now
show that one can derive from $E_{11}$ the projectors on these three
representations. We will focus in particular on the cases of $E_7$
and $E_6$, corresponding to four and five dimensions respectively.
\begin{table}
\begin{center}
\begin{tabular}{|c|c||c|}
\hline \rule[-1mm]{0mm}{6mm} D & G &  ${\bf R_1
\otimes R_{0}}$ \\
\hline \hline \rule[-1mm]{0mm}{6mm} {9} & $SL(2,\mathbb{R})$  &
${\bf 1
\oplus \underline{2} \oplus 2 \oplus \underline{3} \oplus 4 }$\\
\hline \rule[-1mm]{0mm}{6mm} {8} & {$SL(3,\mathbb{R}) \times
SL(2,\mathbb{R})$} &  ${\bf (\underline{3 ,2 }) \oplus (3, 2)
  \oplus (3,4) \oplus (\underline{\overline{6},2}) \oplus (15,2)}$ \\
\hline\rule[-1mm]{0mm}{6mm} {7} & {$SL(5,\mathbb{R})$} & ${\bf 10 \oplus \underline{15} \oplus \underline{40} \oplus 175}$ \\
\hline\rule[-1mm]{0mm}{6mm} {6} & {$SO(5,5)$} & ${\bf \overline{16} \oplus \underline{\overline{144}} \oplus 560}$  \\
\hline\rule[-1mm]{0mm}{6mm} {5} & {$E_{6(+6)}$} & ${\bf
\overline{27} \oplus \underline{\overline{351}} \oplus
\overline{1728}}$
\\
\hline\rule[-1mm]{0mm}{6mm} {4} & {$E_{7(+7)}$} & ${\bf 56 \oplus \underline{912} \oplus 6480}$  \\
\hline\rule[-1mm]{0mm}{6mm} {3} &  {$E_{8(+8)}$} &  ${\bf
\underline{1} \oplus
248 \oplus \underline{3875} \oplus 27000 \oplus 30380}$ \\
\hline
\end{tabular}
\end{center}
\caption{\sl Table giving the irreducible representations that arise
in the product ${\bf R_1 \otimes R_{0}}$ in various dimensions. The
representations to which the embedding tensor and the $D-1$-form
generators belong are underlined.
\label{tablewithembeddingtensorinvariousdim}}
\end{table}

Given the tensor product ${\bf D_\Lambda \otimes Adj}$, the
projectors $\mathbb{P}_{\bf D_\Lambda}$, $\mathbb{P}_{\bf D_1}$ and
$\mathbb{P}_{\bf D_2}$ on the representations of eq.
(\ref{dlambdad1d2appendix}) can be constructed in terms of
$\delta^M_N$, $\delta^\alpha_\beta$ and $D^\alpha_M{}^N$ as
\cite{dWSTgeneral}
  \begin{eqnarray}
  & & \mathbb{P}_{\bf D_\Lambda}{}_{\alpha N}^{M \beta} = {d_\Lambda
  \over d} D^\beta_N{}^P D_{\alpha P}{}^M \nonumber \\
  & & \mathbb{P}_{\bf D_1}{}_{\alpha N}^{M \beta} = a D^\beta_N{}^P D_{\alpha
  P}{}^M + b D_{\alpha N}{}^P D^\beta_{P}{}^M + c \delta^M_N
  \delta^\beta_\alpha \nonumber \\
  & & \mathbb{P}_{\bf D_2}{}_{\alpha N}^{M \beta} = - (a + {d_\Lambda
  \over d})  D^\beta_N{}^P D_{\alpha P}{}^M  -b D_{\alpha N}{}^P
  D^\beta_{P}{}^M
  + (1 -c ) \delta^M_N
  \delta^\beta_\alpha \quad , \label{definitionofprojectorsappendix}
  \end{eqnarray}
where one makes use of eq. (\ref{gDDisdoverdlambdaappendix}) and the
fact that the sum of the projectors is the identity. Note that the
three coefficients $a$, $b$ and $c$ are not specified and will be
given later. We require the projectors to satisfy
  \begin{eqnarray}
  & &  \mathbb{P}_{\bf D_\Lambda}{}_{\alpha N}^{M \beta}
  \mathbb{P}_{\bf D_\Lambda}{}_{\beta P}^{N
  \gamma} = \mathbb{P}_{\bf D_\Lambda}{}_{\alpha P}^{M \gamma} \qquad
  \mathbb{P}_{\bf D_\Lambda}{}_{\alpha N}^{M \beta} \mathbb{P}_{\bf D_i}{}_{\beta
  P}^{N
  \gamma} = 0 \nonumber \\
  & & \mathbb{P}_{\bf D_i}{}_{\alpha N}^{M \beta}
   \mathbb{P}_{\bf D_j}{}_{\beta P}^{N \gamma}
  = \delta_{ij} \mathbb{P}_{\bf D_i}{}_{\alpha P}^{M \gamma} \qquad
  i,j= 1,2 \quad . \label{projectorconditionsappendix}
  \end{eqnarray}
We now show that for $E_7$ and $E_6$ these projectors are determined
using $E_{11}$. In principle the $E_{11}$ derivation of the
projectors can be also carried out for $D_5$, which corresponds to
the six-dimensional case, but it is not needed because in section 6
we have used the explicit form of the structure constants, which
encodes automatically the projectors.

We first consider {\bf the case of $E_7$}. In section 4 we have
shown that the invariant tensor $S^{M\alpha}_A$ resulting from the
commutator of the 1-form and the 2-form satisfies the constraints of
eqs. (\ref{gravitinoconditiononS}) and
(\ref{S+DDS=0infourdimensions}). These constraints follow from the
Jacobi identities of the $E_{11}$ algebra. On the other hand, the
$E_{11}$ algebra imposes that the 3-form generators in four
dimensions belong to the ${\bf 912}$, which implies that the indices
$M\alpha$ must be projected on the ${\bf 912}$. This can be seen
from the index structure of $S^{M\alpha}_A$  because the only way of
building an invariant from tensoring a ${\bf 912}$ index with the
product ${\bf 56 \otimes 133}$ is that this product is indeed
projected on the ${\bf 912}$. By looking at the general form of the
projectors in eq. (\ref{definitionofprojectorsappendix}), we thus
must require that $S^{M\alpha}_A$ satisfies the conditions
  \begin{eqnarray}
  & & \mathbb{P}_{\bf 56}{}_{\alpha N}^{M \beta} S^{N
  \gamma}_A g_{\beta \gamma} = 0 \nonumber \\
  & & \mathbb{P}_{\bf 912}{}_{\alpha N}^{M \beta} S^{N
  \gamma}_A g_{\beta \gamma} = S^{M \gamma}_A g_{\alpha \gamma}
  \nonumber \\
  & & \mathbb{P}_{\bf 6480}{}_{\alpha N}^{M \beta}S^{N
  \gamma}_A g_{\beta \gamma} = 0 \quad ,
  \end{eqnarray}
and comparing these three conditions with eqs.
(\ref{gravitinoconditiononS}) and (\ref{S+DDS=0infourdimensions}) we
determine a constraint on the parameters $a$, $b$ and $c$ in eq.
(\ref{definitionofprojectorsappendix}). Indeed, the first condition
is automatically satisfied because it reproduces eq.
(\ref{gravitinoconditiononS}), while the second and the third
reproduce eq. (\ref{S+DDS=0infourdimensions}) provided that
  \begin{equation}
  {b \over 1-c} = -2 \quad . \label{bover1-c=-2appendix}
  \end{equation}

We now derive the constraints resulting from eq.
(\ref{projectorconditionsappendix}) in case of $E_7$. The first
condition is automatically satisfied, while the second condition
gives
  \begin{equation}
  {19 \over 8} a + {7 \over 8} b +c =0 \quad ,
  \label{abcrelationappendix}
  \end{equation}
where we have made use of eq. (\ref{ff=deltacasimirappendix}) in
which we have substituted the dual Coxeter number and the Dynkin
index for $E_7$, which are listed in table \ref{grouptheorytable}.
In order to derive the other constraints, one makes use of the
identities
  \begin{equation}
  D_{\alpha N}{}^M D_{\beta MP} D^{\beta NQ} = { 7 \over 8}
  D_{\alpha P}{}^Q \quad ,
  \end{equation}
  \begin{equation}
  D_{\beta M}{}^N D^\beta_P{}^Q = {1 \over 12} \delta_M^Q \delta^N_P
  + {1 \over 24} \delta_M^N \delta_P^Q - {1 \over 24} \Omega^{NQ}
  \Omega_{MP} + D_{\beta}^{NQ} D^\beta_{MP} \quad ,
  \end{equation}
and
  \begin{equation}
  (D^\gamma D_\alpha )_Q{}^R D^\beta_P{}^Q D_{\beta R}{}^M = {1
  \over 24} \delta^M_P \delta^\gamma_\alpha - {3 \over 8} (D^\gamma
  D_\alpha )_P{}^M - {5 \over 12 } (D_\alpha D^\gamma )_P{}^M \quad
  ,
  \end{equation}
which can all be proven using the results listed in this appendix
and in section 4. Using these results, the last condition in eq.
(\ref{projectorconditionsappendix}), with $i=j=1$, gives
  \begin{eqnarray}
  & & {19 \over 8} a^2 + {7 \over 4} a b + 2 ac - {3 \over 8} b^2 =
  a  \nonumber \\
  & & 2 bc - {5 \over 12} b^2 = b \nonumber \\
  & & c^2 + {1 \over 24} b^2 = c \quad .
  \end{eqnarray}
Substituting eq. (\ref{bover1-c=-2appendix}) into the last of these
equations gives
  \begin{equation}
  c = {1 \over 7 } \qquad b = - {12 \over 7} \quad ,
  \end{equation}
and from eq. (\ref{abcrelationappendix}) one gets $a= {4 \over 7}$.
One can show that all the other projector conditions are satisfied.
Substituting this in eq. (\ref{definitionofprojectorsappendix})
finally gives the $E_7$ projectors
  \begin{eqnarray}
  & & \mathbb{P}_{\bf 56}{}_{\alpha N}^{M \beta} = {8
  \over 19} D^\beta_N{}^P D_{\alpha P}{}^M \nonumber \\
  & & \mathbb{P}_{\bf 912}{}_{\alpha N}^{M \beta} = {4 \over 7} D^\beta_N{}^P D_{\alpha
  P}{}^M - {12 \over 7}  D_{\alpha N}{}^P D^\beta_{P}{}^M + {1 \over 7 } \delta^M_N
  \delta^\beta_\alpha \nonumber \\
  & & \mathbb{P}_{\bf 6480}{}_{\alpha N}^{M \beta} = - {132
  \over 133}  D^\beta_N{}^P D_{\alpha P}{}^M  +{12 \over 7}D_{\alpha N}{}^P
  D^\beta_{P}{}^M
  + {6 \over 7} \delta^M_N
  \delta^\beta_\alpha \quad . \label{E7projectorsappendix}
  \end{eqnarray}

We now derive from $E_{11}$ the projectors of eq.
(\ref{definitionofprojectorsappendix}) for {\bf the case of $E_6$}.
We first derive the identities that will be needed. Is section 5 we
have introduced the completely symmetric invariant tensors $d^{MNP}$
and $d_{MNP}$, satisfying
  \begin{equation}
  d^{MNP} d_{MNQ} = \delta^P_Q \quad .
  \label{ddisdeltaE6relationinD=5appendix}
  \end{equation}
From eq. (\ref{ddisdeltaE6relationinD=5appendix}) and the condition
that $d^{MNP}$ and $d_{MNP}$ are invariant tensors,
  \begin{equation}
  D^\alpha_M{}^{(N} d^{PQ)M} = D^\alpha_{(M}{}^N d_{PQ)N} = 0 \quad
  , \label{invarianceofdupandddownappendix}
  \end{equation}
one gets
  \begin{equation}
  D^\alpha_M{}^N d^{MPQ} d_{NSQ} = - {1 \over 2} D^\alpha_S{}^Q
  \quad .
  \end{equation}
One can write the product of two generators in the ${\bf 27}$
contracted by the metric $g_{\alpha \beta}$ as
  \begin{equation}
  g_{\alpha \beta} D^\alpha_M{}^N D^\beta_P{}^Q = \alpha \delta^N_P
  \delta_M^Q + \beta \delta_M^N \delta_P^Q + \gamma d^{NQR} d_{MPR}
  \quad , \label{alphabetagammaDDequationappendix}
  \end{equation}
as can be deduced from the fact that the product ${\bf 27 \otimes 27
\otimes \overline{27} \otimes \overline{27}}$ leads to three
different $E_{6}$ invariants, and the three invariant quantities on
the right hand side of eq. (\ref{alphabetagammaDDequationappendix})
are the most general objects one can write down in terms of
$\delta^N_M$, $d^{MNP}$ and $d_{MNP}$. Eq.
(\ref{gDDisdoverdlambdaappendix}), applied to the $E_6$ case in
which $d= 78$ and $d_\Lambda = 27$, is
  \begin{equation}
  g_{\alpha \beta} D^\alpha_M{}^N D^\beta_N{}^P = {26 \over 9}
  \delta_M^P \quad .
  \end{equation}
Contracting $N$ and $P$ in eq.
(\ref{alphabetagammaDDequationappendix}) thus leads to
  \begin{equation}
  27 \alpha + \beta + \gamma = {26 \over 9} \quad ,
  \end{equation}
while contracting $M$ and $N$ gives
  \begin{equation}
  \alpha + 27 \beta + \gamma =0 \quad .
  \end{equation}
A third relation comes from the identity
  \begin{equation}
  g_{\alpha \beta} D^\alpha_M{}^N D^\beta_P{}^Q d_{NQR} = -{13 \over
  9} d_{MPR} \quad ,
  \end{equation}
which can be derived using eq.
(\ref{invarianceofdupandddownappendix}) iteratively, and leads to
  \begin{equation}
  \alpha + \beta + \gamma = - {13 \over 9} \quad ,
  \end{equation}
or alternatively from contracting eq.
(\ref{alphabetagammaDDequationappendix}) with $D^\gamma_N{}^M$,
which leads to
  \begin{equation}
  \alpha - {1 \over 2} \gamma = 1 \quad .
  \end{equation}
The final result is
  \begin{equation}
  \alpha = {1 \over 6} \qquad \beta = {1 \over 18} \qquad \gamma = -
  {5 \over 3} \quad ,
  \end{equation}
so that
  \begin{equation}
  g_{\alpha \beta} D^\alpha_M{}^N D^\beta_P{}^Q = {1 \over 6} \delta^N_P
  \delta_M^Q + {1 \over 18} \delta_M^N \delta_P^Q -{5 \over 3} d^{NQR} d_{MPR}
  \quad . \label{finalDDequationinD=5appendix}
  \end{equation}
Another useful relation is
  \begin{equation}
  (D^\gamma D_\alpha )_Q{}^R d^{QMT} d_{PRT} = - {1 \over 5} (D^\gamma
  D_\alpha )_P{}^M + {3 \over 10} (D_\alpha D^\gamma )_P{}^M + {1
  \over 30} \delta^M_P \delta^\gamma_\alpha \quad ,
  \end{equation}
which can be derived using the relations given in this appendix.

In section 5 we have shown that the invariant tensor $S^{\alpha M,
NP}$ resulting from the commutator of the 1-form and the 3-form
satisfies the constraints of eqs.
(\ref{gravitinoconditiononSinfivedimensions}) and
(\ref{S=-3over2DDS}). These constraints follow from requiring the
closure of the Jacobi identities of the $E_{11}$ algebra. On the
other hand, the form of this invariant tensor is dictated by the
fact that $E_{11}$ imposes that the 4-form generator is in the ${\bf
\overline{351}}$, and therefore the indices $\alpha M$ of ${\bf
\overline{351}}$ must be projected on the ${\bf \overline{351}}$.
This can be seen from the index structure of $S^{\alpha M, NP}$
because the $NP$ antisymmetric indices correspond to the ${\bf 351}$
and the only way of building an invariant from tensoring a ${\bf
351}$ representation with the product ${\bf \overline{27} \otimes
78}$ is that this product is indeed projected on the ${\bf
\overline{351}}$. By looking at the general expressions of eq.
(\ref{definitionofprojectorsappendix}) for the projectors, we
therefore must impose the conditions
  \begin{eqnarray}
  & & \mathbb{P}_{\bf \overline{27}}{}_{\alpha N}^{M \beta} S^{N
  \gamma}_A g_{\beta \gamma} = 0 \nonumber \\
  & & \mathbb{P}_{\bf \overline{351}}{}_{\alpha N}^{M \beta} S^{N
  \gamma}_A g_{\beta \gamma} = S^{M \gamma}_A g_{\alpha \gamma}
  \nonumber \\
  & & \mathbb{P}_{\bf \overline{1728}}{}_{\alpha N}^{M \beta}S^{N
  \gamma}_A g_{\beta \gamma} = 0 \quad ,
  \end{eqnarray}
and comparing these three conditions with eqs.
(\ref{gravitinoconditiononSinfivedimensions}) and
(\ref{S=-3over2DDS}) we determine a constraint on the parameters
$a$, $b$ and $c$ in eq. (\ref{definitionofprojectorsappendix}). In
particular, the first condition is automatically satisfied because
it reproduces eq. (\ref{gravitinoconditiononSinfivedimensions}),
while eq. (\ref{S=-3over2DDS}) implies that the second and the third
equations give the same constraint, that is
  \begin{equation}
  {b \over 1-c} = - {3 \over 2} \quad . \label{bover1-c=-3over2appendixforE6}
  \end{equation}

We now derive the constraints resulting from eq.
(\ref{projectorconditionsappendix}) in case of $E_6$. The first
condition is automatically satisfied, while the second condition
gives
  \begin{equation}
  {26 \over 9} a + {8 \over 9} b +c =0 \quad ,
  \label{abcrelationappendixE6}
  \end{equation}
where we have made use of eq. (\ref{ff=deltacasimirappendix}) in
which we have substituted the dual Coxeter number and the Dynkin
index for $E_6$, which are listed in table \ref{grouptheorytable}.
The last condition in eq. (\ref{projectorconditionsappendix}), with
$i=j=1$, gives
  \begin{eqnarray}
  & & {26 \over 9} a^2 + {16 \over 9} a b + 2 ac + {7 \over 18} b^2 =
  a  \nonumber \\
  & & 2 bc - {1 \over 2} b^2 = b \nonumber \\
  & & c^2 + {1 \over 9} b^2 = c \quad .
  \end{eqnarray}
Substituting eq. (\ref{bover1-c=-3over2appendixforE6}) into the last
of these equations gives
  \begin{equation}
  c = {1 \over 5 } \qquad b = - {6 \over 5} \quad ,
  \end{equation}
and from eq. (\ref{abcrelationappendixE6}) one gets $a= {3 \over
10}$. One can show that all the other projector conditions are
satisfied. Substituting this in eq.
(\ref{definitionofprojectorsappendix}) finally gives the $E_6$
projectors
  \begin{eqnarray}
  & & \mathbb{P}_{\bf \overline{27}}{}_{\alpha N}^{M \beta} = {9
  \over 26} D^\beta_N{}^P D_{\alpha P}{}^M \nonumber \\
  & & \mathbb{P}_{\bf \overline{351}}{}_{\alpha N}^{M \beta} = {3 \over 10} D^\beta_N{}^P D_{\alpha
  P}{}^M - {6 \over 5}  D_{\alpha N}{}^P D^\beta_{P}{}^M + {1 \over 5 } \delta^M_N
  \delta^\beta_\alpha \nonumber \\
  & & \mathbb{P}_{\bf \overline{1728}}{}_{\alpha N}^{M \beta} = -
  {45
  \over 65}  D^\beta_N{}^P D_{\alpha P}{}^M  +{6 \over 5}D_{\alpha N}{}^P
  D^\beta_{P}{}^M
  + {4 \over 5} \delta^M_N
  \delta^\beta_\alpha \quad . \label{E6projectorsappendix}
  \end{eqnarray}

The projectors of eqs. (\ref{E7projectorsappendix}) and
(\ref{E6projectorsappendix}) that we have obtained from $E_{11}$
exactly coincide with the projectors that one obtains from group
theory. In table \ref{grouptheorytable} we list the dual Coxeter
number, the Dynkin index of the fundamental representation, the
dimension of the group $d$, the dimension of the fundamental
representation $d_\Lambda$ and the dimension $d_1$ of the
representation ${\bf D_1}$, as well as the values of the
coefficients $a$, $b$ and $c$ in eq.
(\ref{definitionofprojectorsappendix}), for some simple Lie groups.
One can see in particular that the values of $a$, $b$ and $c$ in the
table for $E_6$ and $E_7$ are exactly those that we have derived
from $E_{11}$.
\begin{table}
\begin{center}
\begin{tabular}{|c||c|c|c|c|c|c|c|c|c|}
\hline \rule[-1mm]{0mm}{6mm} G & $g^\vee$ & $\tilde{I}_\Lambda$ &
$d$ & $d_\Lambda$ & $d_1$ &
$a$ & $b$ & $c$ \\
\hline \hline \rule[-1mm]{0mm}{6mm} $A_r$ & $r+1$  & ${1 \over 2}$ &
$r^2 +2r$
& $r+1$ & ${1 \over 2}(r-1)(r+1)(r+2)$ & $-{1 \over 2r}$ & $- {1\over 2}$ & ${1 \over 2}$\\
\hline\rule[-1mm]{0mm}{6mm} $G_2$ & 4 & 1 & 14 & 7 & 27  & $-{3 \over 14}$  & $-{6 \over 7}$ & ${3 \over 7}$\\
\hline\rule[-1mm]{0mm}{6mm} $F_4$ & 9 & 3 & 52 & 26 &  273 & ${1 \over 4}$  & $-{3 \over 2}$ & ${1 \over 4}$ \\
\hline\rule[-1mm]{0mm}{6mm} $E_6$ & 12 & 3 & 78 & 27 & 351 & ${3 \over 10}$ & $-{6 \over 5}$ & ${1 \over 5}$ \\
\hline \rule[-1mm]{0mm}{6mm} $E_7$ & 18 & 6 & 133 & 56 & 912 & ${4 \over 7}$ & $-{12 \over 7}$ & ${1 \over 7}$\\
\hline
\end{tabular}
\end{center}
\caption{\sl Table giving the dual Coxeter number, the Dynkin index,
the dimension of the adjoint, the fundamental and the ${\bf D_1}$
representations, as well as the parameters $a$, $b$ and $c$
occurring in eq. (\ref{definitionofprojectorsappendix}), for some
simple Lie groups (see also \cite{dWSTgeneral}).
\label{grouptheorytable}}
\end{table}

\section{Field strengths and gauge transformations}
In this appendix we explicitly evaluate the deformed part of the
field strengths up to rank six from eqs. (\ref{2.41}) and
(\ref{2.44}). For the scalar derivative we find
 \begin{equation}
  F_a = g_\varphi^{-1}(\partial_a+ g A_{a, N_1} \Theta^{N_1}_\alpha
  R^\alpha)g_\varphi \quad .
  \end{equation}
We now write down the field strengths for the gauge fields, which by
assumption  have all their Lorentz indices anti-symmetrised and as
discussed in section two we do not explicitly display their scalar
factors converting from a linear representation to a non-linear
representation  of $G$. We first display the massless part of the
field strengths up to rank 6 included, determined using eq.
(\ref{2.41}). The result is
  \begin{eqnarray}
  & & F^{(0)}_{c a_1 , N_1} = 2 \partial_c A_{a_1 , N_1} \\
  & &
  F^{(0)}_{c a_1 a_2, N_2} = 3[ \partial_c A_{a_1 a_2 , N_2} - {1
  \over 2} L_{a_1 , N_2}{}^{N_1} \partial_c A_{a_2 , N_1} ]
  \\
  & &
  F^{(0)}_{c a_1 a_2 a_3, N_3} = 4 [ \partial_c A_{a_1 a_2 a_3 ,
  N_3} - L_{a_1 , N_3}{}^N_2 \partial_c A_{a_2 a_3 , N_2} + {1 \over
  3!}( L_{a_1} L_{a_2} )_{N_3}{}^{N_1} \partial_c A_{a_3 , N_1} ] \\
  & &   F^{(0)}_{c a_1 ...a_4, N_4} = 5 [ \partial_c A_{a_1 ...a_4 ,
  N_4} + L_{a_1 , N_4}{}^{N_3} \partial_c A_{a_2 a_3 a_4 , N_3} + {1
  \over 2} ( L_{a_1} L_{a_2} )_{N_4}{}^{N_2} \partial_c A_{a_3 a_4 ,
  N_2} \nonumber \\
  & & \quad \qquad - {1 \over 2} L_{a_1 a_2 , N_4}{}^{N_2} \partial_c A_{a_3 a_4 ,
  N_2}- {1 \over 4!} (L_{a_1} L_{a_2} L_{a_3} )_{N_4}{}^{N_1}
  \partial_c A_{a_4 , N_1} ] \\
  & & F^{(0)}_{c a_1 ...a_5, N_5} = 6 [ \partial_c A_{a_1 ...a_5 ,
  N_5} - L_{a_1 , N_5}{}^{N_4} \partial_c A_{a_2 ...a_5 , N_4} -
  L_{a_1 a_2 , N_5}{}^{N_3} \partial_c A_{a_3 a_4 a_5 , N_3}
  \nonumber \\
  & & \quad \qquad +{1 \over 2} ( L_{a_1 } L_{a_2} )_{N_5}{}^{N_3} \partial_c A_{a_3 a_4 a_5 ,
  N_3} - {1 \over 3!} ( L_{a_1}  L_{a_2}  L_{a_3} )_{N_5}{}^{N_2}
  \partial_c A_{a_4 a_5 , N_2} \nonumber \\
  & & \quad \qquad + {1 \over 2} (L_{a_1} L_{a_2 a_3})_{N_5}{}^{N_2}
   \partial_c A_{a_4 a_5 , N_2} + { 1 \over 5!} ( L_{a_1}  L_{a_2}  L_{a_3}
   L_{a_4})_{N_5}{}^{N_1} \partial_c A_{a_5 , N_1}] \quad .
  \end{eqnarray}
The order $g$ part of the same field strengths follows from eq.
(\ref{2.44}). The result is
  \begin{eqnarray}
  & & F^{(1)}_{c a_1 , N_1 } =2 g [ W^{N_2}{}_{N_1} A_{c a_1 , N_2}
  + {1 \over 2} X^{M_1 P_1}{}_{N_1} A_{a_1 , P_1} A_{c , M_1} ]\\
  & & F^{(1)}_{c a_1 a_2, N_2 } =3 g [ W^{N_3}{}_{N_2} A_{ca_1 a_2,
  N_3} - L_{a_1 , N_2}{}^{N_1} W^{M_2}{}_{N_1} A_{c a_2, M_2}
  \nonumber \\
  & & \quad \qquad - {1 \over 3!} L_{a_1 , N_2}{}^{P_1} X^{N_1
  M_1}{}_{P_1} A_{a_2 , M_1} A_{c , N_1} ]\\
  & &  F^{(1)}_{c a_1 a_2 a_3, N_3 } =4 g [ W^{N_4}{}_{N_3} A_{ca_1 a_2 a_3,
  N_4} - L_{a_1 , N_3}{}^{N_2} W^{M_3}{}_{N_2} A_{c a_2 a_3 , M_3} \nonumber \\
  & & \quad \qquad -
  {1 \over 2} L_{a_1 a_2 , N_3}{}^{N_1} W^{N_2}{}_{N_1} A_{c a_3 ,
  N_2}  + { 1 \over 2}  (L_{a_1} L_{a_2} )_{N_3}{}^{N_1} W^{N_2}{}_{N_1} A_{c a_3 ,
  N_2} \nonumber \\
  & & \quad \qquad + {1 \over 4! } ( L_{a_1} L_{a_2} )_{N_3}{}^{P_1} X^{N_1
  M_1}{}_{P_1} A_{a_3 , M_1} A_{c , N_1}] \\
  & & F^{(1)}_{c a_1 ... a_4, N_4 } =5 g [ W^{N_5}{}_{N_4} A_{ca_1 ... a_4,
  N_5} - L_{a_1 , N_4}{}^{N_3} W^{M_4}{}_{N_3} A_{c a_2 a_3 a_4, M_4} \nonumber \\
  & & \quad \qquad - L_{a_1 a_2 , N_4}{}^{N_2} W^{N_3}{}_{N_2} A_{c
  a_3 a_4 , N_3} + { 1 \over 2} ( L_{a_1} L_{a_2})_{N_4}{}^{N_2} W^{N_3}{}_{N_2} A_{c
  a_3 a_4 , N_3} \nonumber \\
  & & \quad \qquad + {1 \over 2} ( L_{a_1} L_{a_2 a_3}
  )_{N_4}{}^{N_1} W^{N_2}{}_{N_1} A_{c a_4 , N_2} - {1 \over 3!} ( L_{a_1} L_{a_2}L_{ a_3}
  )_{N_4}{}^{N_1} W^{N_2}{}_{N_1} A_{c a_4 , N_2} \nonumber \\
  & & \qquad\quad - {1 \over 5!} ( L_{a_1} L_{a_2}L_{ a_3}
  )_{N_4}{}^{P_1} X^{N_1 M_1}{}_{P_1}  A_{a_4 , M_1} A_{c , N_1}]
  \\
  & & F^{(1)}_{c a_1 ... a_5, N_5 } =6 g [ W^{N_6}{}_{N_5} A_{ca_1 ... a_5,
  N_6} - L_{a_1 , N_5}{}^{N_4} W^{M_5}{}_{N_4} A_{c a_2 ... a_5, M_5} \nonumber \\
  & & \quad \qquad - L_{a_1 a_2 , N_5}{}^{N_3} W^{N_4}{}_{N_3} A_{c
  a_3 ...a_5, N_4} +{1 \over 2} (L_{a_1 } L_{a_2} )_{ N_5}{}^{N_3} W^{N_4}{}_{N_3} A_{c
  a_3 ...a_5, N_4} \nonumber \\
  & & \quad \qquad - {1 \over 2} L_{a_1 a_2 a_3 , N_5}{}^{N_2}
  W^{N_3}{}_{N_2} A_{c a_4 a_5 , N_3} + (L_{a_1} L_{a_2 a_3} )_{ N_5}{}^{N_2}
  W^{N_3}{}_{N_2} A_{c a_4 a_5 , N_3} \nonumber \\
  & & \quad \qquad - {1 \over 3!} ( L_{a_1 } L_{a_2 } L_{a_3}
  )_{N_5}{}^{N_2}  W^{N_3}{}_{N_2} A_{c a_4 a_5 , N_3} + {1 \over
  3!} (L_{a_1 a_2} L_{a_3 a_4} )_{N_5}{}^{N_1} W^{N_2}{}_{N_1} A_{c
  a_5 , N_2}\nonumber \\
  & & \quad \qquad  - {1 \over 4} ( L_{a_1}L_{ a_2} L_{a_3 a_4} )_{N_5}{}^{N_1} W^{N_2}{}_{N_1} A_{c
  a_5 , N_2} + {1 \over 4!} ( L_{a_1}L_{a_2} L_{a_3}L_{a_4} )_{N_5}{}^{N_1} W^{N_2}{}_{N_1} A_{c
  a_5 , N_2} \nonumber \\
  & & \quad \qquad + {1 \over 6!} ( L_{a_1} L_{a_2}L_{ a_3} L_{a_4}
  )_{N_5}{}^{P_1} X^{N_1 M_1}{}_{P_1}  A_{a_5 , M_1} A_{c , N_1}]
  \quad .
  \end{eqnarray}
The reader can easily evaluate the remaining field strengths.

The rigid transformations of the group element also determine the
gauge transformations of the various fields. We list here the gauge
transformations for all the forms up to the 6-form. The 1-form
transforms as
  \begin{equation}
  \delta A_{a_1 , N_1 } = a_{a_1 , N_1} - g \Lambda_{M_1}
  X_1^{M_1}{}_{N_1}{}^{P_1} A_{a_1 , P_1} \quad ,
  \end{equation}
the 2-form as
  \begin{equation}
  \delta A_{a_1 a_2 , N_2} = a_{a_1 a_2 , N_2} -{1 \over 2} A_{a_1
  ,N_1 } a_{a_2 , M_1} f^{N_1 M_1}{}_{N_2} - g \Lambda_{M_1}
  X_2^{M_1}{}_{N_2}{}^{P_2} A_{a_1 a_2, P_2} \quad ,
  \end{equation}
the 3-form as
  \begin{eqnarray}
  & & \delta A_{a_1 a_2 a_3, N_3} = a_{a_1 a_2 a_3, N_3} - A_{a_1 a_2, N_2} a_{a_3 ,N_1}
  f^{N_2 N_1}{}_{N_3} \nonumber
  \\
  & & \quad \qquad- {1 \over 3!} A_{a_1 ,N_1 } A_{a_2 , M_1}
  a_{a_3 , P_1}  f^{N_1 N_2}{}_{N_3}   f^{M_1 P_1}{}_{N_2}
  - g \Lambda_{M_1}
  X_3^{M_1}{}_{N_3}{}^{P_3} A_{a_1 a_2 a_3, P_3} \quad ,
  \end{eqnarray}
the 4-form transforms as
  \begin{eqnarray}
  & & \delta A_{a_1 ..a_4 , N_4} = a_{a_1 ...a_4 , N_4} - {1 \over
  2} A_{a_1 a_2 , N_2} a_{a_3 a_4 , M_2} f^{N_2 M_2}{}_{N_4} -
  A_{a_1 a_2 a_3 , N_3} a_{a_4 , N_1} f^{N_3 N_1}{}_{N_4} \nonumber
  \\
  & & \qquad \quad -{1 \over 4 ! } A_{a_1 , N_1} A_{a_2 , M_1}
  A_{a_3 , P_1} a_{a_4 , Q_1} f^{N_1 N_3}{}_{N_4} f^{M_1 N_2}{}_{N_3} f^{P_1
  Q_1}{}_{N_2} \nonumber \\
  & & \quad \qquad
  - {1 \over 4} A_{a_1 a_2 ,N_2} A_{a_3 ,N_1} a_{a_4 ,M_1}  f^{N_2 M_2}{}_{N_4}  f^{N_1
  M_1}{}_{M_2} - g \Lambda_{M_1}
  X_4^{M_1}{}_{N_4}{}^{P_4} A_{a_1 ...a_4, P_4} \ ,
  \end{eqnarray}
and the 5-form transforms as
   \begin{eqnarray}
   & & \delta A_{a_1 ..a_5 , N_5} = a_{a_1 ...a_5 , N_5}
   - A_{a_1 ...a_3 , N_3} a_{a_4 a_5, N_2} f^{N_3 N_2}{}_{N_5} -
   A_{a_1 ...a_4 ,N_4} a_{a_5 , N_1} f^{N_4 N_1}{}_{N_5} \nonumber
   \\
   & & \quad \qquad
   - {1 \over
  2} A_{a_1 a_2 , N_2} A_{a_3 a_4 ,M_2 } a_{a_5 ,N_1}  f^{N_2 N_3}{}_{N_5} f^{M_2
  N_1}{}_{N_3}\nonumber \\
  & & \quad \qquad
  -{1 \over 5 ! } A_{a_1 , N_1} A_{a_2 , M_1}
  A_{a_3 , P_1} A_{a_4 , Q_1} a_{a_5 , R_1} f^{N_1 N_4}{}_{N_5} f^{M_1 N_3}{}_{N_4} f^{P_1 N_2}{}_{N_3} f^{Q_1
  R_1}{}_{N_2} \nonumber \\
  & & \quad \qquad
  - {1 \over 3!} A_{a_1 a_2 ,N_2} A_{a_3 ,N_1} A_{a_4 ,M_1} a_{a_5 ,P_1} f^{N_2 N_3}{}_{N_5} f^{N_1 M_2}{}_{N_3}  f^{M_1
  P_1}{}_{M_2}\nonumber \\
  & & \quad \qquad - g \Lambda_{M_1}
  X_5^{M_1}{}_{N_5}{}^{P_5} A_{a_1 ...a_5, P_5} \ .
  \end{eqnarray}
The parameters $a$ are given in eq. (\ref{massiveidentification}) in
terms of the gauge parameters $\Lambda$. The reader can easily
evaluate the gauge transformations for the higher rank fields.

\section{Extended spacetime in four dimensions}
In this appendix we will consider the four dimensional maximal
gauged supergravities using a non-linear realisation of
$E_{11}\otimes_s l_1$. This closely follows the similar derivation
of the maximal gauged supergravities in five dimensions given in
reference \cite{fabiopeterextendedspacetime} to which we refer for
the details of how this method works.

\subsection{The $l_1$ multiplet in four dimensions}
The $l_1$ multiplet can be thought of as the $E_{11}$ representation
that contains the momentum generator $P_c$ as its lowest component.
The $l_1$ multiplet is the representation of $E_{11}$ with  highest
weight $\lambda_1$, where $\lambda_1$ is the fundamental weight
associated with node 1 of the Dynkin diagram of $E_{11}$.  By
definition it  satisfies the relation $(\lambda_1, \alpha_i) =
\delta_{1i}$, where $\alpha_i$ is the simple root associated with
node $i$ on the Dynkin diagram of $E_{11}$. For our derivation we
will need the $l_1$ multiplet of $E_{11}$ suitable to four
dimensions at low levels.
\par
\begin{figure}[h]
\begin{center}
\begin{picture}(350,70)
\multiput(-30,10)(40,0){7}{\circle{10}}
\multiput(250,10)(40,0){3}{\circle{10}} \put(370,10){\circle{10}}
\multiput(-25,10)(40,0){10}{\line(1,0){30}}
\put(290,50){\circle{10}} \put(290,15){\line(0,1){30}}
\put(-32,-8){$\star$} \put(8,-8){$1$} \put(48,-8){$2$}
\put(88,-8){$3$} \put(128,-8){$4$} \put(168,-8){$5$}
\put(208,-8){$6$} \put(248,-8){$7$} \put(288,-8){$8$}
\put(328,-8){$9$} \put(365,-8){$10$} \put(300,47){$11$}
\end{picture}
\caption{\sl The $E_{12}$ Dynkin diagram. \label{E12}}
\end{center}
\end{figure}
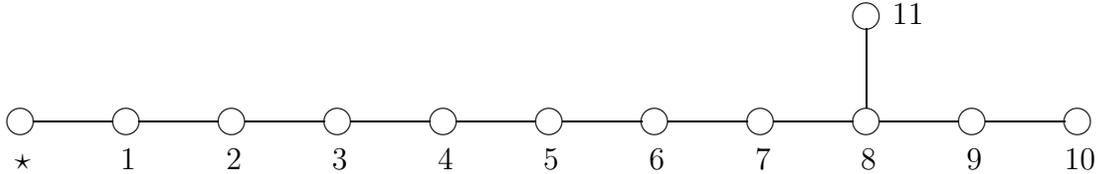

The most straightforward way to find the components of the $l_1$
multiplet as it occurs in four dimensions at low levels is just to
take the $l_1$ multiplet in eleven dimensions
\cite{peterl1multiplet}, carry out the dimensional reduction to four
dimensions by hand, and then collect the result into representations
of the internal symmetry group $E_7$. A more sophisticated method is
to realise that the $l_1$ representation of $E_{11}$ can be obtained
by considering  the adjoint representation of $E_{12}$. The Dynkin
diagram of $E_{12}$ is just that of $E_{11}$, but with one node, the
starred node, added with one line attached to node one as in fig.
\ref{E12}. To find the $l_1$ representation suitable to eleven
dimensions we decompose the adjoint representation of $E_{12}$  into
representations of the $E_{11}$ obtained by deleting the starred
node in the $E_{12}$ Dynkin diagram and keeping only the level one
generators; by level we mean the level associated with  node one
\cite{peterl1multiplet}. Clearly, as the commutation relations
respect the level we must find a representation of $E_{11}$ and it
is in fact the $l_1$ representation. To find the $l_1$
representation in four dimensions one  then carries out the
decomposition $GL(4,\mathbb{R}) \otimes E_7$ corresponding to
deleting in addition node four.

Following either method the  low level elements of the $l_1$
multiplet in four dimensions are found to be  given by
\cite{peteroriginofbranecharges,peterpaul}
  \begin{eqnarray}
& & P_a \ ({\bf 1}) \quad  Z^M \ ({\bf 56}) \quad  Z^{a,\alpha} \
({\bf 133 \oplus 1} ) \quad  Z^{ab,N\beta} \ ({\bf 912 \oplus
56\oplus 1}) \nonumber \\
& &  Z^{abc,[\delta\epsilon]}\ ({\bf 8645\oplus 1539\oplus133
\oplus1})\quad  \cdots
\end{eqnarray}
The indices $a,b,\ldots=0,1,2,3 $ transform under $GL(4,\mathbb{R})
$ in the obvious way while the numbers in brackets indicate the
dimensions of the $E_7$ representations to which the charges belong.

The commutators of $E_{11}$ appropriate to four dimensions  are
given in section four. The commutators of the $E_7$ generators with
the $l_1$ generators are determined by the $E_7$ representation that
the charges belong to and are given by
  \begin{eqnarray}
  & & [R^\alpha, Z^M] = (D^\alpha)_N{}^M Z^N \nonumber \\
  & &  [R^\alpha, Z^{a,\beta}] =
  f^{\alpha\beta}{}_\gamma Z^{a,\gamma} \nonumber \\
  & & [R^\alpha, Z^{ab,N\beta}] = (D^\alpha)_M{}^N Z^{ab,M\beta} +
  f^{\alpha\beta}{}_\gamma Z^{ab,N\gamma} \nonumber \\
  & &  [R^\alpha, Z^{abc,[\delta\epsilon]}] = f^{\alpha\delta}{}_\beta
  Z^{abc,[\beta\epsilon]} + f^{\alpha\epsilon}{}_\beta
  Z^{abc,[\delta\beta]} \quad .
  \end{eqnarray}
\par
The remaining commutators between $E_{11}$ generators and those of
the $l_1$ multiplet can be deduced from their $E_{12}$ origin,  or
just writing down relations compatible with the level assignments,
$GL(4,\mathbb{R})$ character, and using the Jacobi identities. We
may define the way the $l_1$ generators occur in the
$E_{11}\otimes_sl_1$ algebra by the relations
  \begin{eqnarray}
  &  & [R^\alpha,P_a] = 0, [R^{a,N}, P_b] = \delta^a_b Z^N \quad ,
  \qquad
  [R^{ab,\alpha}, P_c] = 2 \delta^{[a}_c Z^{b],\alpha} \nonumber \\
  & & [R^{abc,N\beta}, P_d] = 3 \delta^{[a}_d Z^{bc],N\beta} \quad ,
  \qquad
  [R^{abcd,[\alpha\beta]},P_e] = 4 \delta^{[a}_e
  Z^{bcd],[\alpha\beta]} \quad .
  \end{eqnarray}
The normalisation of the $l_1$ generators is then  fixed by the
choice of coefficients on the right hand side. The fact that the
representation of the charges and the $E_{11}$ generators coincide
on each side of these equations is a consequence of the relationship
that exists between the $l_1$ representation and the adjoint
representation of $E_{11}$ \cite{peteroriginofbranecharges}.
Physically this is the usual relationship between fields and the
charges to which they couple in the Wess-Zumino term of a brane
action.
\par
The remainder of the commutators may be fixed through the  Jacobi
identities.  For example, let us   consider the Jacobi identity
involving $[R^{a,M},[R^{b,N}, P_c]]$. We find, using the $E_{11}$
commutators of eq. (4.9),  that
 \begin{equation}
  [R^{a,M},Z^N] = -D^{MN}_\alpha Z^{a,\alpha} \quad .
  \end{equation}
Now, for convenience we define $Z^{ab,N\beta} = S^{N\beta}_A
Z^{ab,A} $, and the Jacobi relation $[R^{a,M}, [R^{bc,\alpha},
P_d]]$ implies that
 \begin{equation}
[R^{ab,\alpha}, Z^M] = -  Z^{ab,M\alpha} \quad {\rm and} \quad
[R^{a,M},Z^{b,\alpha}] = - Z^{ab,M\alpha} \quad .
\end{equation}
The Jacobi relation  $[R^{ab,\alpha},[R^{cd,\beta}, P_e]]$ leads to
 \begin{equation}
 [R^{ab,\alpha}, Z^{c,\beta}] = Z^{abc,[\alpha\beta]} \quad .
 \end{equation}
The final Jacobi identity $[R^{abc,A}, [R^{d,M}, P_e]]$ implies both
 \begin{equation}
[R^{a,M},Z^{bc,A}] = -c^{MA}_{\alpha\beta} Z^{abc,[\alpha\beta]}
\quad {\rm  and} \quad [R^{abc, A}, Z^M] = -c^{MA}_{\alpha\beta}
Z^{abc,[\alpha\beta]} \quad .
\end{equation}
\par
Thus the commutators between the $E_{11}$ generators and those of
$l_1$ at low levels are given in equations (C.3) to (C.7).

\subsection{Construction of the   four dimensional  gauged maximal
supergravities}

The field strengths and gauge transformations of the massless theory
follow in a straightforward way from the $E_{11}$ algebra, they are
essentially the Cartan forms subject to the appropriate
anti-symmetrisation. They are given by the general formula in eq.
(\ref{2.41}) and more explicitly in section four by setting $g=0$.
We will now derive the field strengths  for the gauged theory
following closely the argument given in
\cite{fabiopeterextendedspacetime} for the five-dimensional case.
\par
We begin by choosing the group element of the non-linear realisation
of $E_{11}\otimes_s l_1$ to be
 \begin{equation}
g = g_{l_1} g_A = e^{x^a P_a} e^{y\cdot Y} e^{A(x) \cdot R} \quad ,
\label{groupelementappC}
 \end{equation}
where
  \begin{equation}
    e^{A(x) \cdot R}=  e^{A_{a_1
  ...a_4, \alpha\beta} R^{a_1 ... a_4, \alpha\beta}} ...
  e^{A_{a
  , M} R^{a , M}} e^{\phi_{\alpha} R^{\alpha}}
  \end{equation}
and
 \begin{equation}
e^{y \cdot Y} = e^{y_M (Z^M + gT^M)} e^{y_{a,\alpha}
(Z^{a,\alpha}+gT^{a,\alpha})} e^{y_{ab,M\alpha} (Z^{ab,M\alpha} + g
T^{ab,M\alpha})} \cdots \quad .
  \end{equation}
The symbols in this latter equation are defined by
 \begin{eqnarray}
& & T^M = \Theta^M_\alpha R^\alpha ,\ \ \ T^{a,\alpha} = W^\alpha_M
R^{a,M} \nonumber \\
& & T^{ab,M\alpha} = {1\over 3} (\Theta^M_\beta
f^{\beta\alpha}{}_\gamma - 2 W^\alpha_N D^{MN}_\gamma) R^{ab,\gamma}
,\ \ \ T^{abc,[\epsilon\delta]} = W^{\delta}_M R^{abc,M\epsilon},
\ldots  \quad .
\end{eqnarray}
The coefficients in these expressions are taken from the results of
section 4. The tensor $W^\alpha_M$ defined here differs from the
tensor denoted in the same way in section 4 by a factor of 2, and
indeed it is defined as
  \begin{equation}
  W_M^\alpha = - {1 \over 2} \Theta^\alpha_M
  \end{equation}
which differs from eq. (4.25). Again, this coefficient is taken to
reproduce the results of section 4.
\par
At first sight it may appear that the above group element contains
all the generators of the $l_1$ multiplet, but this is not the case.
In fact the coordinates $y$ obey projections conditions that mean
that part of the $l_1$ multiplet is absent and plays no role in the
calculation. As discussed in reference
\cite{fabiopeterextendedspacetime}, the part of $l_1$ which is
present is the image of  a map from $E_{11}$ into the $l_1$
representation. It is argued \cite{fabiopeterextendedspacetime} that
demanding certain natural properties of this map leads to the
constraints on $\Theta^M_\alpha$, etc  found as a consequence of
Jacobi identities of the deformed algebra in this paper. These
constraints on $\Theta$ etc are reflected in the projections arising
from the $y$ coordinates. One may hope that such an argument may fix
the coefficients in eqs. (C.11) and (C.12).
\par
For reasons to do with the preservation of the form of the group
element under the action of the group and the required compensating
transformations  the field strengths of the gauged supergravity are
not given in a simple way by the Cartan forms. However, as explained
in reference \cite{fabiopeterextendedspacetime}  we can  calculate
the variation of the fields in the usual way by  taking a group
element $g_0$ and considering its effect on the coset representative
$g_0 g \longrightarrow g'$ and find the field strengths by demanding
that they be invariant under these transformations. In particular
let us consider $g\to g_0 g$ taking $g_0 = e^{b \cdot Z}$ where the
parameter $b$ obeys the same constraints as the $y$ coordinates.  As
discussed in reference \cite{fabiopeterextendedspacetime} $e^{b\cdot
Z} e^{y\cdot Y} = e^{y' \cdot Z} e^{-g b\cdot T}$. However, as the
final field strengths do not depend on the $y$ coordinates  we need
only  calculate the the effect of $e^{-g b\cdot T}$ on the $e^{A
\cdot R}$ term and do not require to know how the $y$ coordinates
change.

Let us first consider  the  transformation $g \rightarrow g_0 g$
with $g_0 = e^{b_M Z^M}$ this leads to the factor $e^{-g
b_M\Theta^M_\alpha R^\alpha}$ acting on the $E_{11}$ coset
representative which  is just the same as an $E_{11}$ transformation
of the ungauged theory, but with parameter
$a_\alpha=b_M\Theta^M_\alpha$. At the next level we consider the
transformation $g_0 = e^{b_{a,\alpha} Z^{a,\alpha}}$, which results
in the  factor $e^{-g b_{a,\alpha} T^{a,\alpha}} = e^{-g
b_{a,\alpha} W^\alpha_M R^{a,M}}$ which is just an $E_{11}$
transformation with the parameter $a_{a,M} = -g b_{a,\alpha}
W^\alpha_M$. Similarly the effect of $g_0 = e^{b_{ab,M\alpha}
Z^{ab,M\alpha}}$ is equivalent to an $E_{11}$ transformation with
parameter $a_{ab,\alpha} = - g b_{ab,M\beta} {1 \over 3}
(\Theta^M_\gamma f^{\gamma\beta}{}_\alpha - 2W^\beta_N D^{MN}_
\alpha)$ and $g_0 = e^{b_{abc,[\epsilon\delta]}
V^{abc,[\epsilon\delta]}}$ is equivalent to an $E_{11}$
transformation with parameter $a_{abc,A} =
-gb_{abc,[\epsilon\delta]} W^\delta_M S^{M\epsilon}_A$. The result
of all  these  transformations on the fields is given by
  \begin{eqnarray}
  & & \delta A_{a,M} = -g b_P \Theta^P_\alpha (D^\alpha)_M{}^N
  A_{a,N} \nonumber \\
   &&
  \delta A_{a_1 a_2,\alpha} = -g b_P \Theta^P_\beta f^{\beta\alpha}{}_\gamma
  A_{a_1 a_2, \alpha} \nonumber \\
  && \delta A_{a_1\dots a_3,M\alpha} = -g
  b_P \Theta^P_\beta (D^\beta_M{}^N A_{a_1 \dots a_3, N\alpha} +
  f^{\beta\gamma}{}_\alpha A_{a_1 \dots a_3, M\gamma} ) \quad .
 \end{eqnarray}
\par
We now consider an $E_{11}$  transformations for the gauged theory,
that is we take $g_0 = e^{a \cdot R}$ to act on the group element of
eq. (\ref{groupelementappC});
  \begin{equation}
  g_0 g = e^{a\cdot R} e^{x^a P_a} e^{y \cdot Y} e^{A \cdot R} =
  e^{x^a P_a + [a \cdot R, x^a P_a]} e^{y \cdot Y + [a \cdot R, y
  \cdot Y]} e^{a \cdot R} e^{A \cdot R} \quad . \end{equation}
As discussed in \cite{fabiopeterextendedspacetime} the
transformations resulting from the $[R,Y]$ in the second term do not
affect the dynamics as the final dynamics does not depend on the $y$
coordinates and so we may ignore this term. The final $a\cdot R$
term has the same effect on the $E_{11}$ fields as the equivalent
transformation in the ungauged theory. In the first factor we find
$e^{x^a P_a + [a \cdot R, x^a P_a]}$ which leads to higher
generators in the $l_1$ multiplet. For example, if we  take
$g_0=e^{a_{a,M} R^{a,M}}$ we find it leads to the term $e^{x^b P_b}
e^{x^a a_{a,M} Z^M}$. This latter factor then acts like a $l_1$
transformation on the rest of the coset representative and, as
discussed above, it  leads to an $x$-dependent  $E_{11}$
transformation.
\par
As a result  we can combine the effect of the $l_1$ transformations
and the $x$ dependence $E_{11}$ transformation together by taking an
$l_1$ transformation with the parameter
  \begin{eqnarray}
  & &
  b_M(x) = b_M + x^a a_{a,M} \nonumber \\
  & &  b_{a,\alpha}(x) = b_{a,\alpha} + x^b
  a_{ba,\alpha} \nonumber \\
  & & b_{ab,M\alpha}(x) = b_{ab,M\alpha} + x^c a_{cab,M\alpha}
  \nonumber \\
  & &  b_{abc,[\delta\epsilon]} = b_{abc,[\delta\epsilon]} + x^d
  a_{dabc,[\delta\epsilon]} \quad . \end{eqnarray}
As noted above we have in addition the usual $E_{11}$
transformations with the $a$ parameters which are related to the
above parameters by
  \begin{eqnarray}
  & &
  a_{a,M} = \partial_a b_M(x) \quad \qquad  a_{ab,\alpha} = {1 \over
  2}
  \partial_a b_{b,\alpha}(x) \nonumber \\
  &  &  a_{abc,M\alpha} = {1 \over 3}
  \partial_a b_{bc,M\alpha}(x) \quad \qquad  a_{abcd,[\delta\epsilon]} = {1
  \over 4} \partial_a b_{bcd,[\delta\epsilon]}(x)  \quad
  .\end{eqnarray}
Thus all the transformations can be expressed in terms of the
$x$-dependent parameters $b(x)$.
\par
The resulting transformations of the fields are given by
 \begin{eqnarray}
  & & \delta A_{a,M} =
  \partial_a b_M(x)
  - g b_M(x) A_{a,N} X^{MN}{}_P - g b_{a,\alpha}(x) W^\alpha_M
  \nonumber \\
  & & \delta A_{a_1 a_2, \gamma} = {1 \over 2}
\partial_{[a_1}b_{a_2],\gamma}(x) + {1 \over 2} \partial_{[a_1|}
b_M(x)A_{|a_2],N}(D_\gamma)^{MN} - g b_M(x) \Theta^M_\alpha
f^{\alpha\beta}{}_\gamma A_{a_1 a_2, \beta} \nonumber \\
& & \quad \qquad  - {1 \over 2} g b_{a_1,\alpha}(x) W^\alpha_M
A_{a_2,N} D^{MN}_\gamma  - {1 \over 3} g b_{a_1 a_2,M\alpha}(x)
(\Theta^M_\beta f^{\beta\alpha}{}_\gamma -
2 W^\alpha_N D^{MN}_\gamma) \nonumber \\
& &  \delta A_{a_1 a_2 a_3, M\alpha} = {1\over 3}
\partial_{[a_1}b_{a_2a_3],M\alpha}(x) +
\partial_{[a_1|}b_M(x)A_{|a_2a_3],\alpha} \ - {1 \over 6}
A_{[a_1|,M}A_{|a_2|,N}\partial_{|a_3]}b_P(x)(D_\alpha)^ {NP}
\nonumber \\
& & \quad \qquad  -gb_N(x) \Theta^N_\beta (D^\beta)_M{}^P A_{a_1
\dots a_3, P\alpha} -gb_N(x) \Theta^N_\beta f^{\beta\gamma}{}_\alpha
A_{a_1 \dots a_3, P \gamma}  -g W^\beta_M b_{a_1,\beta}(x) A_{a_2
a_3, \alpha} \nonumber \\
& & \quad \qquad  -{1 \over 6} g W^\beta_N b_{a_1,\beta}(x)
A_{a_2,P} A_{a_3,M} D^{NP}_\alpha -g b_{a_1 \dots a_3,
[\beta\alpha]}(x) W^{\beta}_M
 \nonumber \\
 & & \delta A_{a_1\dots a_4,[\delta\epsilon]} = {1 \over
4}c^{M,N\beta}_{[\delta\epsilon]}\partial_{a_1} b_M(x) A_{a_2\dots
a_4,N\beta} - {1 \over 24} D^{NP}_\alpha
c^{M,Q\alpha}_{[\delta\epsilon]} A_{a_1,M} A_{a_2,Q} A_{a_3,N}
\partial_{a_4}b_P(x) \nonumber \\
& & \quad \qquad  + {1 \over 4} D^{NP}_\delta
A_{a_1,N}\partial_{a_2}b_P(x) A_{a_3 a_4, \epsilon} + {1 \over 4}
\partial_{a_1}b_{a_2,\delta}(x) A_{a_3 a_4,\epsilon} + {1 \over 4}
\partial_{a_1}b_{a_2 \dots a_4,[\delta\epsilon]}(x) \quad .
\end{eqnarray}
\par
The  field strengths are just the objects which are covariant under
the above  $E_{11}\otimes l_1$  transformations which at order $g^0$
agree with those of the massless theory.  Thus the order $g^1$
variation of $F^{(0)}$, {\it i.e.} the variation of the massless
field strengths,  must cancel the order $g^0$ variation of the order
$g^0$ variation of the order $g^1$ part of the field strength,
$F^{(1)}$. This computation implies that the field strengths of the
gauged theory are given by
  \begin{eqnarray}
  & &  F_{a_1 a_2, M} = 2 \partial_{[a_1} A_{a_2 ],M} + g
  X^{NP}{}_M A_{[a_1 N} A_{a_2 ] P} + 4 g A_{a_1 a_2 \alpha}
  W^\alpha_M \nonumber \\
  & &  F_{a_1 \dots a_3, \alpha} = 3\partial_{[a_1} A_{a_2
  a_3 ],\alpha} + {3 \over 2}(\partial_ {[a_1} A_{a_2,M})
  A_{a_3],N} (D_\alpha)^{MN} \nonumber \\
  & & \quad \qquad  + {1 \over 2} g A_{[a_1, M} A_{a_2, N}
  A_{a_3 ], P} X^{[MN}_Q D_ \alpha^{P]Q} + 6 g A_{[a_1
  a_2, \beta} A_{a_3 ], M} W^\beta_N D_ \alpha^{MN}\nonumber \\
  & & \quad \qquad  + 3g(\Theta^M_\gamma f^{\gamma\beta}{}_\alpha +
  2 g
  W^\beta_N D_\alpha^{MN})A_{a_1 \dots a_3, M\beta} \nonumber \\
  &   &  F_{a_1 \dots a_4, T\eta} = 4\partial_{[ a_1} A_{a_2
  \dots a_4 ], T\eta} - 4(\partial_{[ a_1} A_{a_2 a_3 , \eta})
  A_{a_4],T} \ - {2 \over 3} (\partial_{[a_1}
  A_{a_2 ,N})A_{a_3,P}A_{a_4],T} (D_\eta)^{NP}\nonumber \\
  & & \quad \qquad  -16g W^{\alpha}_T A_{a_1 \dots a_4, [\alpha
  \eta]}   +4g (\Theta^M_\beta f^{\beta \alpha}{}_\eta +  2 g W^\alpha_P
  D^{MP}_ \eta) A_{[a_1 \dots a_3, M\alpha} A_{a_4 ] T} \nonumber \\
  & & \quad \qquad -4g W^\alpha_M A_{[a_1 a_2, \alpha} A_{a_3
  a_4 ], \eta} +4gW^\alpha_P D^{PM}_\eta A_{[a_1 a_2, \alpha}
  A_{a_3, M} A_ {a_4 ], T} \nonumber \\
  & & \quad \qquad + {1 \over 6} g X^{MN}{}_R D_\eta^{RP} A_{[a_1,M}
  A_{a_2, N} A_ {a_3, P} A_{a_4 ], T} \quad .
  \end{eqnarray}
The field strengths and the gauge transformations agree with those
found in section 4, as one can see comparing eqs. (4.31) and (4.32)
with eqs. (C.18) and (C.17), if we identify the gauge parameters of
section 4 as
  \begin{equation}
\Lambda_M = b_M(x) \ \ \  \Lambda_{a,\alpha} = {1 \over 2}
b_{a,\alpha} \ \ \ \Lambda_{ab,M\alpha} = {1 \over 3} b_{ab,M\alpha}
\ \ \ \Lambda_{abc,[\alpha\beta]} = {1 \over 4}
b_{abc,[\alpha\beta]} \quad .
\end{equation}

\end{appendix}

\end{document}